\documentclass[a4paper,11pt]{article}
\usepackage{jheppub} % for details on the use of the package, please see the JINST-author-manual

\usepackage{mathtools}
\usepackage{bm}

\preprint{CERN-TH-2025-142}

\title{\boldmath Gravitational Wave Scattering on Magnetic Fields
}

\author[1]{\small Valerie Domcke,}
\author[2]{\small Camilo Garc\'ia-Cely,}
\author[1]{\small Sung Mook Lee,}

\affiliation[1]{\footnotesize Theoretical Physics Department, CERN, 1 Esplanade des Particules, CH-1211 Geneva 23, Switzerland}
\affiliation[2]{Instituto de F\'isica Corpuscular (IFIC), Consejo Superior de Investigaciones
Cient\'ificas (CSIC) and Universitat de Val\`encia,  C/ Catedratico Jose Beltran 2, E-46980 Paterna, Spain}

\emailAdd{valerie.domcke@cern.ch}
\emailAdd{camilo.garcia@ific.uv.es}
\emailAdd{sungmook.lee@cern.ch}

\abstract{
The conversion of gravitational to electromagnetic waves in the presence of background magnetic fields is known as the inverse Gertsenshtein effect, analogous to the Primakoff effect for axions.
Rephrasing this conversion as a classical electrodynamics problem in the far-field regime of a magnetized region, we derive the angular distribution of the intensity and polarization of the emitted electromagnetic waves.
We discuss the interplay of the internal structure of the magnetic field, the polarization of the gravitational wave and the scattering angle, demonstrating for example that a dipolar field can convert an unpolarized stochastic gravitational wave background into polarized electromagnetic emission, with peak emission intensity along the equator.
We moreover outline how to incorporate medium effects in this framework, necessary for a realistic 3D description of gravitational wave to photon conversion in the magnetosphere of neutron stars.
}

\begin{document}
\maketitle
\flushbottom

\section{Introduction}

\begin{figure}
    \centering
    \includegraphics[width=0.75\linewidth]{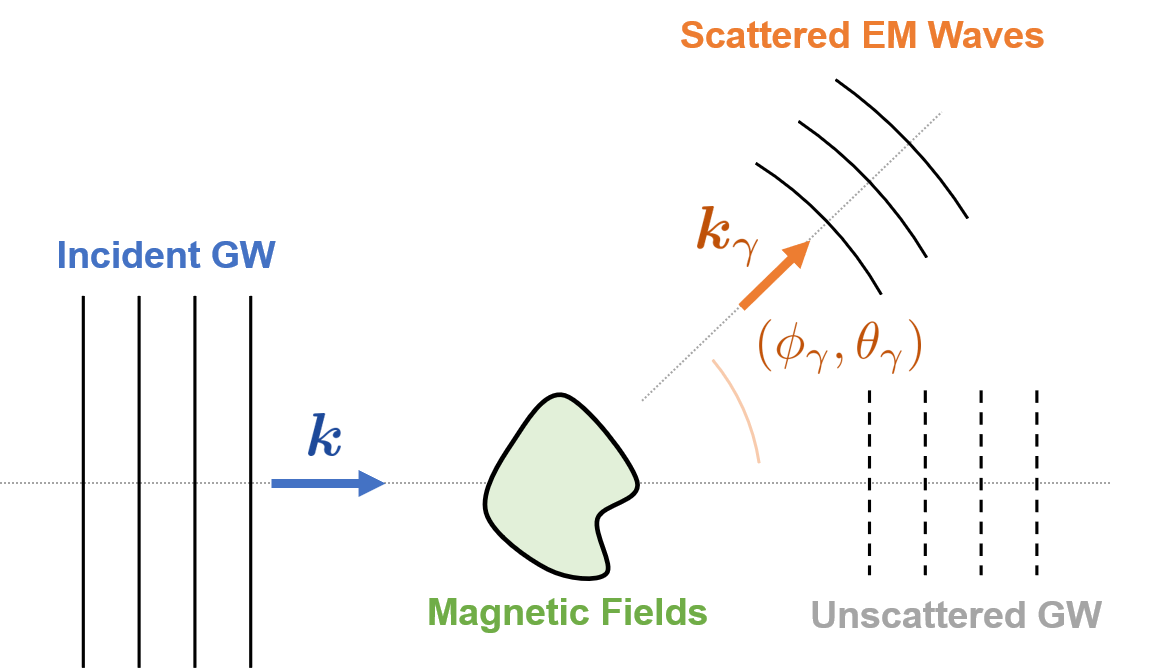}
    \caption{Schematic diagram of the emission of electromagnetic waves sources by a GW passing through a magnetized region.}
    %for the general concepts of the gravitational wave scattering on magnetic fields resulting in the scattered EM waves.}
    \label{fig:concept}
\end{figure}

Classical electromagnetism 
in general relativity predicts the conversion of gravitational to electromagnetic (EM) radiation and vice versa in the presence of background magnetic fields, known as the Gertsenshtein effect~\cite{Gertsenshtein}.
This observation has been used to search for and set limits on gravitational waves (GWs), both in laboratory setups~\cite{Ejlli:2019bqj,Ringwald:2020ist,Berlin:2021txa,Berlin:2023grv,DeMiguel:2023nmz,Domcke:2024eti} and astrophysical environments~\cite{Domcke:2020yzq, Ito:2023fcr, Ito:2023nkq, Dandoy:2024oqg, McDonald:2024nxj,Hong:2024ofh,Kushwaha:2025mia,Matsuo:2025blj}.
This process shares many similarities with the Primakoff effect~\cite{Primakoff:1951iae}, the analogous conversion of axions to photons~\cite{Raffelt:1987im}.
The latter has been used to set strong bounds on axion dark matter~\cite{Darling:2020uyo,Darling:2020plz,Battye:2021yue,Foster:2022fxn,Battye:2023oac} supported by recent theoretical refinements that account for three-dimensional geometric effects~\cite{McDonald:2023shx,McDonald:2023ohd,Gines:2024ekm,McDonald:2024uuh, Beutter:2018xfx}.
This progress has been key to performing robust searches for axions in complex astrophysical environments \cite{Mirizzi:2006zy,Pshirkov:2007st,Wang:2015dil,Masaki:2017aea,Hook:2018iia,Foster:2020pgt,Buckley:2020fmh,Witte:2021arp,Millar:2021gzs,DeMiguel:2022ojb,Dessert:2022yqq,Caputo:2024oqc,Smarra:2024vzg,Fan:2025ixw,Tercas:2025paf}.
The goal of the present paper is to develop similar tools for describing the conversion of GWs to EM radiation in realistic, three-dimensional magnetic field configurations.
Starting from Maxwell's equations in curved spacetime~\cite{Einstein:1916cb, Landau:1975pou}, the question can be reformulated as a classical scattering problem in electromagnetism, see Fig.~\ref{fig:concept}.
This allows us to compute the exact EM wave function sourced by a GW in the presence of a magnetic field.
Evaluating this expression in instructive toy examples of  magnetic field configurations provides insights into the phenomenology of GWs scattering off magnetized structures.
In the limit of a uniform magnetic field, the EM wave emitted in forward direction yields the well-known expression for the GW-to-photon conversion probability~\cite{Raffelt:1987im}, while the back-scattered EM wave is suppressed -- though non-zero.
When GWs scatter off compact magnetized regions -- such as those surrounding neutron stars -- they
induce EM fields with specific angular patterns in intensity and polarization.\footnote{
The role of polarization in the Gertsenshtein effect has recently received renewed interest~\cite{Kushwaha:2024uhh,Li:2025eoo,Chiba:2025odu,Kushwaha:2025mia}.
These papers focus on maximally polarized GWs (pure states in the language below), and discuss the evolution of the GW and EM polarization along the line of sight, as well as a possible impact on the CMB. 
Our study differs in that we solve the full three-dimensional problem, revealing a much richer polarization phenomenology -- most notably the production of polarized EM radiation from an isotropic, unpolarized SGWB.}
The intensity scales with the transverse component of the magnetic field traversed, and the differential cross-section reflects the tensor structure and polarization of the GW.
In general, the polarization of the EM waves depends on the polarization of the incoming GW and the configuration of the background magnetic fields.
In the case of unpolarized GWs, the polarization of the outgoing EM wave only depends on the angle between the incoming GW and outgoing EM wave.
This results in a distinct polarized EM emission when e.g.\ a neutron star is exposed to an isotropic stochastic gravitational wave background (SGWB).

Including plasma effects is crucial when evaluating the GW to EM radiation conversion probability in dense astrophysical environments, as they lead to a non-zero effective photon mass  and thus spoil the resonant conversion between (strictly massless) GWs and EM waves~\cite{Raffelt:1987im}.
We propose a formalism to include an effective photon mass which allows us to solve for the full EM wave function, reproducing limiting cases discussed in the literature within a more general framework.
This work complements earlier studies on GW to photon conversion in neutron stars and related systems.
Ref.~\cite{DeLogi:1977qe} developed a formalism to study the intensity of the EM fields generated in the far field regime, based on  scattering amplitudes, which we briefly summarize in Appendix~\ref{app:S-matrix}.
More recently, Refs.~\cite{Ito:2023fcr} and \cite{Dandoy:2024oqg} study the 1D conversion probability along the GW propagation direction, suppressing the angular dependence of the background magnetic fields.
Ref.~\cite{McDonald:2024nxj} addresses the full 3D problem of the neutron star using ray-tracing techniques, focusing on isotropic SGWBs and taking into account only the conversion on resonance.
On the other hand, the focus of this paper is on the conceptual development of the formalism, highlighting the relevance and the phenomenology of various aspects through the study of simplified toy models.
These tools and insights will contribute to addressing the full neutron star problem, including geometric and plasma effects; however, a complete treatment lies beyond the scope of the present paper.
The remainder of this paper is organized as follows.
Section~\ref{sec:basics} introduces the Gertsenshtein effect as well as our conventions to describe the GWs and EM waves.
Section~\ref{sec:Magnetic_Domain_Model} introduces a very simple toy model of a GW traversing a uniform magnetic field domain, and provides explicit solutions of Maxwell's equations for the reflected and transmitted EM wave.
A more powerful method, based on the well-known Green's function for scattering problems in EM, is introduced in Section~\ref{sec:Green's_Function_Method}, particularly well suited for describing the emission pattern of EM waves sourced by a GW passing through a localized magnetized volume. 
Finally, Section~\ref{sec:Towards_including_Medium_Effects} discusses how to include medium effects within this formalism, before we conclude in Section~\ref{sec:conclusions}. Several Appendices contain technical details. 
Appendix~\ref{app:notation} collects details on our notation and conventions, including our convention to visualize the polarization of electromagnetic and gravitational waves.
Appendix~\ref{app:Derivations of Magnetic Domain Wall Model} provides computational details leading to the results in Sections~\ref{sec:Magnetic_Domain_Model} and \ref{sec:Green's_Function_Method}, while Appendix~\ref{app:S-matrix} summarizes the $S$-matrix approach adopted in Ref.~\cite{DeLogi:1977qe}, demonstrating agreement with the results obtained in this work.
Appendix~\ref{app:axion} applies the formalism developed here to the case of relativistic axions, and finally Appendix~\ref{app:WKB} gives details on the WKB approximation used in Section~\ref{sec:Towards_including_Medium_Effects}.

\section{Gravitational Wave to Photon Conversion in Vacuum}
\label{sec:basics}

In curved spacetime, Maxwell's equations read~\cite{Landau:1975pou}
\begin{align}
 \nabla_{\nu} F^{\mu\nu} = j^\mu_\text{ext} /\sqrt{-g}\,, \quad \nabla_\nu F_{\alpha \beta} + \nabla_\alpha F_{\beta \nu} + \nabla_\beta F_{\nu \alpha} = 0 \,.
\end{align}
Considering small perturbations around flat spacetime due to GWs, we can expand the metric and the electromagnetic field strength tensor as $ g_{\mu\nu} = \eta_{\mu\nu} + h_{\mu\nu} $ and $ F_{\mu\nu} = \bar{F}_{\mu\nu} + F^{h}_{\mu\nu} $, respectively.
Here, $\bar{F}_{\mu\nu} $ is EM field in flat spacetime, sourced by the external current $j^\mu_\text{ext}$
\begin{align}
    \partial_{\nu} \bar{F}^{\mu\nu} = j_{\rm ext}^{\mu}.
\end{align}
On the other hand, $F^{h}_{\mu\nu}$ is the induced EM field which solves Maxwell's equations at linear order in the GW amplitude,
\begin{equation}
    \begin{aligned}
  & \partial_{\nu} F_{h}^{\mu\nu} = \partial_{\nu} \left( -\frac{h}{2} \bar{F}^{\mu \nu}  - \bar{F}^{\nu \alpha}
 {h^{\mu}}_{\alpha}  +   \bar{F}^{\mu \beta} 
 {h^{\nu}}_{\beta}  \right) \equiv j_{\rm eff}^{\mu} \,, \\
 & \partial_\nu F^h_{\alpha \beta} + \partial_\alpha F^h_{\beta \nu} + \partial_\beta F^h_{\nu \alpha} = 0 \,.
 \label{eq:perturbed_maxwell}
\end{aligned}
\end{equation}
As usual, the last equation is satisfied as long as $F^h_{\mu \nu} = \partial_\mu A^h_\nu - \partial_\nu A^h_\mu$.
We have introduced the effective current $j_\text{eff}^\mu$, proportional to the background EM field and the GW tensor, which acts as an effective source term in the inhomogeneous Maxwell equation~\cite{Herman:2020wao,Berlin:2021txa,Domcke:2022rgu,Domcke:2023bat}.
Throughout this paper, we will work with static magnetic fields while employing the transverse traceless gauge for the GWs. 
This corresponds to the situation where the GW frequency is larger than the eigenfrequencies of the system generating the magnetic fields, such that the latter cannot respond to the deforming force of the GW.
In particular, for neutron stars and their magnetosphere, with a speed of sound of about 1/3 of the speed of light, this holds if the magnetic fields can be treated as approximately constant over the wavelength of the GW, which will be the case for situations of interest.
Incidentally, this approximately coincides with the limit of the wave equation in which geometric optics can be applied, namely when the phase of GW and EM waves varies much more rapidly than their amplitude, that is, when the wavelength is much smaller than the characteristic scale of the background through which they propagate.
For a given magnetic field background, we can now compute the effective current according to Eq.~\eqref{eq:perturbed_maxwell}.
Throughout this paper, we will  work in Coulomb gauge, $ \nabla \cdot {\bm A} = 0 $. At first, we will be considering static background magnetic fields in the absence of electric fields and free charges, which implies a vanishing effective GW-induced charge $j_\text{eff}^0 = \rho_\text{eff} = 0$ and consequently $A_h^{0} = 0$.
Hence,
\begin{align}
 \Box A_{h}^{\mu} = - j_{\rm eff}^{\mu} \,.
 \label{eq:Ajeff}
\end{align}

This description holds as long as we can neglect the back-conversion of induced EM fields to GWs, i.e.\ neglect the impact of the induced EM fields on the spacetime metric, which is typically the case for all conversion volumes that do not span cosmological scales.\footnote{See~\cite{Raffelt:1987im} for a discussion of the resulting oscillation probabilities between GWs and EM fields.}
Moreover, we are so far omitting medium effects, which we will return to in Sec.~\ref{sec:Towards_including_Medium_Effects}.
With these preliminaries, the problem of determining the EM field sourced by a GW in the presence of background magnetic (or electric) fields reduces to solving the usual wave equations of the EM fields in the presence of the effective current $j^\mu_\text{eff}$.

In Sec.~\ref{sec:Magnetic_Domain_Model}, we begin by studying a toy model with a single uniform magnetic field domain, deriving the properties of the transmitted and reflected EM waves.
In Sec.~\ref{sec:Green's_Function_Method}, we then generalize this to provide a general expression for the induced EM field sourced by a GW traversing a compact magnetic field volume.
We will recover the standard results for the intensity of the EM wave emitted in forward direction, but focus in particular on the angular distribution of intensity and polarization of the emitted EM waves.
In this section we neglect the presence of any charged plasma, to which we will return in Sec.~\ref{sec:Towards_including_Medium_Effects}.

\paragraph{Effective Current.}
Let us start by setting some notation. For a GW propagating in the direction
\begin{align}
   \bm{k} = \omega \left( \sin \theta \cos \phi, \,
   \sin \theta \sin \phi, \,
   \cos \theta \right)\,, 
\end{align}
with the choice of two transverse unit vectors
\begin{align}
    \bm{v} = (-\sin \phi , \cos \phi , 0 ) \,, &&
    \bm{u} = (\cos \theta \cos \phi , \cos \theta \sin \phi , - \sin \theta ) \,,
\end{align}
we can express the GW in the transverse traceless (TT) frame as
\begin{align}
    h^{\rm TT}_{ij} = \left(  h_{+}  e_{ij}^{+}+ h_{\times}  e_{ij}^{\times} \right) e^{-i (\omega t - {\bm k} \cdot {\bm r}) } \label{eq:GW_TT}
    \,,
\end{align}
with our conventions for the polarization tensors specified in App.~\ref{app:notation}.
For induced EM fields propagating in $ {\bm k}_{\gamma} $ direction we define analogous transverse vectors $ \bm{u}_{\gamma} $ and $ \bm{v}_{\gamma}$ by replacing $ \theta \rightarrow \theta_{\gamma}$ and  $ \phi \rightarrow \phi_{\gamma}$, with $\theta_\gamma$ and $\phi_\gamma$ indicating the polar and azimuthal angle of wave vector.
Then, given a monochromatic GW as in Eq.~\eqref{eq:GW_TT}, the induced current can be written as
\begin{align}
  \bm{j}_{\rm eff} = \sum_{\lambda = +,\times} h_{\lambda} \left[ (i \bm{k} + \nabla) \times (e^{\lambda} \bm{B}_{0}) \right] e^{- i (\omega t - \bm{k} \cdot \bm{r} ) }\,,
\end{align}
where $(e^\lambda {\bm B}_{0})_i = e_{ij}^\lambda B^j$.
In the case that the magnetic field has a uniform field direction (here chosen to be along the $z$-axis) and a slowly varying profile (here along the $x$-axis), this expression simplifies to
\begin{align}
     \bm{j}_{\rm eff}  =  \frac{1}{\sqrt{2}}  e^{- i (\omega t - \bm k \cdot {\bm r} )} s_{\theta} \left[ i \omega B_{0}(x)
 \left(  h_{+} \bm{v} - h_{\times} \bm{u}  \right) - B_{0}^{\prime}(x) \, \bm{n} \times (h_{+} \bm{u} + h_{\times} \bm{v} ) \right] \,.
 \label{eq:jeff_uniform_field_direction}
\end{align}

\paragraph{Observables.}

The relevant observables are the intensity and polarization of the induced EM fields ${\bm E}_{h}$ and  ${\bm B}_{h}$
\begin{align}
    \bm{E}_{h} = - \frac{\partial \bm{A}_{h}}{\partial t} \,, && \bm{B}_{h} = \nabla \times \bm{A}_{h} \, .
\end{align}
They are characterized by the matrix $\langle E_{h i}^{*} E_{h j} \rangle $, 
where $ \langle \, \cdot \, \rangle $ refers to the ensemble average.
On the one hand, the intensity of the wave, $\mathcal{I}_{\gamma}$, is
given by its trace,  which coincides with the magnitude of the Poynting vector
\begin{align}
    \bar{\bm S}  = \frac{1}{4} \left( {\bm E}_{h}^{*} \times {\bm B}_{h} + {\bm E}_{h} \times {\bm B}_{h}^{*} \right) \,,
    \label{eq:Poynting}
\end{align}
i.e. $ \mathcal{I}_{\gamma}  = \vert \langle \bar{\bm S} \rangle \vert $.
On the other hand, the polarization properties of the induced EM fields are
best described by the density matrix
i.e.\ $ \rho_{ij} \equiv \langle E_{h i}^{*} E_{h j} \rangle / \mathcal{I}_{\gamma} $. By construction ${\rm tr} \rho =1$. Furthermore, being transverse to ${\bm k}_\gamma$, this matrix is completely determined by its components in the $\{{\bm u}_\gamma, {\bm v}_\gamma\}$ basis, which read
\begin{align}
    \rho = \frac{1}{2} \begin{pmatrix}
        1 + \xi_{3}  & \xi_{1} - i \xi_{2} \\
        \xi_{1} + i \xi_{2} & 1 - \xi_{3}
    \end{pmatrix}  \,,
    \label{eq:pol_matrix}
\end{align}
Here $ \xi_{i}  \in \mathbb{R}$ ($ i = 1,2,3$) are the usual Stokes parameters,\footnote{Often, $Q$, $U$ and $V$ in units of $\mathcal{I}_{\gamma}$ are used instead of $ \xi_{3}$, $ \xi_{1} $ and $ \xi_{2} $ in the literature, respectively. }
which satisfy $\xi_1^2+\xi_2^2+\xi_3^2 \leq 1$. 
More explicitly, we can write
\begin{align}
    \xi_3 = \frac{ \langle|{\bm A}_{h} \cdot {\bm u}_{\gamma}|^2\rangle - \langle|{\bm A}_{h} \cdot {\bm v}_{\gamma}|^2\rangle }{\langle|{\bm A}_{h} \cdot {\bm u}_{\gamma}|^2\rangle + \langle|{\bm A}_{h} \cdot {\bm v}_{\gamma}|^2\rangle} ,
    &&
    \xi_1 + i \xi_2 = \frac{ 2 \langle ({\bm A}_{h} \cdot {\bm u}_{\gamma})^* {\bm A}_{h} \cdot {\bm v}_{\gamma} \rangle }{\langle|{\bm A}_{h} \cdot {\bm u}_{\gamma}|^2\rangle + \langle|{\bm A}_{h} \cdot {\bm v}_{\gamma}|^2\rangle} \, .
    \label{eq:pol_components}
\end{align}
The Stokes parameters inherit their transformation properties under rotations from polarization vectors~\cite{Weinberg:2008} and are therefore often written in vector notation.
They provide a transparent way to describe the polarization,
as they allow for the construction of Lorentz invariants. In particular, $\xi_2$ and $\sqrt{\xi_1^2 + \xi_3^2}$ remain invariant under the choice of the orthonormal basis ${\bm u}_\gamma$ and ${\bm v}_\gamma$.
The parameter $\xi_2$ vanishes for linearly polarized light and reaches magnitude one for circular polarization.

Unpolarized radiation is characterized by all $\xi_i = 0$. In general, for partially polarized light, the degree of polarization $ p $ is given by $ \det \rho = (1-p^{2}) / 4$, i.e.
\begin{align}
    p = \sqrt{\xi_{1}^{2}+ \xi_{2}^{2}+ \xi_{3}^{2}} \,.
    \label{eq:deg_of_pol} 
\end{align}
Here $ p = 1 $ (hence $ \det \rho = 0 $) corresponds to a pure polarization state while $  p = 0 $ to unpolarized light (or a maximally mixed state).
See App.~\ref{app:notation} for our conventions on visualizing polarization vectors in two- and three-dimensional plots.

\begin{table}[t]
\centering
\setlength{\tabcolsep}{10pt}
\renewcommand{\arraystretch}{1.25}
\begin{tabular}{lccc}
\hline
\textbf{GW State} &
$\boldsymbol{\langle |h_{+}|^{2}\rangle}$ &
$\boldsymbol{\langle |h_{\times}|^{2}\rangle}$ &
\textbf{${\bm\xi}^{\rm GW}$} \\ \hline
Linear $\times$ (pure)            & $\neq 0$ & $0$                       & $(0,0,1)$ \\
Linear $+$ (pure)       & $0$      & $\neq 0$                  & $(0,0,-1)$ \\
Unpolarized (mixed)          & $\langle |h|^{2}\rangle$ & $\langle |h|^{2}\rangle$ & $(0,0,0)$ \\ \hline
\hline
\textbf{EM State} &
$\boldsymbol{\langle |{\bm A}_{h}\!\cdot{\bm u}_{\gamma}|^{2}\rangle}$ &
$\boldsymbol{\langle |{\bm A}_{h}\!\cdot{\bm v}_{\gamma}|^{2}\rangle}$ &
\textbf{${\bm\xi}$} \\ \hline
Linear vertical (pure)   & $\neq 0$ & $0$                       & $(0,0,1)$ \\
Linear horizontal (pure) & $0$      & $\neq 0$                  & $(0,0,-1)$ \\
Unpolarized (mixed)      & $\langle |A|^{2}\rangle$ & $\langle |A|^{2}\rangle$ & $(0,0,0)$ \\ \hline
\end{tabular}
\caption{Summary of GW and EM polarization states and the corresponding Stokes parameters. We do not consider circular polarization and therefore $\langle h_{+}^{*}h_{\times}\rangle=0$ and $\langle ({\bm A}_{h}\!\cdot{\bm u}_{\gamma})^{*} ({\bm A}_{h}\!\cdot{\bm v}_{\gamma})\rangle =0$.}
\label{table:xi}
\end{table}

By analogy, for GWs we define \cite{1974ApJ...189...39A,Seto:2008sr,NA:2017mmh} 
\begin{align}
  \xi_3^{\text{GW}} = \frac{ \langle|h_+|^2\rangle - \langle|h_\times|^2\rangle }{\langle|h_+|^2\rangle + \langle|h_\times|^2\rangle} ,
    &&
   \xi_1^{\text{GW}} + i \xi_2^{\text{GW}} = \frac{ 2\langle h_\times h_+^* \rangle }{\langle|h_+|^2\rangle + \langle|h_\times|^2\rangle} .
\end{align}
For convenience, the cases considered in this work are illustrated in Tab.~\ref{table:xi}.

Before closing the section, let us    
note that the electric and magnetic fields in Eq.~\eqref{eq:Poynting} are not gauge independent quantities in curved spacetime and as such are introduced as convenient auxiliary quantities only.
However, observables such as the power are gauge invariant and can equally be expressed in terms of covariant expressions, see~\cite{Domcke:2023bat} for a more detailed discussion.

\section{Magnetic Domain Model}

\label{sec:Magnetic_Domain_Model}

\begin{figure}
\center
\includegraphics[width=0.5\textwidth]{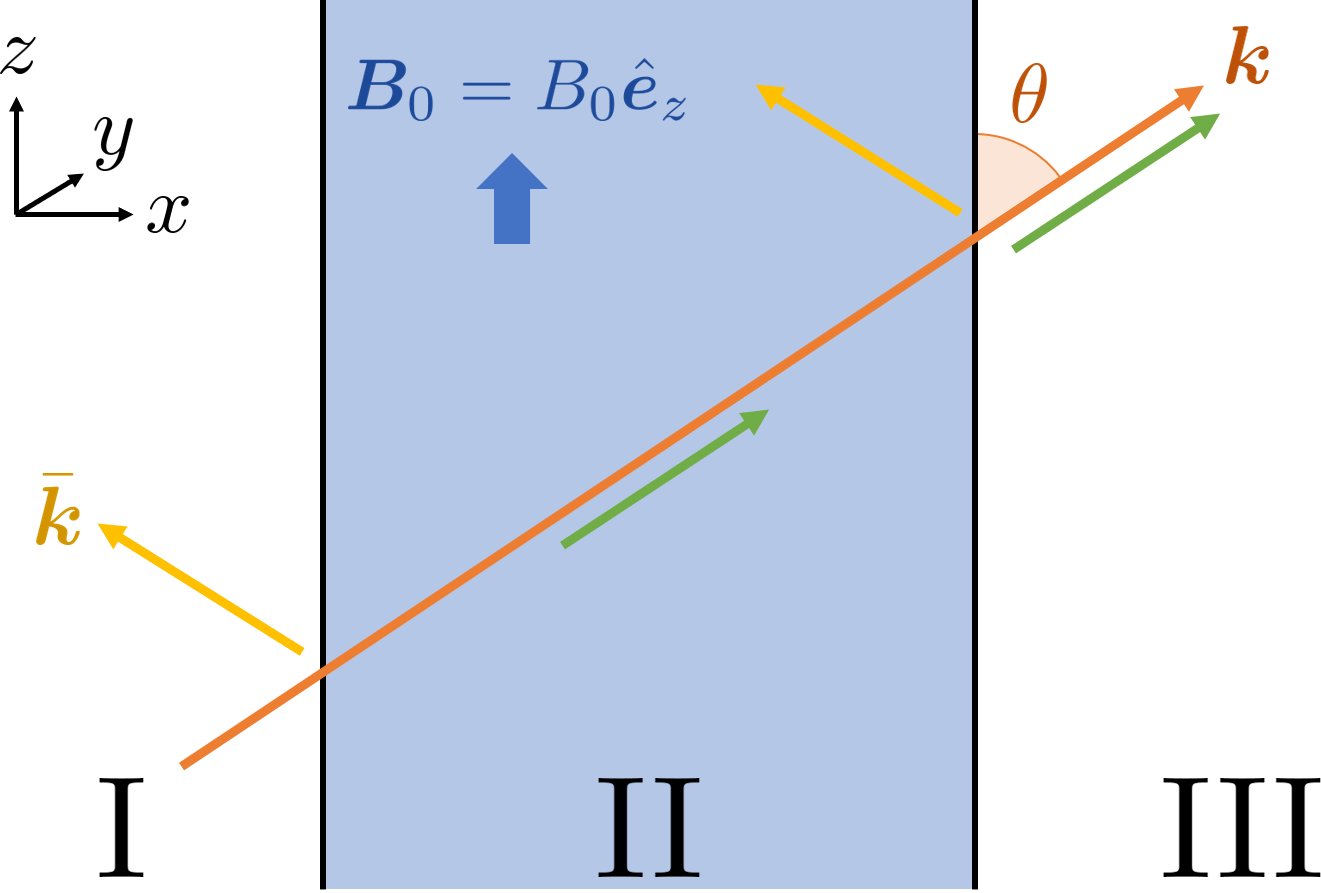}
 \caption{Sketch of the magnetic domain model. The blue colored region indicates a constant magnetic field in $ \hat{\bm{e}}_{z} $-direction. The orange colored arrow denotes the momentum of the GW ($ \bm k $), where we suppressed $ \phi $ direction. The green arrow shows the direction of transmitted EM field parallel to the incident angle of GW, and the yellow arrow indicates the reflected EM wave with the momentum of $ x $-component reversed ($\bar{\bm k} $). }
 \label{fig:magnetic_domain}
\end{figure}

\paragraph{Single Domain.} We first consider a uniform background magnetic field oriented in the $ + \hat{z} $-direction over a domain $ x_{1} < x < x_{2} $, as depicted in Fig.~\ref{fig:magnetic_domain}:
\begin{align}
    \bm{B}_{0} = B_{0} [ \Theta(x-x_{1}) - \Theta(x-x_{2}) ] \hat{\bm e}_{z} \,,
    \label{eq:magnetic_domain}
\end{align}
or equivalently,
\begin{align}
    F^0_{\mu\nu}
    = \begin{pmatrix}
      0 & 0 & 0 & 0 \\
      0 & 0 & B_{0} & 0 \\
      0 & -B_{0} & 0 & 0 \\
      0 & 0 & 0 & 0 
    \end{pmatrix} [ \Theta(x-x_{1}) - \Theta(x-x_{2}) ] \,.
\end{align}
For later reference, let us label the three regions as (I):~$x<x_1$, (II):~$x_1<x<x_2$, (III):~$x>x_2$ and introduce the normal vector of the boundaries as  $ \hat{\bm n} = -\hat{\bm e}_{x} $.
In this case, the effective charge and current in Eq.~\eqref{eq:perturbed_maxwell}, $j^\mu_\text{eff} = (\rho_\text{eff},  {\bm j}_{\rm eff})$, are given by
\begin{equation}
\begin{aligned}
    \rho_{\rm eff} & = 0, \\ 
    {\bm j}_{\rm eff} & = {\bm j}_{b} [\Theta(x-x_{1}) - \Theta(x-x_{2})] + {\bm K}_{1} \delta(x-x_{1}) +  {\bm K}_{2} \delta(x-x_{2}) \,, \label{eq:j_eff_magnetic_domain}
\end{aligned}
\end{equation}
with the effective current receiving contributions from the bulk of the magnetic field volume,
\begin{align}
    {\bm j}_{b} 
      = - i \omega h_{\lambda} \hat{\bm k} \times (e^{\lambda} {\bm B}_{0})  = \frac{i}{\sqrt{2}} \omega B_{0}  e^{- i (\omega t - \bm k \cdot {\bm r} )} s_{\theta} \left(  h_{+} \bm{v} - h_{\times} \bm{u}  \right)  \, , \label{eq:jb}
\end{align}
where $(e^\lambda {\bm B})_i = e_{ij}^\lambda B^j$,
and from its surfaces at $x = x_{1} $ and $ x=x_{2} $~\cite{Domcke:2023bat},
\begin{align}
    {\bm K}_{1}  = -{\bm K}_{2}
       =  - h_{\lambda} \hat{\bm n} \times (e^{\lambda} {\bm B}_{0}) = \frac{1}{\sqrt{2}}  B_{0} e^{- i (\omega t -  {\bm k} \cdot {\bm r} )} s_{\theta} \hat{\bm n} \times \left( h_{+} \bm{u} + h_{\times} \bm{v} \right) \, .
\end{align}

By plugging this into Eq.~\eqref{eq:perturbed_maxwell}, we can solve for the induced EM field subject to the boundary conditions
\begin{equation}
    \begin{aligned}
   x < x_1 ~ : && & \text{no right-moving EM wave}   \\
   x= x_1 ~ :  && & {\bm B}_{h}^{(\text{II})} - {\bm B}_{h}^{(\text{I})} =  \hat{\bm{n}} \times {\bm K}_{1}    \, ,
&&
    {\bm E}_{h}^{(\text{II})} =  {\bm E}_{h}^{(\text{I})} \, ,  \\
 x= x_2 ~ :   &&  & {\bm B}_{h}^{(\text{III})} - {\bm B}_{h}^{(\text{II})} = \hat{\bm{n}} \times {\bm K}_{2}
 \, , &&
     {\bm E}_{h}^{(\text{III})} =  {\bm E}_{h}^{(\text{II})} \, ,   \\
      x > x_2 ~ : && & \text{no left-moving EM wave}
    \label{eq:boundary2}
\end{aligned}
\end{equation}
or equivalently we can express the conditions at the two interfaces as,
\begin{equation}
\begin{aligned}
   x=x_1 ~ : &&  & {\bm A}_{h}^{\rm (II)}   =  {\bm A}_{h}^{\rm (I)}  \,,  && 
   \frac{ \partial {\bm A}_{h}^{(\text{II})}}{\partial x} - \frac{\partial {\bm A}_{h}^{(\text{I})}}{\partial x} = - {\bm K}_{1} \,,
  \\
  x=x_2 ~ : &&  & {\bm A}_{h}^{\rm (III)}   = {\bm A}_{h}^{\rm (II)}   \,,
    &&
    \frac{ \partial {\bm A}_{h}^{{(\text{III})}}}{\partial x} - \frac{\partial {\bm A}_{h}^{(\text{II})}}{\partial x} = - {\bm K}_{2} \,.
    \label{eq:boundary3}
\end{aligned}
\end{equation}
The boundary conditions at $x < x_1$ and $x > x_2$ are dictated by causality, taking into account that we are aiming to compute the EM wave sourced by the passing of the GW.

\paragraph{Transmitted and Reflected Waves.}
Here, we explicitly present the form of the solution outside of the magnetized volume, see App.~\ref{app:Derivations of Magnetic Domain Wall Model} for more details on the derivation.
In Region III ($x>x_{2}$), we find for the transmitted wave,
\begin{align}
\label{eq:AT}
  {\bm A}_{h}^{(\text{III})} =
       -  \frac{B_{0}}{ 2\sqrt{2} c_{\phi}} (x_2 - x_1)
        e^{- i(  \omega t - {\bm k} \cdot {\bm r} )} \left( h_+  {\bm v} - h_\times {\bm u} \right)\,.
\end{align}
The transmitted EM wave is proportional to the bulk effective current~\eqref{eq:jb} whereas the contributions from the surface currents cancel.
The direction of polarization depends on both the incoming GW polarization and the orientation of the magnetic field, with the latter implicit in Eq.~\eqref{eq:AT} through the choice of ${\bm u}$ and ${\bm v}$. 
Moreover, notably, we also obtain a reflected wave in Region I,
\begin{equation}
\begin{aligned}
    \label{eq:AR}
         {\bm A}_{h}^{(\text{I})} &  =
           \frac{B_{0} e^{-i (\omega t- \bar{\bm k}\cdot  {\bm r}) }}{4 \sqrt{2} i \omega s_\theta}
            \left[- 2 c_\theta t_\phi( h_\times \bar{\bm v} +  h_+ \bar{\bm u}) 
           + (t_\phi^2 - 1) h_\times \bar{\bm u}  + (c_{2 \theta} - t_\phi^2) h_+ \bar{\bm v}  \right]    \\
           & \hspace{8.5cm}  \times \left( e^{2i x_2 \bar{\bm k}\cdot {\bm n}} -  e^{2i x_1 \bar{\bm k}\cdot {\bm n}} \right) \\
&\propto  
(
\begin{array}{cc}
 1 & 0 
\end{array})
{\cal T}
\left(
\begin{array}{c}
 h_+ \\ h_\times
\end{array}
\right)
\bar{\bm u}
+(
\begin{array}{cc}
 0 & 1
\end{array})
{\cal T}
\left(
\begin{array}{c}
 h_+ \\ h_\times
\end{array}
\right)\bar{\bm v}
\,, ~\text{with} ~~ {\cal T}  =\left(
\begin{array}{cc}
 2 c_{\theta_\gamma}t_{\phi_\gamma} & t_{\phi_\gamma}^2-1 \\
 c_{2 \theta_\gamma}-t_{\phi_\gamma}^2 & 2 c_{\theta_\gamma}t_{\phi_\gamma} \\
\end{array}
\right)  \,,
%\label{eq:AasofT}        
\end{aligned}
\end{equation}
where the bars on $\bar{\bm k}$, $\bar{\bm u}$ and $\bar{\bm v}$ indicate a reflection of the wave vector ${\bm k}$ at the surface of the magnetic field, i.e.\ at the $yz$-plane and $\theta_{\gamma} = \theta$, and $ \phi_{\gamma} = \pi - \phi $. 
In particular,
\begin{align}
 \bar{\bm v}  = (-\sin \phi , -\cos \phi , 0 ) \,, \quad
 \bar{\bm u}  = (- \cos \theta \cos \phi, \, \cos \theta \sin \phi, \, - \sin \theta )\,,
\end{align}
see also App.~\ref{app:notation}. 
The matrix ${\cal T}$ introduces a compact notation capturing the polarization structure of the reflected wave, and as we discuss in App.~\ref{app:S-matrix}, it is related to the scattering matrix associated with graviton-photon conversion.

The reflected wave obtains contributions both from the bulk and surface currents.
The apparent divergences at $\phi = \pm \pi/2$ and $\theta = \{0, \pi\}$ correspond to an incident GW parallel to the surface of the magnetic domain.
In this case the reflected and transmitted waves can interfere, which results in a finite amplitude of the full wave. Note also that in our toy model, the magnetic domain extends infinitely in the $y-z$ direction. 
In any real scenario the apparent divergence in the expressions above will be regularized by the finite volume of the magnetic domain.

Contrary to the transmitted EM wave, the amplitude of the reflected wave is not proportional to the magnetic domain size $L$ for $\omega L > 1$, indicating the absence of resonant conversion.
In most scenarios of interest, the reflected EM wave is thus suppressed compared to the transmitted wave, and has largely been ignored in the literature.
In particular, the seminal work~\cite{Raffelt:1987im} introduced the approximation $ \omega^{2} + \partial^{2} = (\omega + i \partial ) (\omega - i \partial ) \simeq 2 \omega  (\omega - i \partial )  $ to reduce Maxwell's equations to first order differential equations, which leads to the omission of the reflected wave.
We highlight that the reflected wave has, however, been pointed out in Ref.~\cite{Boccaletti:1970pxw} using essentially the same methods as discussed above, as well as in~\cite{DeLogi:1977qe} using the $S$-matrix approach discussed below and in Appendix~\ref{app:S-matrix}.
Our results agree with these earlier works.
Importantly, as we will see below, for specific spatial configurations of the background magnetic field, the reflected EM wave can become larger than the transmitted one.

\paragraph{Intensity \& Conversion Probability.}

The gauge invariant observables in this process are the power of the transmitted and reflected EM waves. For the transmitted wave, we obtain
\begin{equation}
    \begin{aligned}
    \mathcal{I}_{\rm \gamma}^{T} 
    & = \frac{1}{8} \omega^{2} B_{0}^{2} L^{2} ( \langle \vert h_{+} \vert^{2} \rangle + \langle \vert h_{\times} \vert^{2} \rangle )  \sec^{2} \phi
     \\
    &  = \frac{1}{8} \omega^{2} B_{0}^{T 2} D^{2} ( \langle \vert h_{+} \vert^{2} \rangle + \langle \vert h_{\times} \vert^{2} \rangle )\,,
    \label{eq:conversion}
    \end{aligned}
\end{equation}
where in the second line we express the result in terms of the propagation length $ D = L/(\cos \phi \sin \theta ) $ of the GW through the magnetic domain and the magnitude of the magnetic field component transverse to the direction of the GW, $ B_{0}^{T} =  B_{0} \sin \theta $.
Eq.~\eqref{eq:conversion} is the well-known expression for resonant GW to EM conversion~\cite{Gertsenshtein,Raffelt:1987im,Domcke:2020yzq}, displaying an amplitude of the EM wave that grows linearly with the size of the conversion volume.

For the reflected EM wave we find
\begin{equation}
    \begin{aligned}
    \mathcal{I}_{\gamma}^{R}  & = \frac{1}{8} B_{0}^{2}  \sin^{2} \left( \omega L c_{\phi} s_{\theta} \right) \left[ \langle \vert h_{+} \vert^{2}  \rangle ( - 4 c_{\theta}^{2} + s_{\theta}^{-2} c_{\phi}^{-4} ) + \langle \vert h_{\times} \vert^{2} \rangle ( s_{\theta}^{-2}  c_{\phi}^{-4} - 4 t_{\phi}^{2}  )
   \right. \\
   & \hspace{8 cm} \left. + 4 (\langle h_{+}^{*} h_{\times} \rangle + \langle h_{\times}^{*} h_{+} \rangle ) c_{\theta} t_{\phi} \right] \, .
    \end{aligned}
\end{equation}
The $  \sin^{2} \left( \omega L c_{\phi} s_{\theta} \right) $ prefactor indicates a suppression of the reflected wave compared to the transmitted wave due to destructive interference of EM waves emitted in different regions of the magnetic domain.
In astrophysical contexts, we will typically be dealing with GWs forming stochastic GW backgrounds (SGWBs) or wave packets emitted by individual sources. In the following we will distinguish cases in which the ensemble of GWs traversing a magnetic region is linearly polarized ($\langle |h_+|^{2} \rangle \gg \langle |h_\times|^{2} \rangle $ or vice versa, found e.g.\ for an edge-on black hole binary) or in which the ensemble of GWs is unpolarized.
In both cases, we can neglect the term proportional to $ \langle h_{+}^{*} h_{\times} \rangle$.

The transition or reflection probability can be obtained by considering the ratio between the intensity of the emitted EM waves and the one stored in the GWs. In particular, for the transmitted wave this yields
\begin{align}
    \mathcal{P}_{h \rightarrow \gamma}^{T} =  \frac{\mathcal{I}_{\gamma}^{T}}{\mathcal{I}_{\rm GW}} = 4 \pi G B_{0}^{T 2}  D^{2} \, , 
    \label{eq:transition_prob}
\end{align}
where the intensity of the incident GW is given by
\begin{align}
   \mathcal{I}_{\rm GW}
   = \frac{\omega^{2}}{32 \pi G} \left( \langle \vert h_{+} \vert^{2} \rangle + \langle \vert h_{\times} \vert^{2} \rangle \right) \,,
\end{align}
with $G$ denoting Newton's constant.

\paragraph{Polarization.}
For the transmitted wave, we explicitly find
\begin{equation}
    \bm{\xi}^{(\text{III})} = \left(
\begin{array}{ccc}
 -1 & 0 & 0 \\
 0 & 1 & 0 \\
 0 & 0 & -1 \\
\end{array}
\right) \bm{\xi}^{\text{GW}} \, .
\end{equation}
Thus, the transmitted electromagnetic wave inherits the polarization properties of the GW.
In particular, linearly (circularly) polarized GWs give rise to linearly (circularly) polarized electromagnetic radiation, while unpolarized GWs lead to unpolarized transmitted radiation.\footnote{When GWs are converted to EM waves in extended magnetized regions, the impact of Faraday rotation on the propagating EM waves may be relevant~\cite{Li:2025eoo}.}
The orientation of the polarization vector of the transmitted EM field is given by the direction of the effective current \eqref{eq:jb} and depends on the orientation of the magnetic field.

\begin{figure}
    \centering
    
\includegraphics[width=0.45\linewidth]{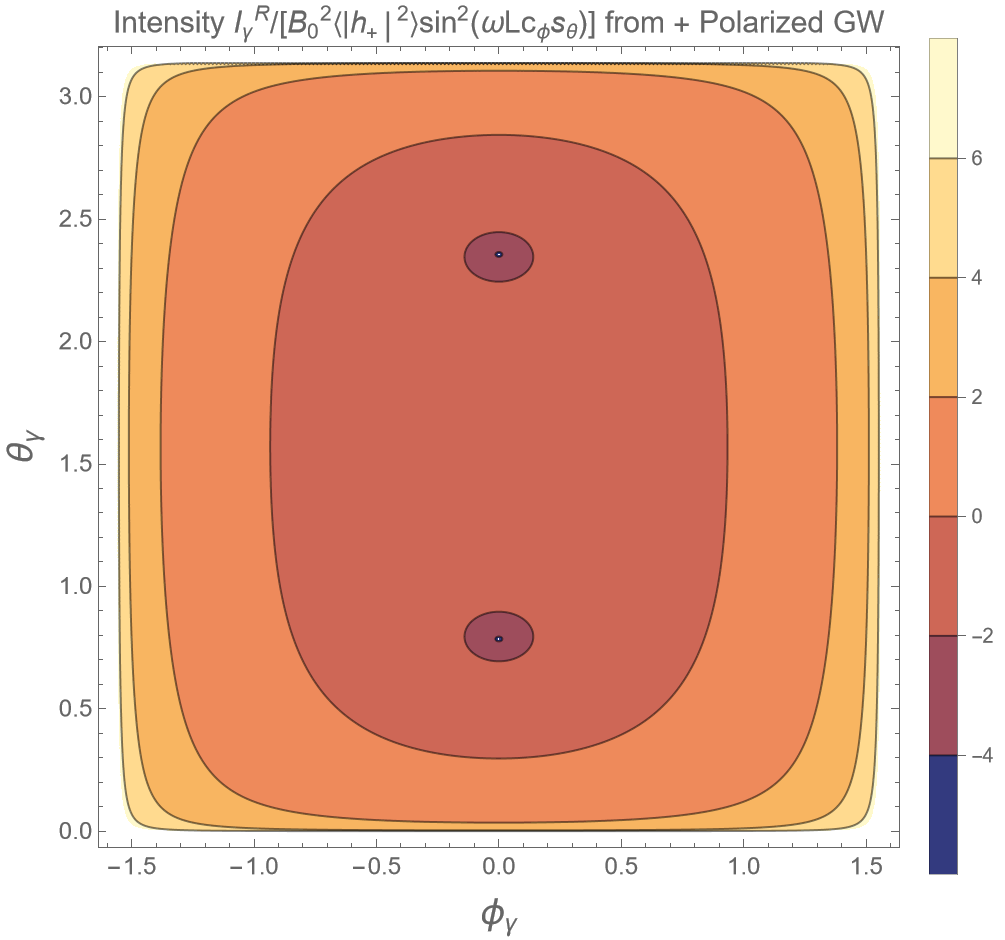}
\includegraphics[width=0.4\linewidth]{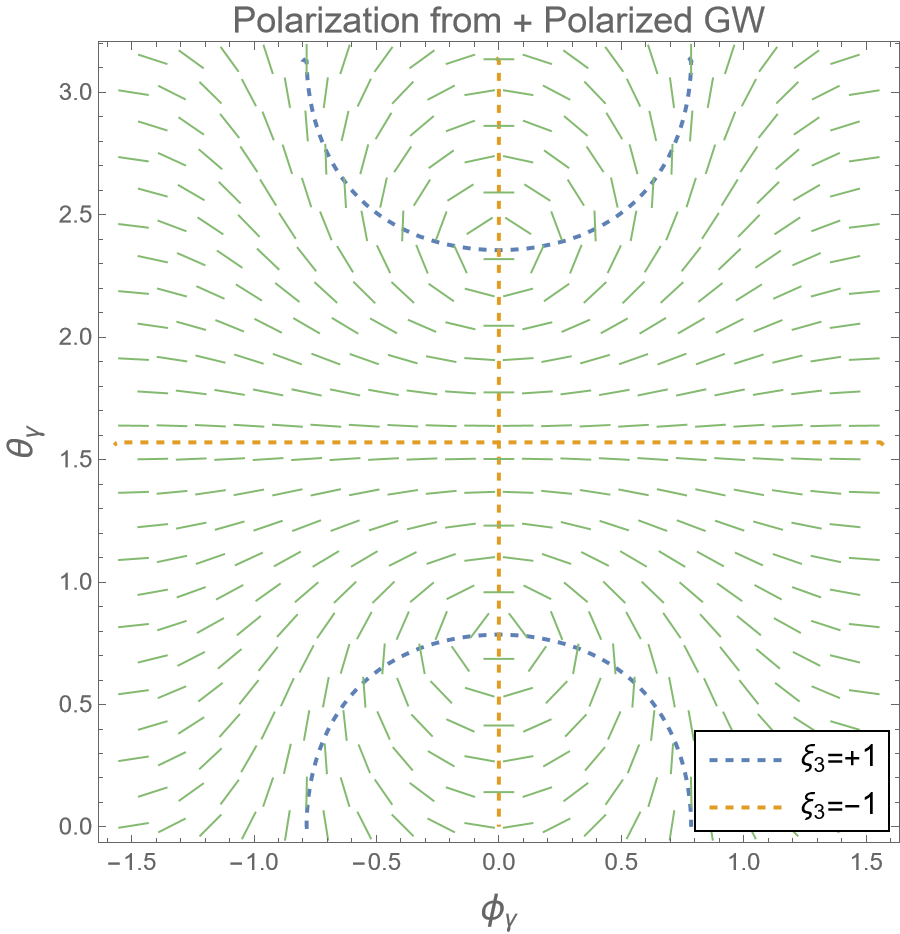}

\includegraphics[width=0.45\linewidth]{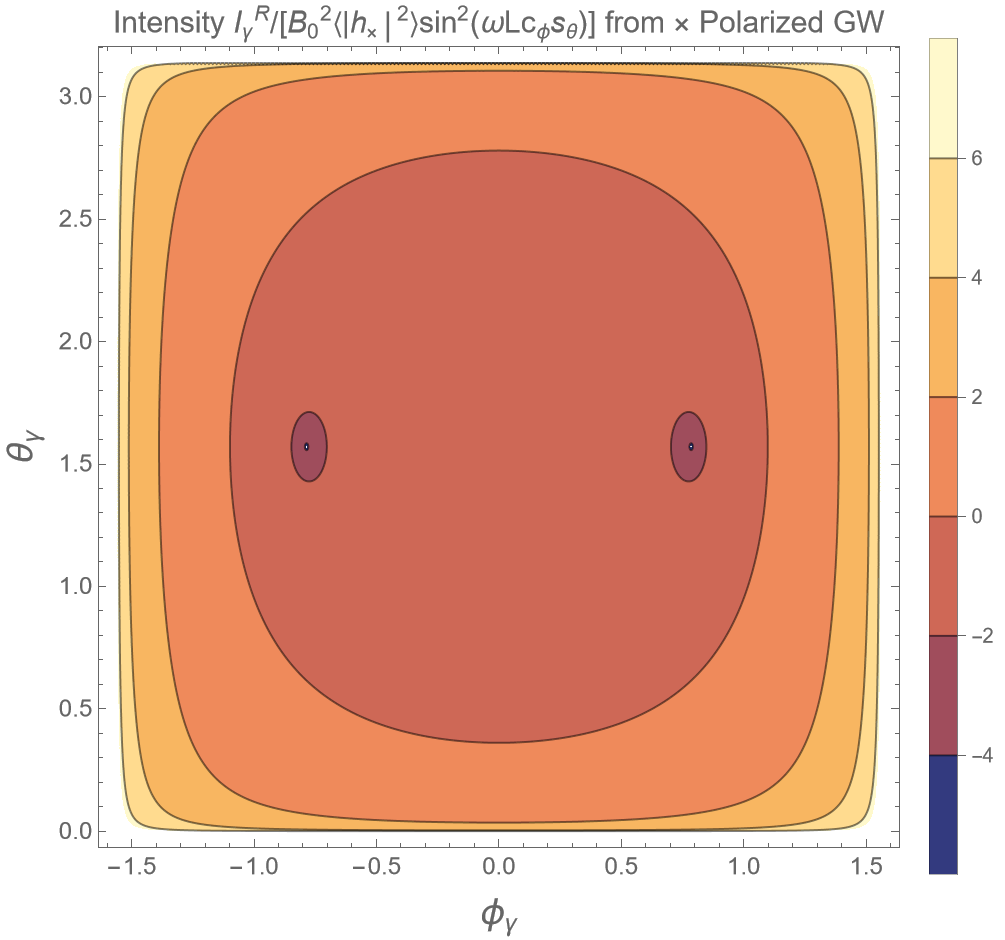}
\includegraphics[width=0.4\linewidth]{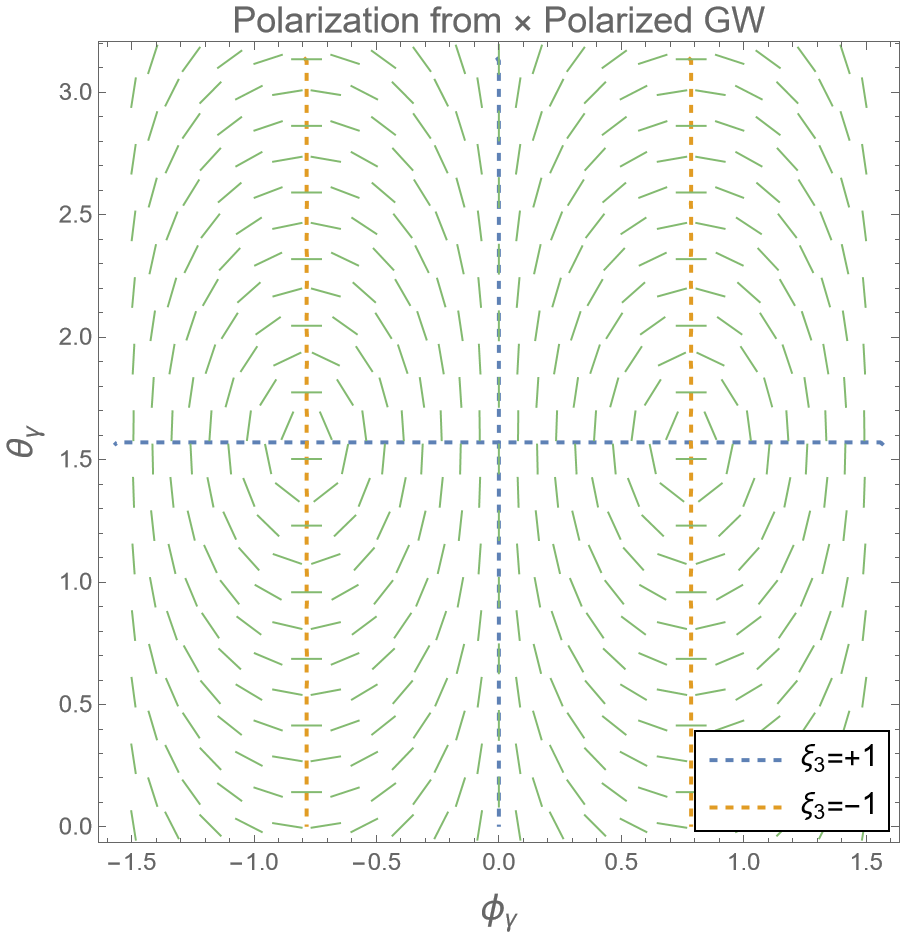}

    \caption{
    Angular distribution of intensity (left column, $\sin^2(\omega L c_{\phi_\gamma} s_{\theta_\gamma})$ factored out) and the polarization (right column) for waves reflected off a single magnetic domain for a maximally polarized GW with $h_{+}$ ($ \xi_{3}^{\rm GW} = +1 $) in the first row, $h_{\times}$ ($ \xi_{3}^{\rm GW} = -1 $) in the second row with the outgoing EM wave in the direction determined by $ \theta_{\gamma} $ and $ \phi_{\gamma} $. For intensity, we draw
    $ \text{Log} \left[ \mathcal{I}_{\gamma}^{R}/ [ B_{0}^{2} \langle \vert h_{+/\times} \vert^{2} \rangle \sin^{2}(\omega L c_{\phi_{\gamma}} s_{\theta_{\gamma}})  ] \right] $.
    For the polarization, the induced EM fields are pure states for a maximally polarized incoming GW, and the direction is indicated by the orientation of the lines, see App.~\ref{app:notation}.
    }
    \label{fig:Ref_Domain_pol}
\end{figure}

\begin{figure}
    \centering
    
\includegraphics[width=0.45\linewidth]{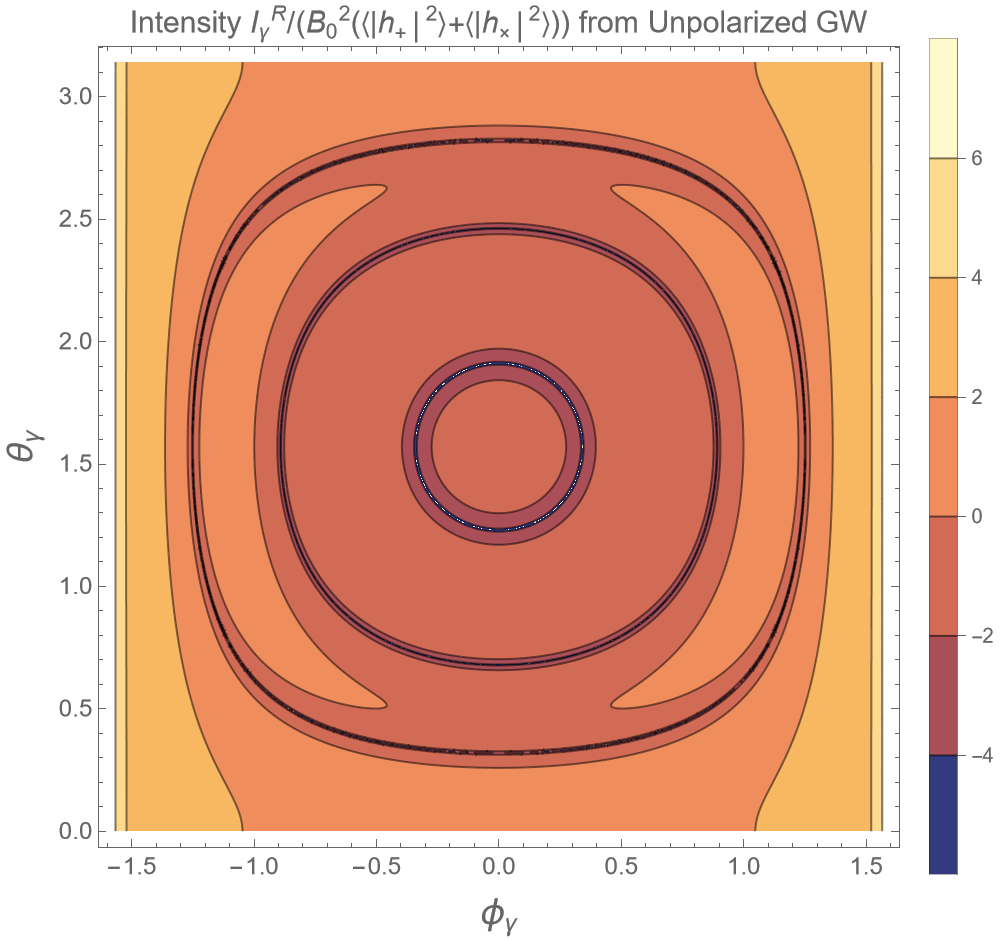}
\includegraphics[width=0.45\linewidth]{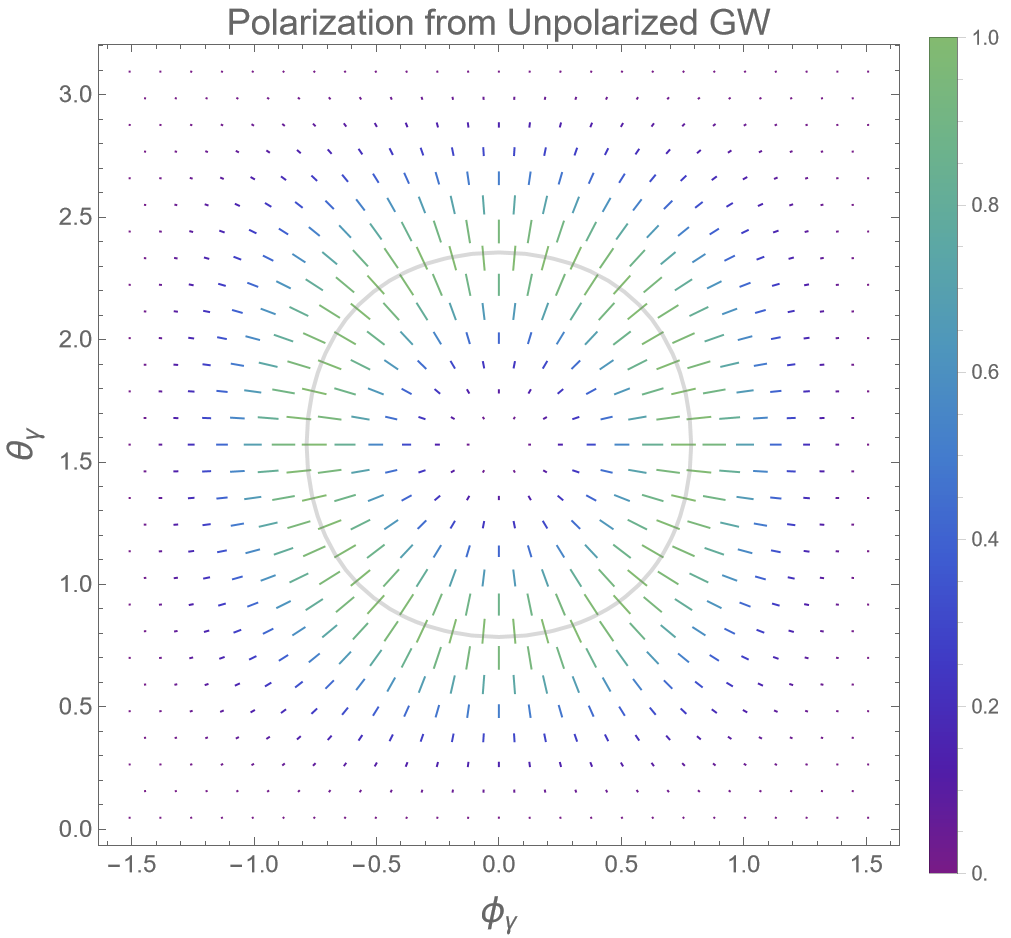}

\caption{Angular distribution of intensity (normalized by $ B_{0}^{2} \left( \langle \vert h_{+} \vert^{2} \rangle + \langle \vert h_{\times} \vert^{2} \rangle \right)$) (left) and the polarization (right) for waves reflected off a single magnetic domain for an unpolarized GW with outgoing EM wave in the direction determined by $ \theta_{\gamma} $ and $ \phi_{\gamma} $.
The left panel displays contours of vanishing intensity from destructive interference given by the $ \sin^{2}(\omega L c_{\phi_{\gamma}} s_{\theta_{\gamma}}) $ term.
Here, $ \omega (x_{2} - x_{1}) = 10 $ is chosen for illustration. 
For the polarization, the color and the length of the line denote the degree of polarization and the direction is indicated by the orientation of the lines, see App.~\ref{app:notation}. The gray contour corresponds to the angles where the reflected waves are pure states.}
\label{fig:Ref_Domain_unpol}
\end{figure}

The situation differs for reflection.
In this case, the Stokes parameters are given by
\begin{align}
\label{eq:xiasofT}
  \xi^{(\text{I})}_i = \frac{\text{tr} \left\{ \sigma^i \left( {\cal T} {\cal T}^T + {\cal T} \, (\bm{\xi}^{\text{GW}} \cdot \bm{\sigma}) \, {\cal T}^T \right) \right\}}
{\text{tr} \left\{ {\cal T} {\cal T}^T + {\cal T} \, (\bm{\xi}^{\text{GW}} \cdot \bm{\sigma}) \, {\cal T}^T \right\}}   \,,
\end{align}
where the matrix ${\cal T}$ is given in Eq.~\eqref{eq:AR} and $ \bm{\sigma} = (\sigma_{1}, \sigma_{2}, \sigma_{3})$ are the Pauli matrices. 
For the specific case of unpolarized GWs ($\xi^{\text{GW}}_i = 0$) this reads
\begin{equation}
    \begin{aligned}
     \bm{\xi}^{\rm (I)} = \left(  
          - \frac{ 8 c_{\theta_{\gamma}}
          s_{\theta_{\gamma}}^{2}
          t_{\phi_{\gamma}}   }{  ( c_{2\theta_{\gamma}} + t^{2}_{\phi_{\gamma} } )^{2} + c_{\phi_{\gamma}}^{-4} } , \,  0 , \, \frac{ 
1 - c_{\phi_{\gamma}}^{-4} + s_{2\theta_{\gamma}}^{2} + 2 c_{2\theta_{\gamma}} t_{\phi_{\gamma}}^{2} + t_{\phi_{\gamma}}^{4}
}{  ( c_{2\theta_{\gamma}} + t^{2}_{\phi_{\gamma} } )^{2} + c_{\phi_{\gamma}}^{-4} } \right) \,,
\end{aligned}
\end{equation}
illustrating that the reflection process induces polarization even in the absence of GW polarization. The reflected wave is always linearly polarized, circular polarization does not arise. Intensity and polarization maps for reflected waves are shown in Fig.~\ref{fig:Ref_Domain_pol} for maximally polarized GWs ($ \langle \vert h_{+} \vert^{2} \rangle =  0$ or $ \langle \vert h_{\times} \vert^{2} \rangle =  0$) and in Fig.~\ref{fig:Ref_Domain_unpol} for unpolarized GWs.

For maximally polarized GWs (with $ \xi_{3}^{\rm GW} = \pm  1$ and $ \xi_{1}^{\rm GW} = \xi_{2}^{\rm GW} = 0 $), the induced EM waves are also maximally polarized as shown in Fig.~\ref{fig:Ref_Domain_pol}.
As detailed in App.~\ref{app:notation}, for maximally polarized EM waves, the polarization with $ \xi_{3} = 1 $ is parallel to $ \bm{u}_{\gamma} $ corresponding to vertically oriented polarization vectors in the $ ( \phi_{\gamma}, \theta_{\gamma} ) $ plane in our convention.
This corresponds to the blue line in Fig.~\ref{fig:Ref_Domain_pol}. 
Similarly, along the contour with $ \xi_{3} = - 1 $, the polarization vectors are horizontally aligned, shown by the orange line.
At the intersection of two lines, the intensity vanishes (corresponding blue dots in the intensity map in the left panels of Fig.~\ref{fig:Ref_Domain_pol}) and the polarization map becomes concentric near these points.
For arbitrary $ h_{+} $ and $ h_{\times} $, $ \bm{A}_{h}^{\rm (I)} = \bm{0} $ happens for
\begin{align}
    \phi_{\gamma} = \arctan \left( \sqrt{ \frac{2h_{+}^{2} + h_{\times}^{2}}{h_{+}^{2} + h_{\times}^{2} } } , \frac{h_{\times}}{\sqrt{h_{+}^{2} + h_{\times}^{2} }}  \right) \,, &&
    \theta_{\gamma} = -\arccos \left( \frac{h_{+}}{\sqrt{2 h_{+}^{2} + h_{\times}^{2} }} \right)\,,
\end{align}
which satisfies 
\begin{align}
    \bm{A}_{h}^{\rm (I)} = \bm{0} \quad:\quad \sin^{2} \theta_{\gamma} \sin^{2} \phi_{\gamma} + \cos^{2} \theta_{\gamma} = 1/2 \,.
\end{align}
In Cartesian coordinates, this corresponds to the intersection between $ x^{2} + y^{2} + z^{2} = 1 $ with $ x = 1/ \sqrt{2}$.
Therefore, although the usage of $ h_{+} $ and $ h_{\times} $ is coordinate-dependent, the physically invariant statement is that one of the polarization stated is not reflected at an angle of $\pi/4$ with respect to the $yz$-plane. 
This can be seen as the GW analogy of \textit{Brewster angle} at which a light with a particular polarization is perfectly transmitted without reflection.
A key difference here is the inefficiency of the GW to photon conversion process, implying that the vast majority of the transmitted energy remains in the GW.

In Fig.~\ref{fig:Ref_Domain_unpol}, which shows the angular distribution of the intensity and polarization of the reflected EM waves in the case of an unpolarized GW source, this corresponds to the contours where the reflected EM waves become fully polarized.

In summary, the resonantly enhanced transmitted wave simply inherits its polarization from the effective current and its intensity depends on the incident angle only by selecting the component of the magnetic field orthogonal to the GW propagation direction.
This is a reflection of the constructive interference (i.e.\ resonant conversion) of transmitted waves across the magnetized region.
On the other hand, due to interference effects, the reflected wave displays a more complex intensity and polarization pattern, depending on the orientation and size of the magnetic field domain as well as the polarization and incident angle of the GW. 
Interestingly, when the angle between the incoming GW and the reflected EM wave is $\pi/2$ (i.e.\ the angle of the incoming GW with respect to the normal vector of the plane is $\pi/4$), the magnetic domain acts as a polarizer, converting unpolarized GWs to maximally polarized EM waves.

\paragraph{Multiple Domains and Continuum Limit.}
These results can be generalized to an arbitrary static background magnetic field in the geometric optics limit, that is, when such background varies very little with respect to the EM/GW wavelengths.
For a given magnetic field configuration (sourced by an astrophysical or laboratory system), one can discretize the magnetic field profile along the line-of-sight of the GW.
Using the expressions above, one can compute the transmitted and reflected EM wave sourced within each infinitesimal domain of uniform magnetic field.
The total reflected and transmitted wave are then simply obtained as a linear combination of these individual contributions.

Let us consider for example the case of a magnetic field profile with varying amplitude but uniform direction $\hat{\bm e}_z$ and a GW incoming along the $x$-axis,
\begin{align}
    {\bm B}_0 = \sum_{i=1}   B_{0}(x_{i}) \left[ \Theta(x-x_{i}) -  \Theta(x-x_{i+1})  \right] \hat{\bm e}_{z}.
\end{align}
From Eq.~\eqref{eq:AT} we note that the amplitude of the transmitted wave scales as
\begin{align}
 \sum_i B_0(x_i) \Delta x_i  \quad \rightarrow \quad \int B_0(x) dx\,,
\end{align}
where $\Delta x_i = x_{i+1} - x_i$ and the arrow indicates the continuum limit. We thus obtain for the total transmitted wave,
\begin{align}
\label{eq:Gen_T}
  {\bm A}_{h}^{T} =
       - \frac{e^{- i \omega ( t - x )}}{ 2 \sqrt{2} } \left( \int B_0(x) \, dx \right)
     \left( 0, h_{+}, h_{\times} \right) \,.% \left( h_+ {\bm v} - h_\times  {\bm u} \right)\,.
\end{align}

For the reflected wave, we note from Eq.~\eqref{eq:AR} that the amplitude is proportional to
\begin{align}
 \sum_i  B_0(x_{i} ) \left( e^{2 i x_{i+1} \bar{\bm k}\cdot\hat{\bm n}} - e^{2 i x_{i} \bar{\bm k}\cdot\hat{\bm n}} \right) 
 \quad \rightarrow \quad 2 i (\bar{\bm{k}} \cdot \hat{\bm n}) \int B_0 (x) e^{2 i x \bar{\bm k} \cdot \hat{\bm n}} dx \,.
\end{align}
and $\bar{\bm k} \cdot \hat{n} = \omega x$ so that we obtain
\begin{align}
     {\bm A}_{h}^{R}  = &
      - \frac{e^{-i \omega (t + x) }}{2\sqrt{2}}
         \left( \int B_0 (x) e^{ 2 i \omega x } dx \right) 
         \left( 0, h_{+}, h_{\times} \right)
       \label{eq:Gen_R}
\end{align}
For the contribution from the effective bulk current, the contributions from the regions labeled by $i$ simply add up.
The contributions from the effective surface current cancel between neighboring domains if the magnetic field amplitude is identical, but yield a net contribution when the magnetic field changes.
This is fully accounted for by the expression above.

For magnetic field profiles $B_0(x)$ which vary slowly along the line of sight compared to the GW frequency, a prerequisite of the domain model approach, we see that the exponential contributes a rapidly oscillating function to the integral, which leads to a suppression of the amplitude of the reflected wave.
In other words, while the transmitted EM waves interfere constructively in this limit, the reflected waves interfere largely destructively.\footnote{
We note that based on these results, we cannot confirm the claim in Ref.~\cite{Addazi:2024kbq} regarding an additional resonance enhancement mechanism.}

\paragraph{Example: Antisymmetric Magnetic Field Configuration.}
For a homogeneous magnetic field, the amplitude of the transmitted wave is resonantly enhanced by a factor $(\omega L)$, whereas for the reflected wave no such enhancement occurs for $\omega L > 1$.
However, for a particular spatial structure of the background magnetic field, the transmitted wave may be suppressed so that the dominant signal arises from the reflected wave.
To illustrate this, consider a GW propagating in $x$-direction through a magnetic field profile with
\begin{align}
    {\bm B}_0 = \begin{dcases}
    B_{0} \sin \left(2 \pi  x / L  \right) \hat{\bm e}_{z} & (0 \leq x \leq L ) \\
    0 & (\text{otherwise})
    \end{dcases} \,,
\end{align}
implying an angular frequency of the background $ \omega_{b} = 2 \pi / L $.
In this case, Eq.~\eqref{eq:Gen_T} implies that the amplitude of the transmitted wave vanishes, ${\bm A}_{T} = {\bm 0}$, simply because the EM waves generated in the region with negative magnetic field cancel those generated in the region with positive magnetic field.
On the contrary, for the reflected wave Eq.~\eqref{eq:Gen_R} yields
\begin{align}
   {\bm A}_{R} \simeq \frac{- i \pi B_0 e^{- i \omega (t-x)}  }{2 \sqrt{2} \omega^2 L} e^{i \omega L} \sin(\omega L)   \left( 0, h_{+}, h_{\times} \right) \,,
\end{align}
where we have employed the geometric-optics limit $\omega \gg \omega_b = 2 \pi/L$. This amounts to a reflected intensity of
\begin{align}
  \mathcal{I}_{\gamma}^{R}
  \simeq
 \frac{ \pi^2}{8 \omega^{2} L^{2}} B_{0}^{2} ( \langle \vert h_{+} \vert^{2} \rangle + \langle \vert h_{\times} \vert^{2} \rangle) \sin^2(\omega L) \,.
 \label{eq:sin_R}
\end{align}
The prefactor in Eq.~\eqref{eq:sin_R} is always less than unity, suppressing the amplitude of the reflected wave as anticipated in Eq.~\eqref{eq:Gen_R}.
Nevertheless, in this specific example, power in the reflected wave exceeds the power in the transmitted wave.

\section{Green's Function Method}
\label{sec:Green's_Function_Method}

The magnetic domain model discussed in the previous section allows us to compute the transmitted and reflected EM waves for an arbitrary magnetic field profile, assuming a uniform direction of the magnetic field.
Extending this approach to a more general spatial structure of a static magnetic field is possible by performing appropriate rotations to account for the relative angles between the incoming GW and the magnetic field in each domain.
In practice, this approach rapidly becomes cumbersome. 
Instead, in this section we present another method borrowed from classical electrodynamics, to efficiently determine the GW to EM wave conversion for such configurations.
This approach relies on determining the Green's function which solves the left-hand side of Eq.~\eqref{eq:Ajeff}, and then convoluting this with the source, i.e.\ the effective current.

In Coulomb gauge,
with no effective charge density and with the boundary condition that induced fields vanish at the spatial infinity, the general solution is formally given by
\begin{equation}
    \begin{aligned}
    \bm{A}_h(t,\bm{r}) = - \int d^3\bm{r}' \int dt' \, G(\bm{r}, t; \bm{r}', t') \, \bm{j}_{\rm eff}( t',\bm{r}')\,,
    \\   \text{with} \hspace{0.5cm}  G(\bm{r}, t; \bm{r}', t') = - \frac{1}{4\pi}\frac{\delta\left(t' - t + |\bm{r} - \bm{r}'| \right)}{|\bm{r} - \bm{r}'|}\,,
    \end{aligned}
\end{equation}
leading to
\begin{equation}
\bm{A}_h( t, \bm{r}) =  \frac{1}{4\pi}\int \frac{\bm{j}_{\rm eff}( t_r, \bm{r}')}{|\bm{r} - \bm{r}'|} \, d^3\bm{r}' \quad \text{with} \quad t_r \equiv t - |\bm{r} - \bm{r}'| \,.
\end{equation}
For a plane GW, this can be simplified to
\begin{align}
 \bm{A}_{h} (t, \bm{r} ) = -\int d^{3}\bm{r}^{\prime} ~  G^{(3)}(\bm r - \bm r^{\prime}) \bm{j}_{\rm eff} ( t, \bm r^{\prime}) \,, &&
    G^{(3)} (\bm {r} - \bm{r}^{\prime}) = -  \frac{e^{i \omega \vert \bm{r} - \bm{r}^{\prime} \vert }}{ 4\pi \vert {\bm r} - {\bm r}^{\prime} \vert} \,,
    \label{eq:sol_green}
\end{align}
where the Green function $ G $ is the solution to
\begin{align}
    (\omega^{2} + \nabla^{2}) G^{(3)} (\bm {r} - \bm{r}^{\prime}) = \delta^{(3)} (\bm {r} - \bm{r}^{\prime})\,.
\end{align}

In this section, we first demonstrate that the Green's function method successfully reproduces the results for the magnetic domain example discussed in Sec.~\ref{sec:Magnetic_Domain_Model}.
We then show how this approach can be straightforwardly generalized to solve the case of a GW incident at an arbitrary angle, propagating through a magnetic field profile with a uniform magnetic field direction. 
Finally, we discuss localized sources, where the magnetic field direction is not uniform.

\subsection{Magnetic Domain Example Revisited}
\label{sec:magnetic_domains_2}

As a warm-up, let us consider the case where $ \bm{k} = \omega \hat{\bm e}_{x} $ while allowing for a smoothly varying magnetic field $ \bm{B}_{0}(x) $ pointing in the $z$-direction, as in Sec.~\ref{sec:Magnetic_Domain_Model}.
In this setup, the effective current reads
\begin{align}
     \bm{j}_{\rm eff}  =  \frac{i}{\sqrt{2}} \omega B_{0}(x)  e^{- i \omega (t - x)} ( 0, h_{+}, h_{\times}) \,,
     \label{eq:effective_current_mag_domain}
\end{align}
and the solution is given by Eq.~\eqref{eq:sol_green}
\begin{align}
     \bm{A}_{h}(t, \bm r)
     & =  \frac{i e^{- i \omega t}}{ \sqrt{2}  }  \omega \,( 0, h_{+}, h_{\times})   \int d^{3} r^{\prime}   B_{0}  (x^{\prime}) e^{i \bm{k} \cdot \bm{r}^{\prime} }  \frac{e^{i \omega \vert \bm{r} - \bm{r}^{\prime} \vert} }{ 4 \pi \vert \bm{r} - \bm{r}^{\prime} \vert } \,.
     \label{eq:MD_Greens_1}
\end{align}

In this particular setup, the effective current only depends on $ x^{\prime} $, so one can perform the integration along the $ yz$-plane  transverse to the incoming GW direction.
Explicitly, using
\begin{align}
     \int_{-\infty}^{\infty} dy^{\prime} dz^{\prime}  
    ~
    \frac{e^{i \omega \vert \bm{r} - \bm{r}^{\prime} \vert} }{ 4\pi  \vert \bm{r} - \bm{r}^{\prime} \vert } 
     =  
 \int_{0}^{\infty}  d\rho
    ~  e^{i \omega x^{\prime} }  \frac{\rho e^{i \omega \sqrt{ (x-x^{\prime})^{2} + \rho^{2 }   } } }{ 2 \sqrt{ (x-x^{\prime})^{2} + \rho^{2 }  } }  = - \frac{e^{i \omega \vert x - x^{\prime} \vert}}{2 i \omega} \,, \label{eq:3Dto1D}
\end{align}
where $ \rho = \sqrt{(y^{\prime} - y)^{2} + (z^{\prime} - z)^{2} }$, we have
\begin{align}
   \bm{A}_{h} & = -  \frac{ e^{- i \omega t}}{2  \sqrt{2} } ( 0, h_{+}, h_{\times})  \int_{x_{1}}^{x_{2}}{dx^{\prime}} ~  B_{0}  (x^{\prime}) 
    ~  e^{i \omega x^{\prime} }  e^{i \omega \vert x - x^{\prime} \vert} \nonumber \\
    & =  -  \frac{ e^{- i \omega t}}{2  \sqrt{2}  }
    ( 0, h_{+}, h_{\times}) 
    \begin{dcases}
        e^{+ i \omega x} \int_{x_{1}}^{x_{2}}{dx^{\prime}} ~  B_{0}  (x^{\prime}) & (x > x_{2}) \\
        e^{- i \omega x} \int_{x_{1}}^{x_{2}}{dx^{\prime}} ~  e^{2 i \omega x^{\prime} }   B_{0}  (x^{\prime})  & (x < x_{1}) 
    \end{dcases} \,.
    \label{eq:Gen_TR_green}
\end{align}
Note that $  G^{(1)}(x-x^{\prime}) \equiv e^{i \omega \vert x - x^{\prime} \vert}/ (2 i \omega) $ solves
$ (\omega^{2} + \partial_{x}^{2}) G^{(1)} (x-x^{\prime}) = \delta(x-x^{\prime}) $.
These expressions coincide with the result found in Eqs.~\eqref{eq:Gen_T} and \eqref{eq:Gen_R}.

\paragraph{Generalization to Arbitrary GW Direction.}
This result can be straightforwardly generalized to an arbitrary incident GW direction. The main results are summarized in this section, with more details given in App.~\ref{app:Derivations of Magnetic Domain Wall Model}.

The general expression for the effective current arising from a GW with wave vector $\bm{k} = \omega  ( \sin \theta \cos \phi, \sin \theta \sin \phi, \cos \theta )$ and a magnetic field oriented in the $z$-direction with a slowly varying profile is given by Eq.~\eqref{eq:jeff_uniform_field_direction}. The formal solution is immediately obtained as
\begin{align}
     \bm{A}_{h} 
     & =  \frac{e^{- i \omega t}}{ \sqrt{2}  }  s_{\theta}    \int d^{3} \bm{r}^{\prime}   \, \left[ 
 \left(  h_{+} \bm{v} - h_{\times} \bm{u}  \right) i \omega B_{0}(x)  - \bm{n} \times (h_{+} \bm{u} + h_{\times} \bm{v} )  B_{0}^{\prime}(x) \right]  e^{i \bm{k} \cdot \bm{r}^{\prime} }  \frac{e^{i \omega \vert \bm{r} - \bm{r}^{\prime} \vert} }{ 4 \pi \vert \bm{r} - \bm{r}^{\prime} \vert }.
 \label{eq:MD_Greens_2}
\end{align}
Contrary to the example above, this effective current depends also on $y'$ and $z'$. The integration over these dimensions can be performed by introducing the line of sight variable $ \bm{k} \cdot \bm{r}^{\prime} = \omega \ell^{\prime} $, see App.~\ref{app:Derivations of Magnetic Domain Wall Model} for details, yielding
\begin{align}
&\int  d^{3}r^{\prime} ~B_{0} (x^{\prime}) e^{i \omega \ell^{\prime}} \frac{e^{i \omega \vert \bm{r} - \bm{r}^{\prime} \vert} }{ 4 \pi \vert \bm{r} - \bm{r}^{\prime} \vert } = - \int d\ell^{\prime}   ~B_{0}(\ell^{\prime} s_{\theta} c_{\phi} ) e^{i \omega \ell^{\prime}}  \frac{e^{i \vert \bm{k} \cdot \bm{r} - \omega \ell^{\prime} \vert  }}{  2 i \omega } \,,
\label{eq:integration}
\end{align}
where we use the fact that $ B_{0} $ only depends on $ x $. For $ \theta = \pi/2$ and $ \phi = 0$, this reduces to Eq.~\eqref{eq:3Dto1D}.
Therefore, we obtain the general formula
\begin{equation}
\begin{aligned}
    \bm{A}_{h} 
    &  = 
-\frac{e^{- i \omega t}}{2\sqrt{2}}  s_{\theta} \int_{\ell_{1}}^{\ell_{2}} d\ell^{\prime}  \left[  \left(  h_{+} \bm{v} - h_{\times} \bm{u}  \right)  B_{0}(\ell^{\prime} s_{\theta} c_{\phi}) \right. \\
& \hspace{4.5cm} \left.
- \frac{1}{i \omega}  \hat{\bm{n}} \times \left(  h_{+} \bm{u} + h_{\times} \bm{v}  \right)  B_{0}^{\prime}(\ell^{\prime} s_{\theta} c_{\phi})  \right] 
e^{i \omega \ell^{\prime}}  e^{i \vert \bm{k} \cdot \bm{r} - \omega \ell^{\prime} \vert  } 
\,.
\end{aligned}
\label{eq:general_sol_domain}
\end{equation}
This expression can be further simplified when considering the solution outside the magnetic domain, i.e. for transmitted and reflected waves.
For the transmitted wave,
\begin{equation}
    \begin{aligned}
    \bm{A}_{h}^{T} 
      & =
-\frac{e^{- i (\omega t - \bm{k} \cdot \bm{r})}}{2\sqrt{2}}    \left(  h_{+} \bm{v} - h_{\times} \bm{u}  \right) \left( \int_{\ell_{1}}^{\ell_{2}} d\ell^{\prime} \, B_{0}^T(\ell^{\prime} s_{\theta} c_{\phi} ) \right)  \\
  & =
-\frac{e^{- i (\omega t - \bm{k} \cdot \bm{r})}}{2\sqrt{2}}   c_{\phi}^{-1}  \left(  h_{+} \bm{v} - h_{\times} \bm{u}  \right) \left( \int_{x_{1}}^{x_{2}} dx^{\prime} \, B_{0}(x^{\prime}) \right) \,,
\label{eq:AT_domain_revisited}
\end{aligned}
\end{equation}
with $B_0^T = s_\theta B_0$ the magnetic field transverse to the GW propagation and we assumed the background magnetic field gradually decays so that $ B_{0} = 0 $ for $ x \leq x_{1}$ and $ x \geq x_{2} $.
On the other hand, for the reflected wave,
\begin{equation}
    \begin{aligned}
    \bm{A}_{h}^{R} 
    = 
    -\frac{e^{- i (\omega t - \bar{\bm{k}} \cdot \bm{r})}}{2\sqrt{2} }   c_{\phi}^{-1} \left[  \left(  h_{+} \bm{v} - h_{\times} \bm{u}  \right)  + 2 s_{\theta} c_{\phi} \hat{\bm{n}} \times \left(  h_{+} \bm{u} + h_{\times} \bm{v}  \right)  \right] \left( \int_{x_{1}}^{x_{2}}  dx^{\prime} ~ B_{0}(x^{\prime}) e^{2 i \omega s_{\theta} c_{\phi} x^{\prime}} \right)  \,.
    \end{aligned}
\end{equation}
In the limit of an incoming GW along the $x$-axis, this reduces to Eqs.~\eqref{eq:Gen_T} and \eqref{eq:Gen_R} or, equivalently, Eq.~\eqref{eq:Gen_TR_green}.

\subsection{Localized Sources}
\label{sec:localized_sources}

In this section, we will derive general expressions for GW to photon conversion in compact, localized magnetic field regions.
One possible application is GW to photon conversion in the dipolar magnetic field of neutron stars, which we will discuss in more detail below.

For a given localized magnetic field configuration and an observer at a distance $r$, we can take the far-field approximation
\begin{equation}
    \begin{aligned}
    \label{eq:A_h_3D}
        {\bm A}_{h}(t,{\bm r} ) & = \frac{1}{4\pi} \int d^{3} { r}^{\prime} \, \frac{{\bm j}_{\rm eff} (  t  - \vert {\bm r} - {\bm r}^{\prime} \vert, {\bm r}^{\prime}) }{\vert {\bm r} - {\bm r}^{\prime} \vert}  \\ & \simeq \frac{1}{4\pi r} \int d^{3} { r}^{\prime} \, {\bm j}_{\rm eff} (  t  - r + r^{\prime} (\hat{\bm r} \cdot \hat{\bm r}^{\prime}, {\bm r}^{\prime}) ) \,,
    \end{aligned}
\end{equation}
where we used
\begin{align}
    \vert {\bm r} - {\bm r}^{\prime} \vert \simeq r - r^{\prime} (\hat{\bm r} \cdot \hat{\bm r}^{\prime})  + \cdots \,,
\end{align}
with $ r \gg r^{\prime} $.
The use of retarded time, $t - \vert {\bm r} - {\bm r}^{\prime} \vert$, reflects causality, thus implementing the boundary conditions which enforce only outgoing EM waves.
Inserting this into the harmonic function contained in the effective current, $\bm j_\text{eff}(\bm r, t) = \bm J_\text{eff}(\bm r) \exp(- i \omega t + i \bm k \cdot \bm r)$, we obtain
\begin{align}
\label{eq:j_eff_3D}
 \bm j_\text{eff}(t  - r + \hat{\bm r} \cdot \bm r', \bm r') = \bm J_\text{eff}(\bm r') 
 e^{-i \omega (t - r)} 
 e^{- i ( \bm{k}_{\gamma} - \bm{k} ) \cdot \bm{r}^{\prime}}
 \,.
\end{align}
Denoting the momentum transfer as $\bm q = \bm{k}_{\gamma} - \bm k$,
 the integral in Eq.~\eqref{eq:A_h_3D} can be performed explicitly,
 \begin{align}
    \int d^{3} r^{\prime} ~ \bm{J}_{\text{eff}} ({\bm r}^{\prime} ) \, e^{- i \bm q \cdot \bm r'}
     = \tilde{ \bm{J} }_{\text{eff}} (\bm q) =  - i  
     \sum_{\lambda = + , \times }  
     h_{\lambda} \bm{k}_{\gamma} \times ( e^{\lambda} \tilde{ \bm{B} }_{0} )
      \,,
     \label{eq:J_eff_localized}
 \end{align}
where the tilde indicates a Fourier transform with respect to the momentum transfer $\bm q$, $\tilde X(\bm q) = \int d^3r' X(\bm r') \exp(- i \bm{q} \cdot \bm{r}^{\prime})$.
Since we have taken the background magnetic field to be static, the frequency of any emitted EM waves matches that of the incoming wave, but the spatial structure of the magnetic field allows for scattering and corresponding momentum transfer.
We thus find
\begin{equation}
    \begin{aligned}
     {\bm A}_{h}(t, {\bm r} ) & \simeq 
    \frac{e^{- i \omega (t - r)}}{4\pi r} \int d^{3} {\bm r}^{\prime} ~ {\bm J}_{\rm eff} ({\bm r}^{\prime} ) e^{  - i ( \bm{k}_{\gamma} - {\bm k} ) \cdot {\bm r}^{\prime} } 
     =  \frac{e^{-i \omega (t - r)}}{4 \pi r} \tilde{ \bm{J} }_{\text{eff} } ( \bm{q}) \, .
    \end{aligned}
   \label{eq:A_GF_sol}
\end{equation}
From this, we can obtain intensity  and polarization properties as before, and the cross section per solid angle as
\begin{align}
\label{eq:cross_section}
     \frac{d\sigma}{d\Omega} =  r^{2} \frac{ \mathcal{I}_{\gamma}  }{ \mathcal{I}_{\rm GW} } \,.
\end{align}

Similar to Eq.~\eqref{eq:AR}, we cast the induced gauge field as\footnote{In more detail, by plugging  Eq.~\eqref{eq:J_eff_localized} in Eq.~\eqref{eq:A_GF_sol}, we obtain $ \bm{A}_{h} \cdot \bm{u}_{\gamma} \propto h_{\lambda}  \bm{u}_{\gamma} \cdot (\bm{k}_{\gamma}  \times e^{\lambda} \tilde{\bm{B}}_{0}  ) = h_{\lambda}  (\bm{u}_{\gamma} \times \bm{k}_{\gamma} ) \cdot ( e^{\lambda} \tilde{\bm{B}}_{0} ) = - h_{\lambda} \bm{v}_{\gamma}^{T} e^{\lambda} \tilde{\bm{B}}_{0}  $, and similarly $\bm{A}_{h} \cdot \bm{v}_{\gamma} \propto h_{\lambda} \bm{u}_{\gamma}^{T} e^{\lambda} \tilde{\bm{B}}_{0}   $.
Hence
\begin{equation}
\begin{aligned}
   {\bm A}_{h}(t, {\bm r} ) & = - \frac{ i \omega e^{- i \omega (t-r)} }{4\pi r}  \sum_{\lambda = +, \times} h_{\lambda} \left[ - \left( \bm{v}_{\gamma}^{T} e^{\lambda} \tilde{\bm{B}}_{0}   \right) \bm{u}_{\gamma} + \left( \bm{u}_{\gamma}^{T} e^{\lambda} \tilde{\bm{B}}_{0}   \right) \bm{v}_{\gamma}  \right]
\,,
\end{aligned}
\label{eq:Ainfootnote}
\end{equation}
which leads to Eq.~\eqref{eq:AasofT2}.
}

\begin{equation}
\label{eq:AasofT2}
\begin{aligned}
   {\bm A}_{h}(t, {\bm r} ) & = - \frac{i\omega e^{-i \omega (t-r) } }{4 \pi r}
\left[
\left(
\begin{array}{cc}
 1 & 0 
\end{array}\right)
{\cal T}
\left(
\begin{array}{c}
 h_+ \\ h_\times
\end{array}
\right)
\bm{u}_\gamma
+\left(
\begin{array}{cc}
 0 & 1
\end{array}\right)
{\cal T}
\left(
\begin{array}{c}
 h_+ \\ h_\times
\end{array}
\right)\bm{v}_\gamma
\right]
, \,
\end{aligned}
\end{equation}
with
\begin{eqnarray}
{\cal T} \equiv
\frac{1}{\sqrt{2}} \left(
\begin{array}{cc}
 \bm{v}_\gamma \cdot \bm{v}   & -\bm{v}_\gamma \cdot \bm{u}  \\
 - \bm{u}_\gamma \cdot \bm{v}   & \bm{u}_\gamma \cdot \bm{u}
\end{array}
\right)
\left(
\begin{array}{cc}
 \tilde{\bm{B}}_0 (\bm{q}) \cdot \bm{v} & - \tilde{\bm{B}}_0 (\bm{q}) \cdot \bm{u} \\
 \tilde{\bm{B}}_0 (\bm{q}) \cdot \bm{u}  & \tilde{\bm{B}}_0 (\bm{q}) \cdot \bm{v}
\end{array}
\right)\,.
\label{eq:TmatN}
\end{eqnarray}

The absence of the $\tilde{\bm{B}}_0 (\bm{q}) \cdot \bm{k} $ component is a manifestation of the fact that only the transverse components of the fields couple to the incoming GW.
In this way, the intensity reads
\begin{equation}
\begin{aligned}
\mathcal{I}_\gamma &= \frac{\omega^4}{16 \pi^2 r^2}
{\rm tr} \left\{
{\cal T}
\left(
\begin{array}{cc}
 \langle |h_+|^2 \rangle &\langle h_+ h_\times^* \rangle \\
 \langle h_\times h_+^* \rangle & \langle |h_\times|^2 \rangle
\end{array}
\right)
{\cal T}^T
\right\} \\
&=
\frac{\omega^4}{32 \pi^2 r^2} \left( \langle |h_+|^2 \rangle + \langle |h_\times|^2 \rangle \right)
\text{tr} \left\{ {\cal T} {\cal T}^T + {\cal T} \, (\bm{\xi}^{\text{GW}} \cdot \bm{\sigma}) \, {\cal T}^T \right\}
\label{eq:Igamma}
\,,
\end{aligned}
\end{equation}
while the Stokes parameters are given by the following generalization of Eq.~\eqref{eq:xiasofT}
\begin{align}
\xi_i = \frac{\text{tr} \left\{ \sigma^i \left( {\cal T} {\cal T}^T + {\cal T} \, (\bm{\xi}^{\text{GW}} \cdot \bm{\sigma}) \, {\cal T}^T \right) \right\}}
{\text{tr} \left\{ {\cal T} {\cal T}^T + {\cal T} \, (\bm{\xi}^{\text{GW}} \cdot \bm{\sigma}) \, {\cal T}^T \right\}} \, .
\end{align}

As reviewed in Appendix~\ref{app:S-matrix}, the entries of ${\cal T}$ coincide --apart from an overall normalization -- with the $S$-matrix elements for graviton–photon scattering in an external magnetic field; thus the effective-current formulation reproduces the quantum-mechanical result.\footnote{For a recent  discussion on quantum-mechanical aspects of the Gertsenshtein effect, see~\cite{Ikeda:2025uae}.  }

\paragraph{Example: Unpolarized GWs with a Fixed Direction.}

For the case of unpolarized GWs, the previous expressions simplify to  
\begin{equation}
    \begin{aligned}
    & \mathcal{I}_{\gamma} 
    =  \frac{\omega^{4} \langle \vert h \vert^{2} \rangle}{32\pi^{2} r^{2}}   |\tilde{ \bm{B} }_{0}^{T}  |^{2}   \left( 1 + (\hat{ \bm k}\cdot \hat{ \bm k}_{\gamma})^{2} \right) \,,
    \end{aligned}
\end{equation}
where $ \tilde{ \bm{B} }_{0}^{T} \equiv \tilde{ \bm{B} }_{0} - (\tilde{ \bm{B} }_{0} \cdot \hat{\bm k}) \hat{\bm k}  $ is the component of the background magnetic field transverse to the direction of the GW. The scaling of ${\cal I}_\gamma$ with the square of the size of the magnetic domain, characteristic of a resonant conversion process, is implicitly contained through the Fourier transform of the magnetic field $\tilde {\bm B}_{0}$.
For the Stokes parameters, we find
\begin{align}
    \bm{\xi} = \frac{1 - (\hat{ \bm k}\cdot \hat{ \bm k}_{\gamma})^{2} }{1 + (\hat{ \bm k}\cdot \hat{ \bm k}_{\gamma})^{2}}\left( \sin \varphi \,, 0 \,,  \cos \varphi\right) 
    \label{eq:pol_localized}
\end{align}
where $\varphi$ is an angle that depends on the choice of coordinate system.
If this is chosen so that $\phi = \phi_\gamma$ -- which is always possible given arbitrary vectors $\hat{\bm k}$ and $\hat{\bm k}_\gamma $ -- then $\varphi=0$.
This includes the specific case in which the $z$-axis is aligned with the GW direction.
Furthermore, when $\varphi=0$, the only non-vanishing component is  $ \xi_{3}$, which is positive, implying that the photon is polarized parallel to $ \bm{u}_{\gamma}$.\footnote{In a coordinate-invariant manner, this polarization corresponds to the direction perpendicular to $\hat{\bm{k}}_{\gamma}$ that lies within the plane defined by the directions of the GW and the photon.
This can be seen from the fact that, when $\phi = \phi_\gamma$, we have ${\bm v} = {\bm v}_\gamma \propto \hat{\bm{k}} \times \hat{\bm{k}}_{\gamma}$ and hence $ \bm{u}_{\gamma} =  {\bm v}_\gamma\times  \hat{\bm{k}}_{\gamma} \propto  \hat{\bm{k}}- (\hat{\bm{k}}_{\gamma} \cdot \hat{\bm{k}} ) \hat{\bm{k}}_{\gamma}  $}.

We note that the photon polarization does not depend on the magnetic field configuration and in particular does not depend on the orientation of the background magnetic field.
This is evident from the fact that ${\cal T}{\cal T}^{T}$ -- the only B-dependent combination entering in the formulae for the intensity and the Stokes parameters when $\bm{\xi}^{\text{GW}} =0 $ -- is independent of the orientation of the magnetic field. 
Then, for a given GW direction, 
the photon polarization
only depends on the angle between the GW and the induced EM wave, or $ \hat{\bm k} \cdot \hat{\bm k}_{\gamma} $.

\paragraph{Example: Magnetic Dipole.}
A key application of GW to EM wave conversion are neutron stars.  In preparation for a more detailed discussion in Sec.~\ref{sec:neutron_star}, we consider here a  magnetic dipole $ \bm m $ with the magnetic field and vector potential given by
\begin{align}
    {\bm B}_0
    & = \frac{1}{4\pi r^{3}} (3 \hat{\bm r} ({\bm m} \cdot \hat{\bm r}) - {\bm m} )  \,, \qquad     {\bm A}_0 = \frac{{\bm m} \times {\bm r}}{4 \pi r^{3}} \,,
\end{align}
where we set the direction of the dipole $ \hat{\bm m} = ( \sin \alpha \cos \beta  , \sin \alpha \sin \beta  ,  \cos \alpha) $  and take the propagation of the GW to be in $\hat{\bm e}_{z}$ direction. 

To evaluate Eq.~\eqref{eq:A_GF_sol},  we note that the corresponding magnetic field and vector potential in Fourier space are given by
 \begin{align}
    \tilde{ \bm B }_0 (\bm{q}) = i {\bm q} \times \tilde{{\bm A}}_0 (\bm{q}) \,, \qquad       \tilde{{\bm A}}_0 (\bm{q})  = - \frac{ {\bm m} \times \hat{{\bm q}}}{ q } \,,
\end{align}
which allows us to compute the emitted EM wave in the far field regime. 

From Eq.~\eqref{eq:A_h_3D} with $\bm{j}_{\rm eff}$ given in Eq.~\eqref{eq:j_eff_3D} we obtain $ {\bm A}_{h} $, and thus the cross section in Eq.~\eqref{eq:cross_section}). For the specific case $ \beta = 0 $
\begin{equation}
    \begin{aligned}
    \frac{d\sigma}{d\Omega} 
    = \frac{G \omega^{2} m^{2}}{4 \pi} 
       & \left[ \left\{  (c_{\alpha} s_{\theta_{\gamma}} + 2 s_{\alpha} s_{ \theta_{\gamma}/2 }^{2} c_{\phi_{\gamma}} ) s_{2\phi_{\gamma}} - 2 s_{\alpha} s_{\phi_{\gamma}} c_{2\phi_{\gamma}} \right\}^{2}  \right. \\
        & \hspace{0.5cm} \left. + c_{\theta_{\gamma}}^{2} \left\{ (c_{\alpha} s_{\theta_{\gamma}} + 2 s_{\alpha} s_{ \theta_{\gamma}/2 }^{2} c_{\phi_{\gamma}} ) c_{2\phi_{\gamma}} + 2 s_{\alpha} s_{\phi_{\gamma}} s_{2\phi_{\gamma}} \right\}^{2} \right] \,,
\end{aligned}
 (\text{$+$ polarized GW})
\end{equation}
and
\begin{equation}
    \begin{aligned}
    \frac{d\sigma}{d\Omega} 
    = \frac{G \omega^{2} m^{2}}{4 \pi} 
       & \left[ \left\{  (c_{\alpha} s_{\theta_{\gamma}} + 2 s_{\alpha} s_{ \theta_{\gamma}/2 }^{2} c_{\phi_{\gamma}} ) c_{2\phi_{\gamma}} + 2 s_{\alpha} s_{\phi_{\gamma}} s_{2\phi_{\gamma}} \right\}^{2}  \right. \\
        & \hspace{0.5cm} \left. + c_{\theta_{\gamma}}^{2} \left\{ (c_{\alpha} s_{\theta_{\gamma}} + 2 s_{\alpha} s_{ \theta_{\gamma}/2 }^{2} c_{\phi_{\gamma}} ) s_{2\phi_{\gamma}} - 2 s_{\alpha} s_{\phi_{\gamma}} c_{2\phi_{\gamma}} \right\}^{2} \right] \,.
\end{aligned}
 (\text{$\times$ polarized GW})
\end{equation}
This matches the result obtained in Ref.~\cite{DeLogi:1977qe} using the $S$-matrix approach.
The prefactor $m^2 \omega^2$ indicates a resonant conversion of the typical length scale set by the magnitude of the magnetic dipole $m$.

\begin{figure}
    \centering
    \includegraphics[width=0.45\linewidth]{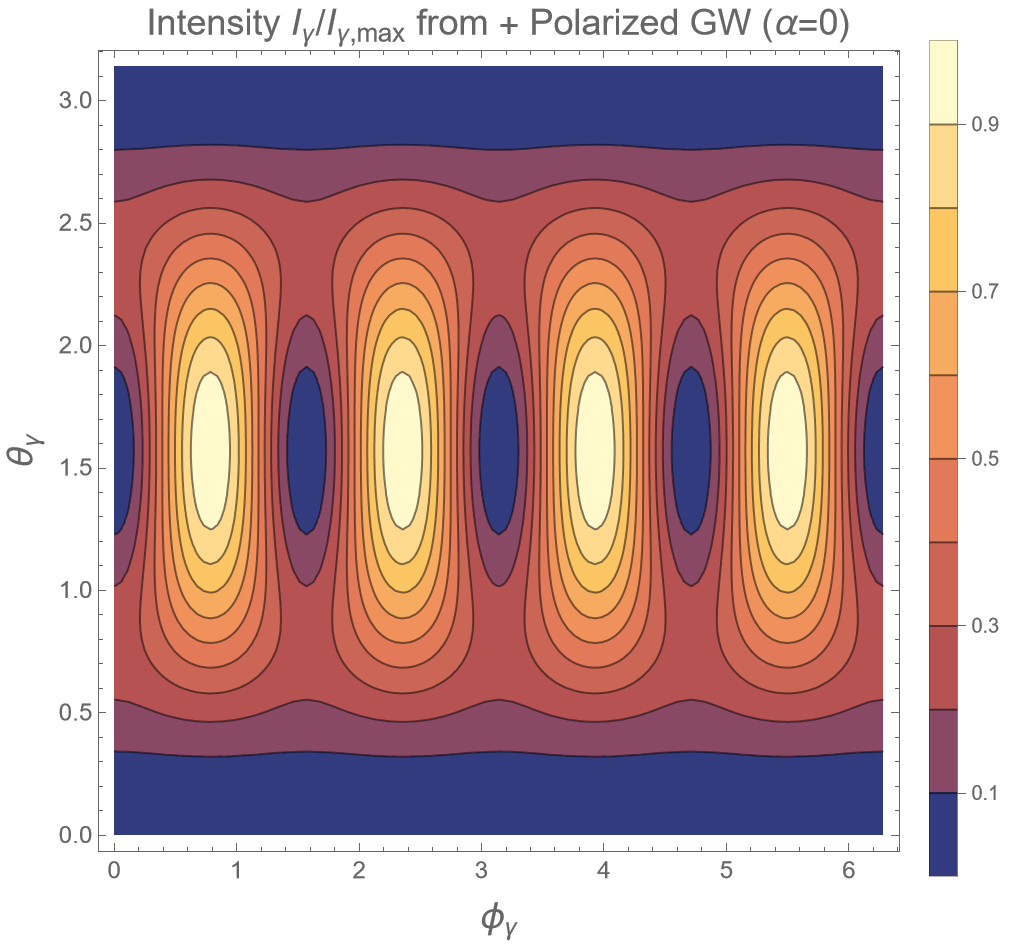}
    \includegraphics[width=0.40\linewidth]{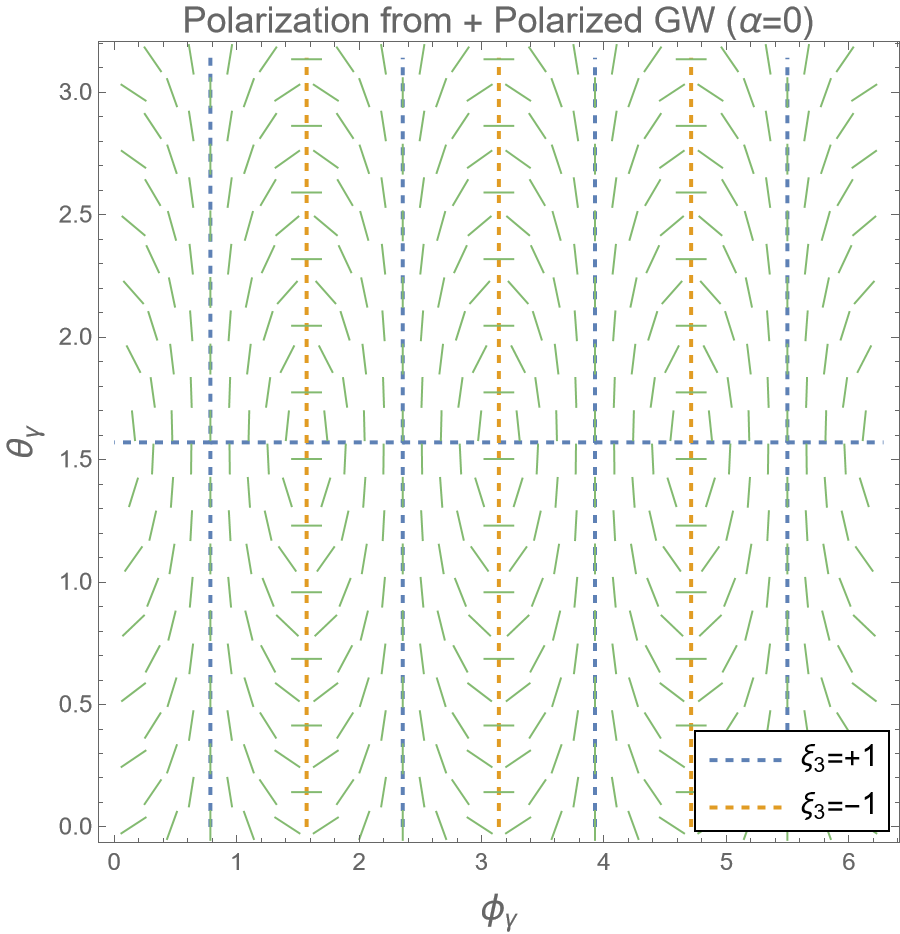}
    \includegraphics[width=0.45\linewidth]{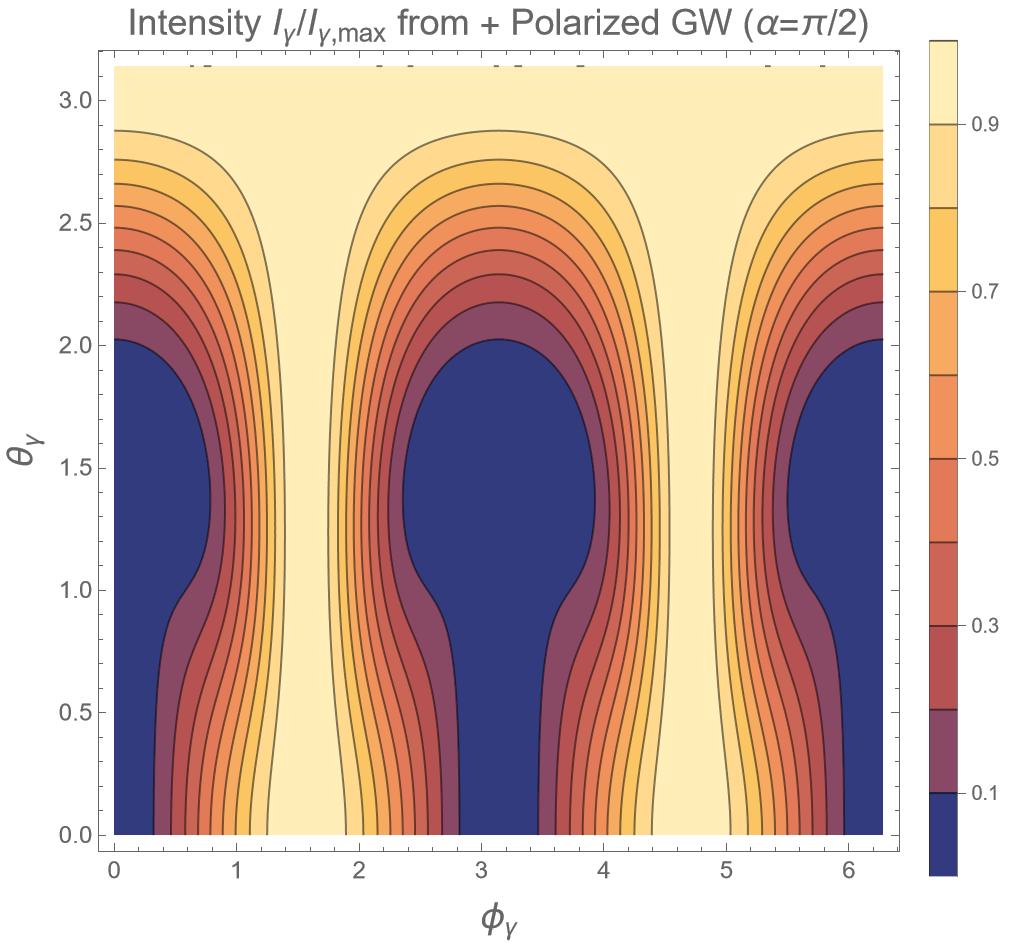}
    \includegraphics[width=0.40\linewidth]{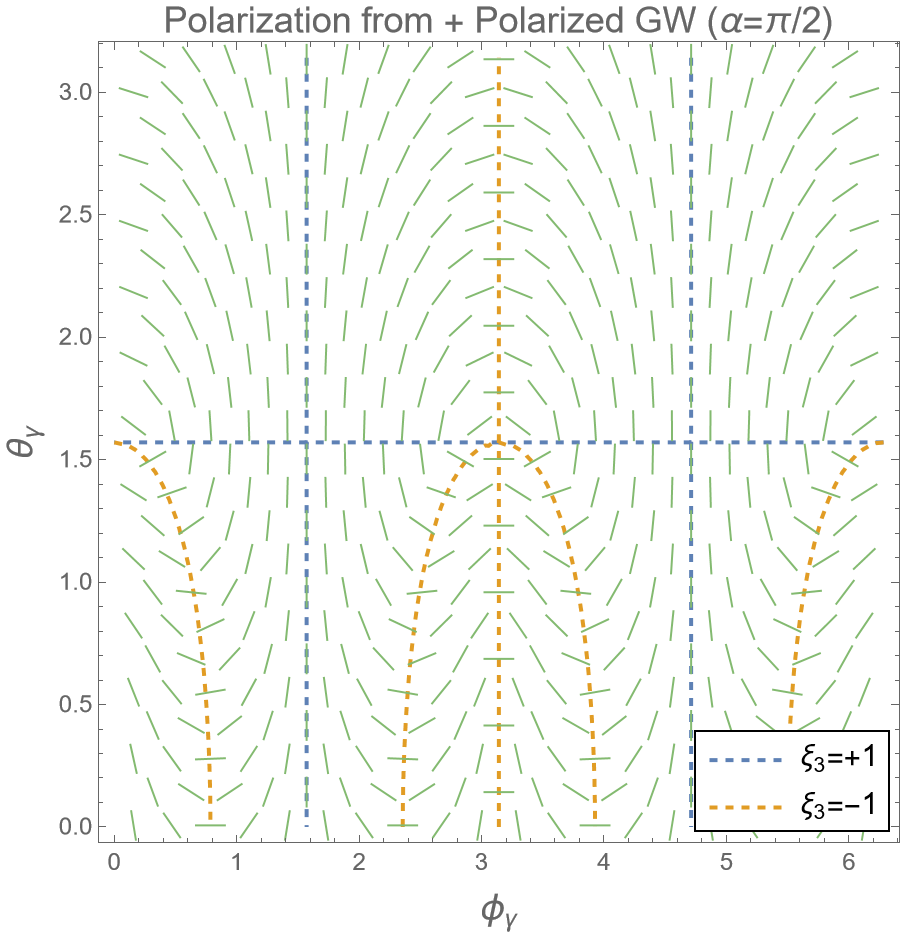}

    \caption{ Intensity $ \mathcal{I}_{\gamma} / \mathcal{I}_{\gamma,\max} $ (left) and polarization (right) of the induced EM fields for maximally $+$ polarized GW with the orientation of the magnetic field given by $ \alpha  = 0 $ (top) and $ \alpha = \pi/ 2$ (bottom), respectively, relative to the GW propagating in $ \hat{\bm{e}}_{z}$ direction. }
    \label{fig:dipole_polarized}
\end{figure}

\begin{figure}
    \centering
     \includegraphics[width=0.45\linewidth]{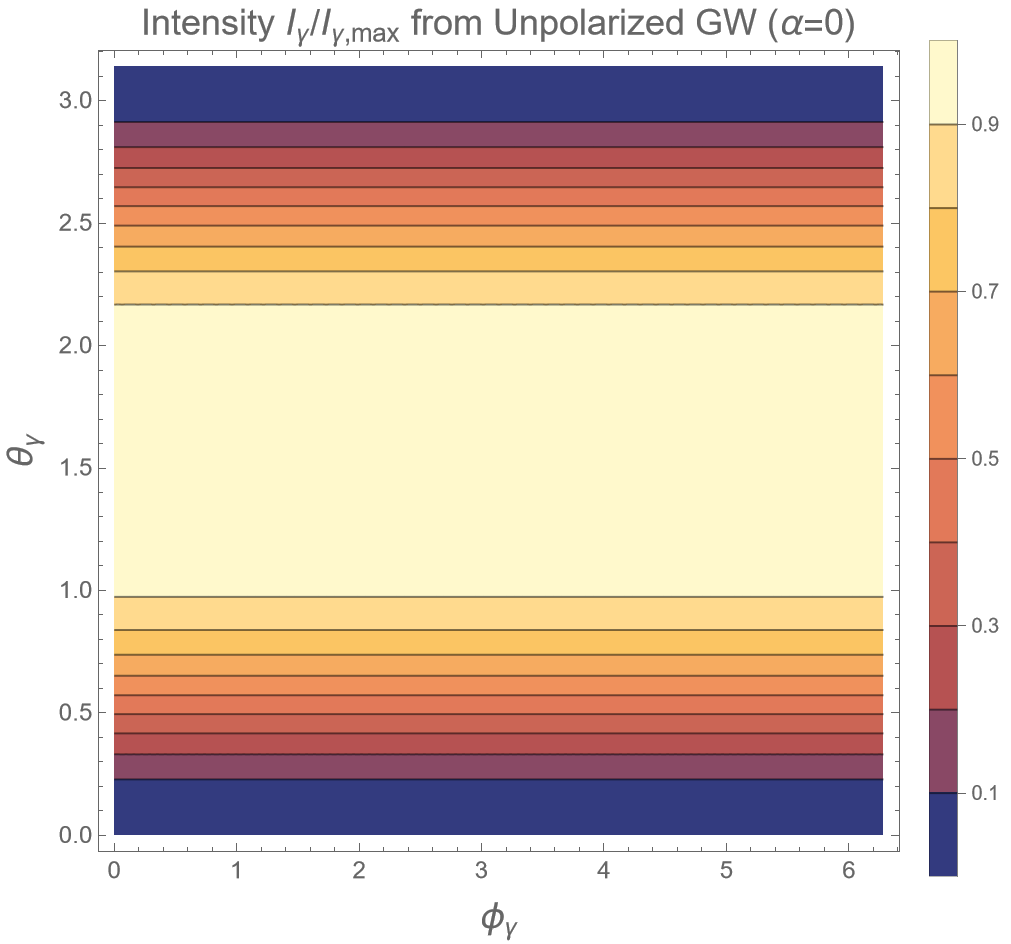}
    \includegraphics[width=0.45\linewidth]{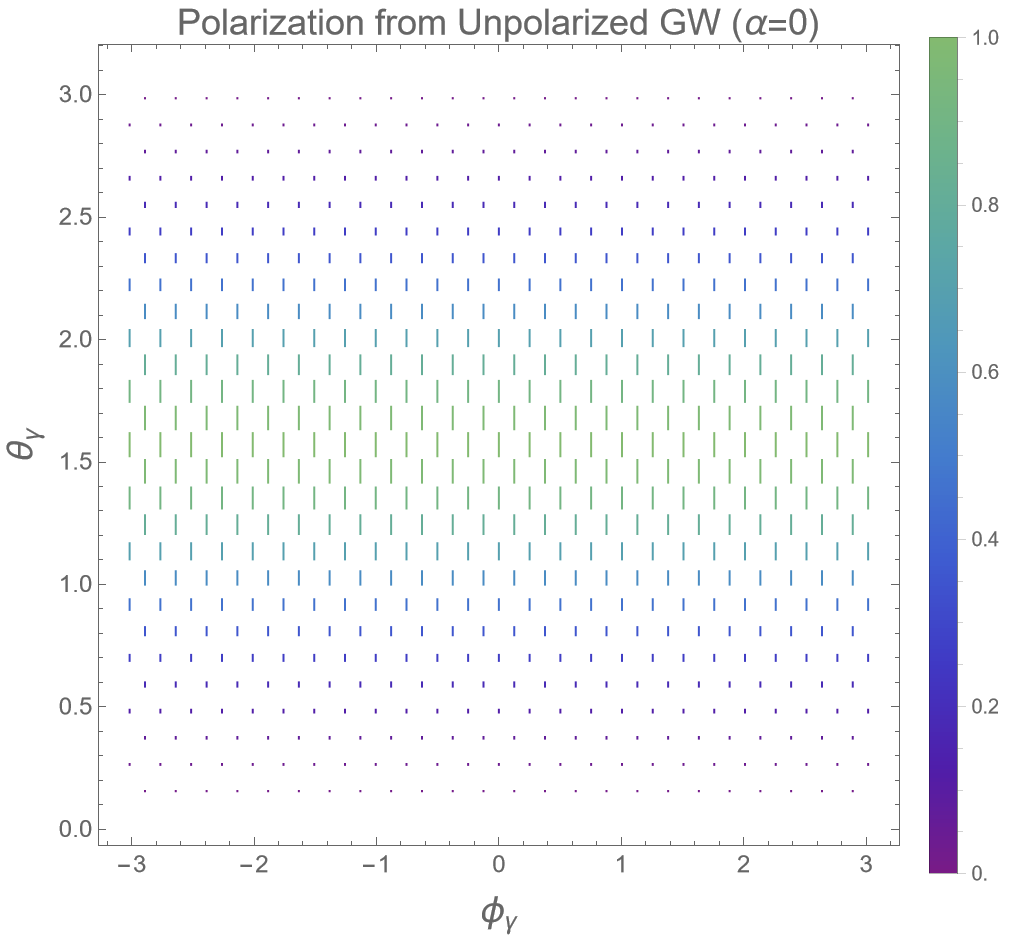}
    \includegraphics[width=0.45\linewidth]{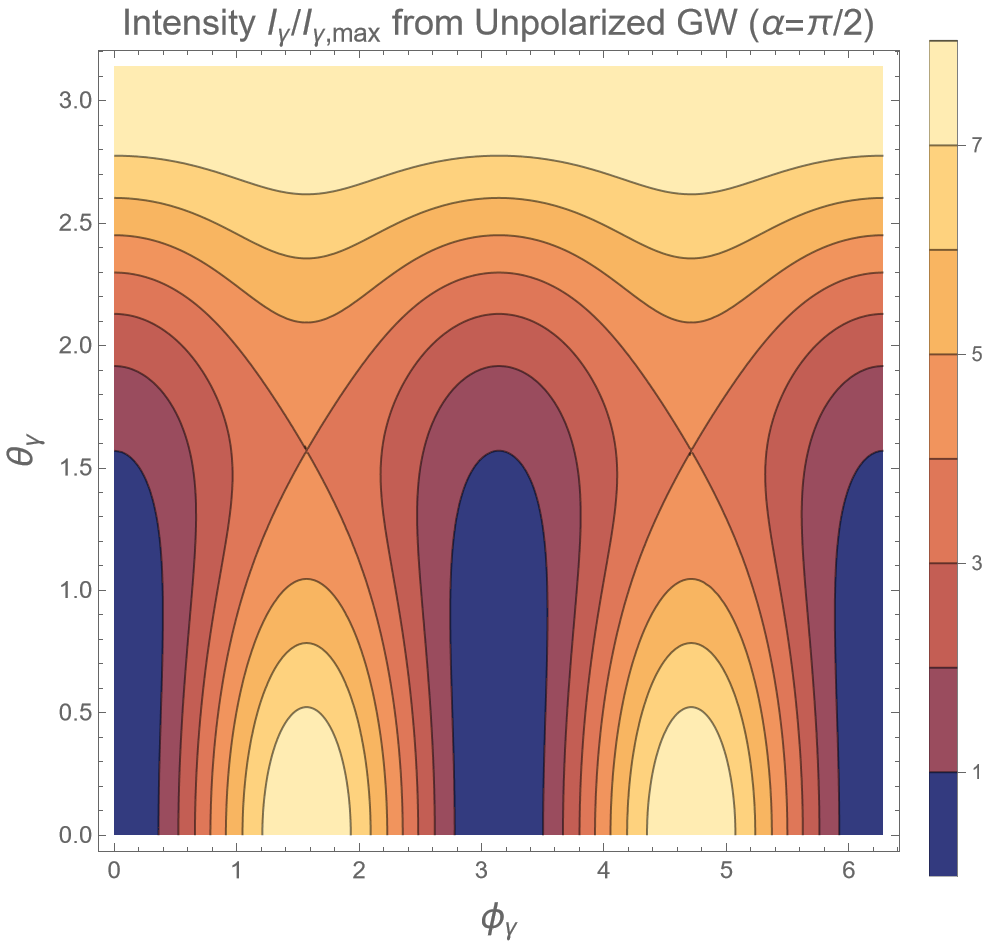}
    \includegraphics[width=0.45\linewidth]{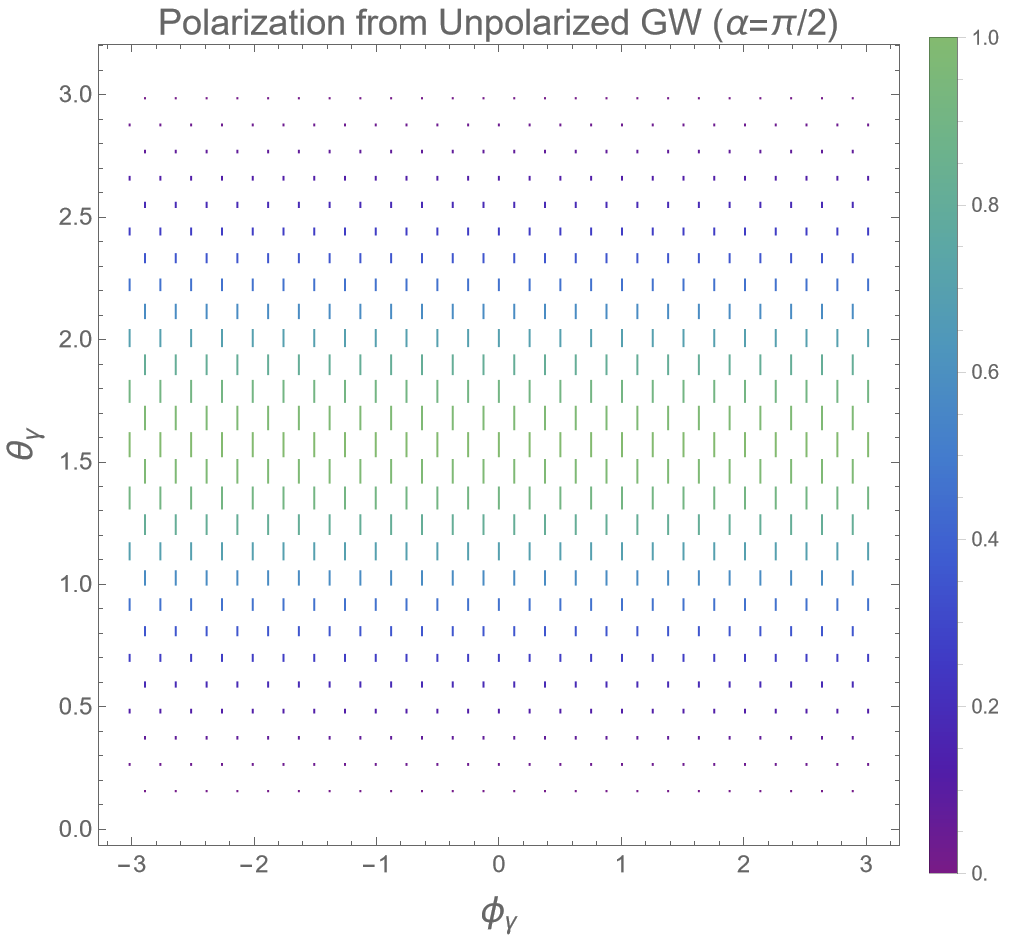}
    
    \caption{ Intensity $ \mathcal{I}_{\gamma} / \mathcal{I}_{\gamma,\max} $ (left) and polarization (right) of the induced EM fields for an unpolarized GW propagating in $ \hat{\bm{e}}_{z}$ direction
    in the presence of a magnetic dipole with $ \alpha = 0 $, $ \beta = 0 $ (top) and $ \alpha = \pi/2$, $ \beta = 0 $ (bottom), respectively. }
   \label{fig:dipole_unpolarized}
\end{figure}

The non-trivial angular dependence of the intensity and polarization for a + polarized incoming GW is illustrated in Fig.~\ref{fig:dipole_polarized} taking $ \alpha = 0 $ and $ \alpha = \pi/2 $, respectively,  both with $\beta = 0$.
On the other hand, intensity and polarization maps for unpolarized GW setting $ \langle \vert h_{+} \vert^{2} \rangle = \langle \vert h_{\times} \vert^{2} \rangle$ are shown in Fig.~\ref{fig:dipole_unpolarized}.
The imprint of the azimuthal symmetry of the magnetic field and the quadrupole structure of the GW are particularly visible in the upper panel of Figs.~\ref{fig:dipole_polarized} and \ref{fig:dipole_unpolarized}.
It is noteworthy that maximal GW to EM conversion happens at $ \theta_{\gamma} = \pi/2 $ which shows the nontrivial directional dependence which only can be caught by considering the full 3D configuration.
From Fig.~\ref{fig:dipole_unpolarized} we note that while the intensity map depends significantly on the orientation of the magnetic field, parametrized by $ \alpha $, the polarization map does not, as discussed below  Eq.~\eqref{eq:pol_localized}.

\paragraph{Example: Magnetic dipole in a Stochastic GW Background.}
As discussed above, for unpolarized GWs with a fixed incident direction, the polarization of the outgoing EM wave depends only on the relative angle to the incoming GW, i.e.\ $\hat{\bm k} \cdot \hat{\bm{k}}_\gamma$, in contrast to the intensity, which  depends on the orientation of the magnetic field. This raises the question of the radiation pattern sourced by a magnetic dipole exposed to an (isotropic, unpolarized) SGWB.

For this consideration, let us take $ \hat{\bm k}_{\gamma} = (0,0,1) $. 
Integrating with respect to the direction of incoming GW (again in the $\{{\bm u}_\gamma, {\bm v}_\gamma\}$ basis),
\begin{equation}
    \begin{aligned}
    %& \langle {E_i E_j}  \rangle  =
     \int d\Omega \,   \langle {E_i E_j} \rangle  = \frac{m^{2} \omega^{4} \langle \vert h \vert^{2} \rangle }{960 \pi r^{2}} \begin{pmatrix}
        33 - 21 c_{2 \alpha} - 7 s^{2}_{\alpha}  c_{2 \beta} & - 7 s_{\alpha}^{2} s_{2\beta} \\
         - 7 s_{\alpha}^{2} s_{2\beta} & 33 - 21 c_{2 \alpha} + 7 s^{2}_{\alpha} c_{2 \beta} 
    \end{pmatrix}  \,, 
\end{aligned}
\end{equation}
we obtain 
\begin{align}
    \frac{d\sigma}{d\Omega}  = \frac{1}{10} Gm^{2} \omega^{2} \left( 11 - 7 c_{2 \alpha} \right) %&& \text{(dipole, SGWB)} 
    \,,
&&
\text{and}
&&
    \bm{\xi} =  -  \left( \frac{ 7 s^{2}_\alpha }{33 - 21 c_{2 \alpha}} \right)  \left( s_{2 \beta} \,,0, c_{2 \beta} \right) %&& \text{(dipole, SGWB)}
    \,.
\end{align}

\begin{figure}
    \centering
\includegraphics[width=0.5\linewidth]{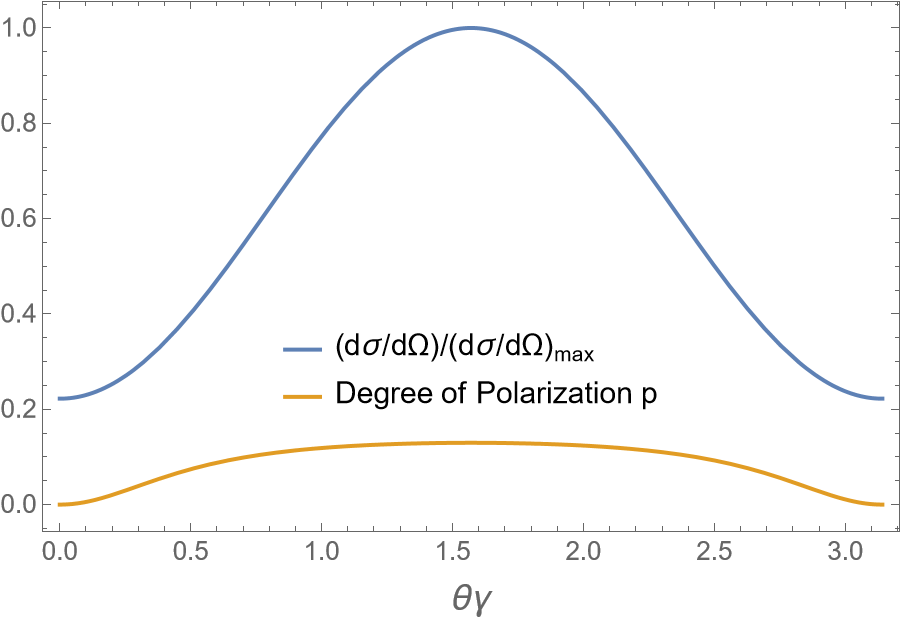}
\includegraphics[width=0.45\linewidth]{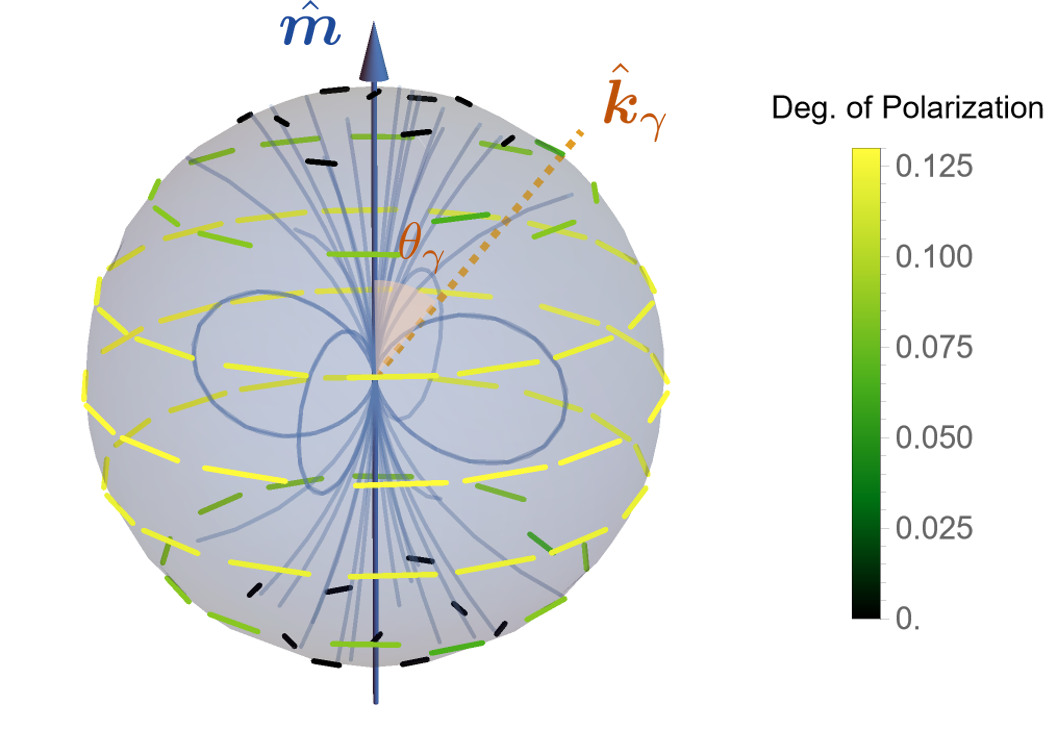}
   \caption{(Left) Ratio of the cross section for the photon conversion with respect to its maximum value (blue) and the degree of polarization (orange) of the induced EM wave from an isotropic stochastic gravitational background scattering off a magnetic dipole.
   (Right) 3D configuration of the polarization map from SGWB induced from the dipole. In both panels, we set $ \hat{\bm{m}} = \hat{\bm{e}}_{z}$.
   }
    \label{fig:SGWB_dipole}
\end{figure}

We depict the corresponding intensity and degree of polarization, and their 3-dimensional configuration with respect to the dipole axis in Fig.~\ref{fig:SGWB_dipole}.
The degree of polarization, $ p = \vert \bm{\xi} \vert$, takes a maximal value $ 7/54 \approx 0.13$ at $ \alpha = \pi / 2$, i.e. when the axis of the dipole is perpendicular to the line of sight of the observer. 
If we use the coordinate system in which the direction of the dipole to be $z$-axis, $ \alpha $ is replaced by $ \theta_{\gamma} $. 

Notably, even an isotropic SGWB results in a net polarization of the EM radiation, depending on the angle between the line-of-sight of the observer and the magnetic dipole axis.
This can be understood as a result of the intensity variation which depends on both the magnetic field direction and the incoming GW direction.
While for each individual GW, the polarization of the produced EM wave only depends on $\hat{\bm k} \cdot \hat{\bm{k}}_\gamma$ and notably not the magnetic field direction, the latter modulates the intensity of the EM waves sourced by GW waves arriving from different directions, thus leading to a net polarization of the outgoing EM radiation.

This example illustrates the rich structure of GW scattering on background magnetic fields, extending the usual GW to EM conversion probability evaluated along the GW propagation direction. The formulation of the impact of the GW in terms of an effective current is particularly useful to reduce the question to a classical EM problem which can solved with the extensive classical EM toolbox.

This formalism immediately be applied also to the case of relativistic axions converting into photons in magnetized regions. We give the corresponding expressions in App.~\ref{app:axion}, recovering some well-known properties of axion to photon conversion.

\section{Towards including Medium Effects}
\label{sec:Towards_including_Medium_Effects}

In many environments relevant for GW to photon conversion, photons possess a non-trivial dispersion, which can be modeled by  an effective mass $\mu$, in general position-dependent. In this case, Eq.~\eqref{eq:Ajeff} generalizes to
%e
\begin{align}
    \left[ \Box - \mu^{2}(\bm{x}) \right] A_{h}^{\mu} = - j_{\rm eff}^{\mu}\,.
    \label{eq:eom_mass}
\end{align}
One example is the plasma mass induced by a thermal plasma along the photon trajectory.
The corresponding dispersion relation is $ \omega^{2} = k^{2} + \mu^{2} $, where $ \omega$ is the photon energy in vacuum.  We will also use the notation $ k = \omega + \Delta $ with $ \Delta \approx - \mu^{2} / (2\omega) $ for small $ \mu \ll \omega $.

The goal of this section is to demonstrate how to include this effect in the formalism developed in the previous sections, focusing on the case where the effective mass changes adiabatically compared to the wavelength of the GW.

\subsection{Adiabatic and Stationary Phase Approximation}

\label{sec:Adiabatic Case_WKB}

Let us come back to the magnetic domain example of Sec.~\ref{sec:Magnetic_Domain_Model} with a GW propagating in $ \hat{\bm e}_{x} $ direction, now including a constant, but small mass compared to the energy of the GW, i.e. $ 0 < \mu  \ll \omega $.
From Eq.~\eqref{eq:eom_mass} we read off that this corresponds to the replacement $\omega \mapsto \sqrt{\omega^2 - \mu^2}$ in the Green's function, i.e.\ in the integrand of Eq.~\eqref{eq:MD_Greens_1}, evaluated in Eq.~\eqref{eq:3Dto1D}:
\begin{equation}
\begin{aligned}
    \bm{A}_{h}^{T}(t,x) & = - \int_{0}^{L} dx^{\prime} \frac{ e^{i \sqrt{\omega^{2} - \mu^{2}} ( x - x^{\prime} ) } }{ 2 i \sqrt{\omega^{2} - \mu^{2}} } \bm{j}_{\rm eff}(x^{\prime})  \\
    & \simeq - \frac{1}{ 2\ \sqrt{2} }  e^{-i \omega (t-x) } \int_{0}^{L} dx^{\prime} B_{0}(x^{\prime}) \exp \left( i \frac{\mu^{2}}{2 \omega} x^{\prime} \right) (0, h_+, h_\times) \,,
\end{aligned}
\end{equation}
where we used the form of the effective current given in Eq.~\eqref{eq:effective_current_mag_domain}.
This result can be generalized to the regime where $\mu$ and $ B_{0} $ are no longer constant, but vary much more slowly than $\omega^{-1}$,
employing a WKB approximation\footnote{
To be precise, the WKB approximation holds when $\vert k^{\prime} / k \vert , \vert  B_{0}^{\prime} / B_{0} \vert  \ll k $ where $ k^{2} = \omega^{2} - \mu^{2} $, with the prime indicating a spatial derivative along the line of sight (see App.~\ref{app:WKB}).
Under the assumption of a small plasma mass $ \mu \ll \omega $, these conditions can be written as
$ (\mu^{2})^{\prime} \ll \omega^{3}$ and $ B_{0}^{\prime} / B_{0} \ll \omega$.
Note that the former condition is weaker than the adiabaticity of $ \mu $ itself.
For example, in the case of neutron star magnetospheres, these conditions are all well satisfied.
}
\begin{align}
    \bm{A}_{h}^{T} \simeq - \frac{1}{ 2\sqrt{2} }  e^{-i \omega (t-x) } \int_{0}^{L} dx^{\prime} \, B_{0}(x^{\prime}) \exp \left( i \int_{0}^{x^{\prime}} dx^{\prime\prime} \frac{\mu^{2}(x^{\prime\prime})}{2 \omega}  \right) (0, h_+, h_\times) \,,
    \label{eq:Ah_WKB}
\end{align}
and accordingly the conversion probability is given by
\begin{align}
    \mathcal{P}_{h \rightarrow \gamma}
    = 4 \pi G \left\vert \int_{0}^{L} dx^{\prime} \,  B_{0}(x^{\prime}) \exp \left( i \int_{0}^{x^{\prime}} dx^{\prime\prime} \, \frac{\mu^{2}(x^{\prime\prime})}{2 \omega}   \right) \right\vert^{2}\,.
    \label{eq:Ph_WKB} 
\end{align}
This matches the results obtained in \cite{Raffelt:1987im}. For a GW with an arbitrary incident angle, $B_{0}$ and $dx^{\prime}$ in the above equation should be replaced by $B_{0}^{T}$, the magnitude of the magnetic fields transverse to the direction of the gravitational wave, and $d\ell^{\prime} $, the line element in the direction of GW propagation, as can be seen from Eq.~\eqref{eq:AT_domain_revisited}.

We note, however, that this expression no longer holds if the orientation of the magnetic field traversed is not uniform.
Since the polarization vector of the transmitted EM wave is given by the effective bulk current~\eqref{eq:jb}, which depends on the orientation of the magnetic field, a change in the direction of the magnetic field leads to the generation of EM waves with different polarization vectors, which no longer add up fully coherently.

Starting from Eq.~\eqref{eq:Ah_WKB}, a further approximation commonly employed is the stationary phase approximation~\cite{Hook:2018iia,McDonald:2024nxj,Long:2024qvd}. Identifying $f(x') = \int dx'' \mu^2(x'')/(2 \omega)$ as a rapidly varying phase in the integral over $x'$, the dominant contribution to the latter arises around $x_\text{res}$ with $f'(x') = 0$ at $x^{\prime} = x_\text{res}$, corresponding to $\mu^2(x_\text{res}) = 0$, while contributions from other points interfere destructively.
This yields
\begin{align}
  \int_0^L dx' B_0(x') e^{ i f(x')} \approx B_0(x_\text{res})  e^{ i f(x_\text{res})} \left( \frac{2 \pi}{|f''(x_\text{res})|} \right)^{1/2} \,,
\end{align}
and hence
\begin{align}
    \mathcal{P}_{h \rightarrow \gamma} \approx \left. 
    \frac{ 16 \pi^{2} G \omega B_{0}^{2} (x) }{ \vert (\mu^{2}(x))^{\prime}\vert } \right\vert_{x = x_{\rm res}}  \,  \qquad (\text{stationary phase approximation}) .
    \label{eq:saddle_point_approx}
\end{align}
The physical interpretation of this conversion probability is an immediate generalization of the results obtained in the previous section: For a vanishing photon mass and a magnetic field with a uniform direction (and sign), the EM waves generated at different points in the magnetic region interfere constructively leading to resonant conversion (hence the subscript `\emph{res}').
A finite photon mass induces a phase shift between the EM waves generated and the GW phase, with the latter inherited by EM waves generated at a later point.
This leads to partially destructive interference of the EM waves.

The stationary phase approximation is valid if the characteristic width of the region of resonant conversion, $ L_{\rm res} \sim 1 / \sqrt{ f^{\prime\prime} (x_{\rm res}) } =  \left( (\mu^{2})^{\prime} / 2\omega \right)^{-1/2} $, is smaller than the typical length scale of the magnetic fields, $ L_{\rm mag} \sim (B_{0}^{\prime} / B_{0})^{-1} $, i.e.
\begin{align}
    \left( \frac{B_{0}^{\prime}}{B_{0}} \right)^{2} \ll \left( \frac{\mu^{2}}{2 \omega} \right)^{\prime} \,.
    \label{eq:adiabativity_magnetic_field}
\end{align}
We note that the conversion probability scales as $\omega^2 L_\text{res}^2$, as expected for resonant conversion. 

\subsection{Application:  neutron stars
}

\label{sec:neutron_star}

Neutron stars are well-known for having strong magnetic fields, which may provide an intriguing chance of observing GW to photon conversion~\cite{Ito:2023fcr, Ito:2023nkq, Dandoy:2024oqg, McDonald:2024nxj}.
In this section, we discuss the implications of the effective mass corrections to photons in neutron stars modeling them using  the Goldreich-Julian (GJ) framework \cite{Goldreich:1969sb}, in which  their magnetosphere is approximated as a dipole (see Sec.~\ref{sec:localized_sources})
\begin{align}
    \bm{B}_{0} = \frac{B_{\rm max}}{2} \left(  \frac{r_{\rm NS}}{r} \right)^{3} \left( 3 \hat{\bm{r}} ( \hat{\bm{m}} \cdot \hat{\bm{r}} ) - \hat{\bm{m}}  \right)\,, \qquad (r > r_{\rm NS} )\,.
\end{align}
Here $r_\text{NS}$ is the neutron star radius and the maximal amplitude $B_{\rm max}$ is obtained at ${\bm r} = r_\text{NS} \, \hat{\bm m}$.
The electron density is found by the requirement of a self-consistent solution to Maxwell’s equations, in which particles confined to magnetic field lines corotate with the star. Ignoring relativistic corrections,\footnote{
In more detail, consistency requires the electron density to follow from the electric field: $\nabla \cdot \mathbf{E} =-en_e $. Furthermore, taking the divergence of the non-relativistic Lorentz force we obtain a relation between the electron velocity and the magnetic field
    \begin{align}    \label{eq:f10}
    m_e  \frac{d}{dt}\nabla \cdot \mathbf{v}= -e (-e n_e +\nabla \cdot(\mathbf{v} \times \mathbf{B}))\,,&&\text{where}&& \mathbf{v} = \boldsymbol{\Omega} \times \mathbf{r}\,.
    \end{align}
The last expression is  the corotating hypothesis. Applying ordinary identities to this, one obtains  $\nabla \cdot \mathbf{v} =0 $, $\nabla  \times \mathbf{v} = 2 \mathbf{\Omega} $ together with  $\nabla \cdot(\mathbf{v} \times \mathbf{B}) =  ( \nabla  \times \mathbf{v}) \cdot \mathbf{B} -\mathbf{v} \cdot ( \nabla  \times \mathbf{B})$. Since Ampère's Law gives  $\nabla  \times \mathbf{B} = -e n_e \mathbf{v}$, Eq.~\eqref{eq:f10} reads $ 0 = -e n_e +2 \mathbf{\Omega}\cdot \mathbf{B}+ e n_e v^2$, leading to
\begin{align}    
     n_e &= \frac{2\, \mathbf{B} \cdot \boldsymbol{\Omega}}{e \left(1 - \frac{1}{2} v^2\right) } \approx  \frac{2}{e } \mathbf{B} \cdot \boldsymbol{\Omega} \,.
\end{align}
This argument also shows that $E \sim B v$ in the GJ framework, hence conversions induced by the electric field remain subdominant as long as $r \ll c/\Omega$, i.e., within the model’s domain of validity.
} this leads to
\begin{equation}
    \begin{aligned}
   n_{e} & = \frac{4 \pi}{e}\frac{ \hat{\bm{\Omega}} \cdot \bm{B}_{0} }{  T} \\
   & \simeq 3.5 \times 10^{11} \left( \frac{r_{\rm NS}}{r} \right) \left( \frac{B_{\rm max}}{10^{13} \, \text{G}} \right) \left( \frac{1~\text{s}}{T} \right) \left\vert 3 \cos \theta_{\gamma} (\hat{\bm{m}} \cdot \hat{\bm{r}} ) - \cos \alpha  \right\vert~\text{cm}^{-3} \,,
    \end{aligned}
\end{equation}
where $ T $ is the period of the NS rotation and we take the rotation axis to be $ \hat{\bm \Omega} = \hat{\bm{e}}_{z}$, 
which can differ from the axis of the magnetic dipole, $ \bm{m} = m (\sin \alpha, 0, \cos \alpha) $.
For concreteness, we will be assuming photons propagating radially such that $\hat{\bm k}_\gamma = \hat{\bm r}$.

There are two contributions to the effective mass, $ \Delta \simeq - \mu^{2} / (2\omega) = \Delta^{\rm vac} + \Delta^{\rm pla}$, namely a vacuum polarization contribution  coming from higher order corrections in QED and the plasma mass\footnote{The effective mass depends on the polarization state of the photon. In this section, for definiteness, we will consider the transverse $ || $ polarization. For the longitudinal $\perp $ polarization,  $ \Delta^{\text{vac}} = \frac{4 \alpha \omega}{90\pi} \left( \frac{B_{0}}{B_{\rm crit}} \right)^{2} $ and $ \Delta^{\text{pla}} = - \frac{\mu_{\rm pla}^{2} \omega^{2} }{2 \omega (\omega^{2} - \omega_{c}^{2})} $ with cyclotron frequency $\omega_{c} = \frac{e B_{0}}{m_{e}}  \simeq 1.7 \times 10^{20} \frac{B_{0}}{10^{13} \, \text{G}} \, \text{Hz}$.
}
\begin{align}
    \Delta^{\text{vac}} = 
     \frac{7 \alpha \omega}{90\pi} \left( \frac{B_{0}}{B_{\rm crit}} \right)^{2}  \,, && \Delta^{\text{pla}} = - \frac{ \mu_{\rm pla}^{2} }{2 \omega} \quad \text{with} \quad \mu_{\rm pla}^{2} =  - \frac{ 2 \pi \alpha n_{c}}{ \omega m_{c}}\,,
\end{align}
where  $ B_{\rm crit} \equiv m_{e}^{2} / e \simeq 4.4\times 10^{13} \, \text{G} $, $\alpha \simeq 1/137$ is the fine structure constant, and
$ n_{c} $ and $m_{c}$ are the number density and mass of the relevant charged particles of the plasma, taken here to be electrons, i.e.\ $ n_{c} \simeq n_{e} $ and $ m_{c} \simeq m_{e} $. 
Note that this expression for $ \Delta^{\rm vac} $ is valid when $ B_{0} = |{\bm B}_0| \ll B_{\rm crit}$, i.e.\ as long as non-linear QED effects are negligible.

We note that this remains a simplified model of the neutron star.
In particular,  the dipole approximation of the magnetosphere is only expected to be valid at $ r \gg r_{\rm NS} $, while the magnetosphere near the surface of the neutron star requires a more detailed modeling \cite{Melrose:2016kaf}.

\paragraph{Resonant Conversion.} The two  contributions to $ \Delta $ have opposite signs, implying there can exist a region where the two contributions cancel each other such that $ \Delta(r_{\rm res}) = 0 $, and for a given frequency $f =\omega/2\pi$, this happens at
\begin{align}
    r_{\rm res} \simeq 25 \left( \frac{B_{\rm max}}{10^{13} \, \text{G}} \right)^{1/3} \left( \frac{f }{10^{13} \, \text{Hz}} \right)^{2/3} \left( \frac{T}{1 \, \text{s}} \right)^{1/3} \left( \frac{r_{\rm NS}}{10\, \text{km}} \right)
    \left\vert \frac{ 3 ( \hat{\bm{m}} \cdot \hat{\bm{r}} )^{2} + 1  }{ 3 \cos\theta_{\gamma} (\hat{\bm{m}} \cdot \hat{\bm{r}}) - \cos \alpha } \right\vert^{1/3}
    \, \text{km} \, .
\end{align}
On the one hand, from the requirement $ r_{\rm res} \geq r_{\rm NS}$, we obtain a minimum frequency for the resonant conversion in the neutron star magnetosphere,
\begin{align}
    f  > f_{\rm res,min} \simeq 2.5 \times 10^{12}  \left( \frac{B_{\rm max}}{10^{13} \, \text{G}} \right)^{-1/2} \left( \frac{T}{1\,\text{s}} \right)^{-1/2} \left\vert \frac{ 3 \cos\theta_{\gamma} (\hat{\bm{m}} \cdot \hat{\bm{r}}) - \cos \alpha }{3 ( \hat{\bm{m}} \cdot \hat{\bm{r}} )^{2} + 1  } \right\vert^{1/2} \, \text{ Hz}   \, . \label{eq:f_min}
\end{align}
Also, as noted above, our expression for $ \Delta_{\rm vac} $ holds only when $ B_{0} \ll B_{\rm crit} $. Requiring this holds at $ r=r_{\rm res}$, we find
\begin{align}
    f \gg 8.3 \times 10^{11} \left( \frac{T}{1 ~ \text{s}} \right)^{-1/2} \left\vert \frac{ ( 3 \cos \theta_{\gamma} (\hat{\bm{m}} \cdot \hat{\bm{r}}) - \cos \alpha )^{2} }{3 ( \hat{\bm{m}} \cdot \hat{\bm{r}} )^{2} + 1  } \right\vert^{1/4} ~\text{Hz} \,.
\end{align}

\begin{figure}
    \centering
    \includegraphics[width=0.75\linewidth]{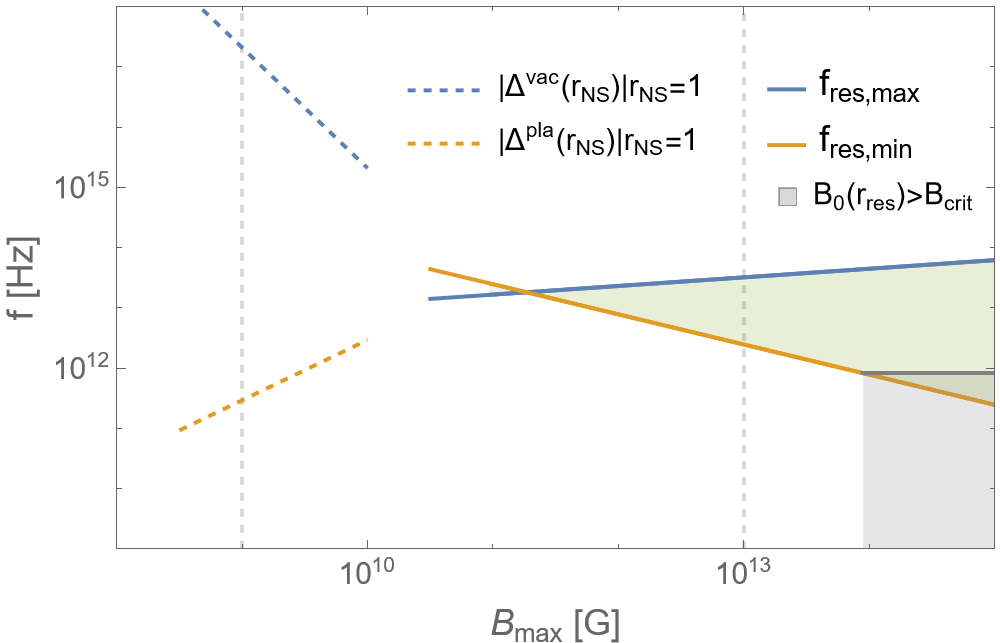}
    \caption{Frequency range of resonant conversion (shaded green, see also Fig.~\ref{fig:neutron_BM1}), ranging from $f_{\rm res,min}$ (solid orange) to $f_{\rm res,max}$ (solid blue) as a function of $ B_{\rm max} $, the magnitude of the magnetic field at the surface of the neutron star, for $ T=1\,\text{s}$ and $ r_{0} = 10 \, {\rm km}$. The gray line and the correspondingly shaded rectangular region denote $ B_{0}(r_{\rm res}) > B_{\rm crit} $. The blue (orange) dashed line corresponds to $ \vert \Delta^{\rm vac (pla)} (r_{\rm NS}) \vert r_{\rm NS} = 1  $, indicating the range of the plateau in Fig.~\ref{fig:neutron_BM2}. The two vertical gray lines show the benchmarks used in Figs.~\ref{fig:neutron_BM1} and \ref{fig:neutron_BM2}, respectively.
    }
    \label{fig:neutron_freq}
\end{figure}

On the other hand, at high frequencies, the stationary phase approximation holds only when the background magnetic field remains approximately constant within the length scale of the resonance, $ L_{\rm res} \sim 1 / \sqrt{ f^{\prime\prime}(x_{\rm res})  } \sim \sqrt{ (\mu^{2}/2\omega)^{\prime} }$,
to avoid destructive interference, i.e.
\begin{align}
    \left( \frac{B_{0}^{\prime}}{B_{0}} \right)^{2} \ll \left( \frac{ \mu^{2} }{2 \omega} \right)^{\prime} \, ,
\end{align}
so that we obtain the condition
\begin{equation}
    \begin{aligned}
    \label{eq:f_max}
    f < f_{\rm res,max} \simeq  3.2 \times 10^{13} ~\text{Hz}   \left( 
    \frac{B_{\rm max}}{10^{13} \, \text{G}} \right)^{1/7} & \left( \frac{T}{1 \, \text{s}} \right)^{-5/7} \left( \frac{r_{0}}{10\,\text{km}} \right)^{3/7}  \\
    & \times \left\vert \frac{ ( 3 \cos \theta_{\gamma} (\hat{\bm{m}} \cdot \hat{\bm{r}}) - \cos \alpha )^{5} }{ (3 ( \hat{\bm{m}} \cdot \hat{\bm{r}} )^{2} + 1  )^{2}} \right\vert^{1/7}  \,.
    \end{aligned}
\end{equation}
See Fig.~\ref{fig:neutron_freq} for the ranges of $f_{\rm res,min}$ and $ f_{\rm res,max} $, 
illustrating that resonant conversion requires a sufficiently large magnetic field of the neutron star,
\begin{align}
    B_{\rm max} > 1.9 \times 10^{11} ~\text{G} \left( \frac{r_{\rm NS}}{10~\text{km}} \right)^{-2/3} \left(  \frac{T}{1~\text{s}} \right)^{1/3}  \left\vert  (3 ( \hat{\bm{m}} \cdot \hat{\bm{r}} )^{2} + 1  ) ( 3 \cos \theta_{\gamma} (\hat{\bm{m}} \cdot \hat{\bm{r}}) - \cos \alpha )  \right\vert^{-1/3} \!\!.
    \label{eq:Bthreshold}
\end{align}
which is satisfied in many young pulsars and magnetars with strong magnetic fields, with $ B_{\rm max} \sim 10^{11-15} \text{G} $. On the other hand, for many millisecond pulsars with $ T \sim 1~\text{ms}$, the necessary minimum magnetic field decreases, but at the same time typical magnitudes of the magnetic fields are also small $ \sim 10^{8-9}~\text{G}$, and there is typically no resonant conversion.

\paragraph{Low- and High-Frequency Approximations.} Given the expression of the graviton to photon conversion probability in Eq.~\eqref{eq:Ph_WKB}, two more approximations can be done: in the low and high frequency limits, we can neglect the contribution from $ \Delta^{\rm vac}$ and $ \Delta^{\rm pla}$ to the effective mass respectively.

For illustrative purposes, we choose the optimal case where the rotation axis of the pulsar is aligned with the direction of the dipole axis (i.e.\ $\alpha = 0$), and the gravitational wave incident angle is perpendicular to this axis (i.e.\ $ \hat{\bm{m}} \cdot \hat{\bm{r}} = 0 $). Also, we focus on the trajectory of GW which pass through the center of the neutron star, while neglecting conversion inside of it. Then, we can take $ B_{0}^{T} = (B_{\rm max}/2) (r_{\rm NS}/r)^{3} $ in Eq.~\eqref{eq:Ph_WKB} with $B_{0} \rightarrow B_{0}^{T}$, and obtain~\cite{Dandoy:2024oqg}
\begin{align}
    \mathcal{P}_{h \rightarrow \gamma}(f) 
     \simeq \begin{dcases}
        \left. \frac{ 8 \pi G B_{\rm max}^{2}  }{ (\Delta^{\rm pla})^{2} } \left\vert \sin\left( \frac{\Delta^{\rm pla} r }{4} \right) \right\vert^{2}  \right\vert_{r=r_{\rm NS}} \propto f^{2} & (\text{low freq. limit}) \\
        \left. \frac{ 2 \pi G B_{\rm max}^{2} r^{2} }{5^{6/5} (\Delta^{\rm vac} r)^{4/5}} \left\vert \Gamma\left(  \frac{2}{5}\right) - \Gamma\left(  \frac{2}{5}, - i \frac{\Delta^{\rm vac} r}{5}\right) \right\vert^{2}  \right\vert_{r=r_{\rm NS}} \!\!\!\!\!\!\! \propto f^{-4/5} & (\text{high freq. limit})
    \end{dcases}
    \label{eq:low_high_approx}
\end{align}
where $ \Gamma(x) $ is the Gamma function, and $ \Gamma(x,a) $ is the incomplete Gamma function.

\begin{figure}
    \centering
    \includegraphics[width=0.75\linewidth]{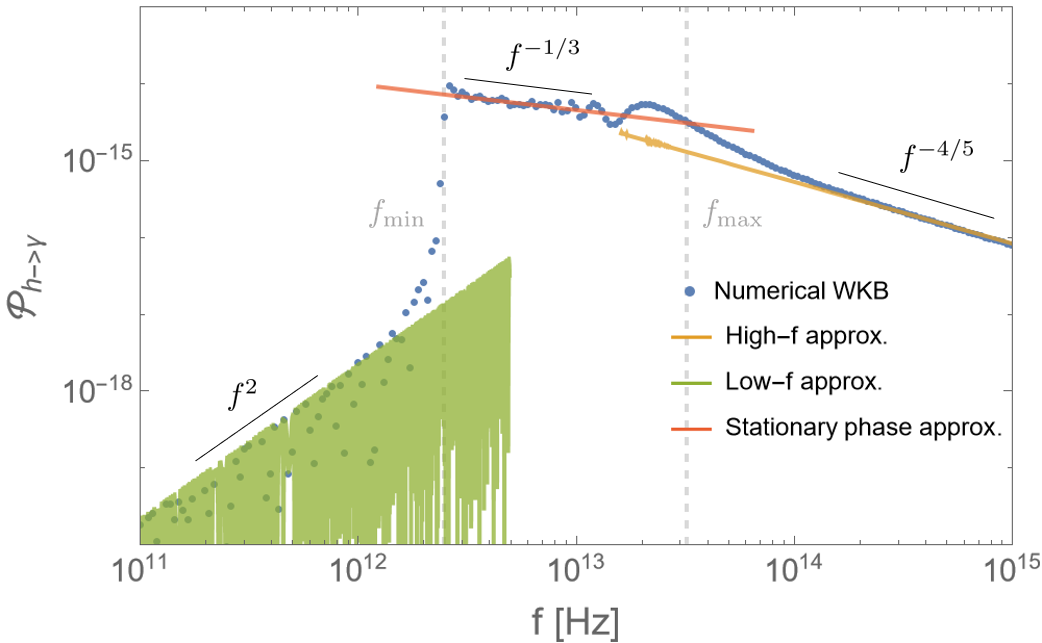}

    \caption{Probability for GW to photon conversion as a function of the frequency  within a large magnetic field in the simplified neutron star model described in the main text, as a function of frequency, for $T= 1 \, \text{s}$, $r_{0} = 10\,\text{km}$ and $ B_{\rm max} = 10^{13} \,\text{G} $. We show the numerically integrated WKB expression in Eq.~\eqref{eq:Ph_WKB} (blue dotted), together with three different approximation schemes which capture well the behavior in the different frequency regimes (see text).
    Resonant conversion in the neutron star magnetosphere occurs between the two vertical dashed gray lines.
    }
    \label{fig:neutron_BM1}
\end{figure}

\begin{figure}
    \centering
    \includegraphics[width=0.75\linewidth]{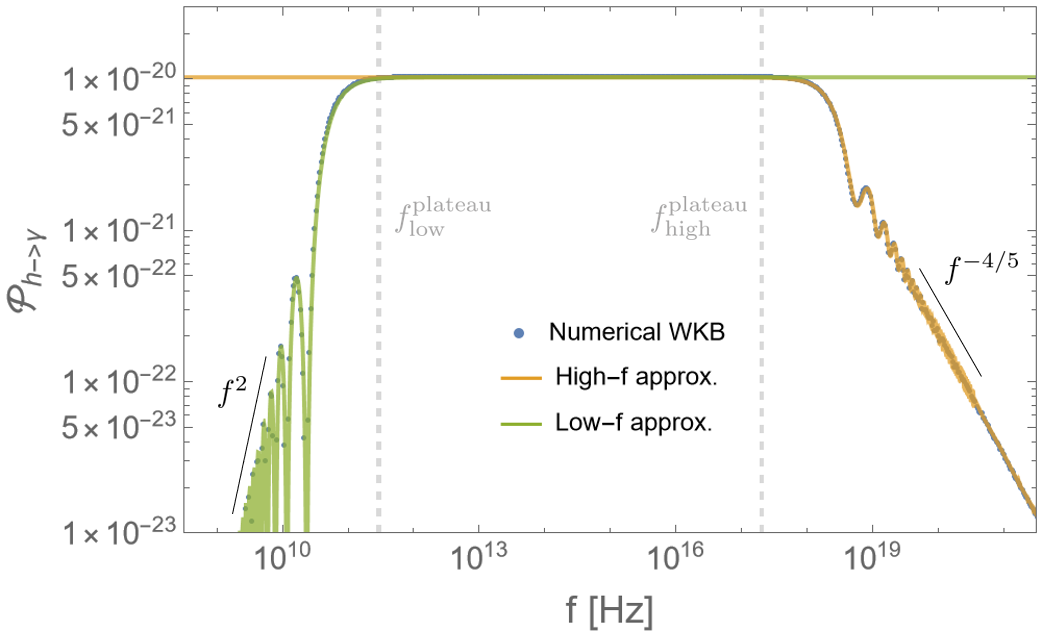}
    
    \caption{
    Same as in Fig.~\ref{fig:neutron_BM1} but with $ B_{\rm max} = 10^{9} \,\text{G} $.
    There is no resonant conversion, 
    though the plateau between the vertical dashed lines indicates quasi-resonant conversion, controlled by the magnetic field strength at the surface of the neutron star (see text for details).
    }
    \label{fig:neutron_BM2}
\end{figure}

\paragraph{Conversion Probability.} We depict two examples of the frequency dependence of the conversion probability in Figs.~\ref{fig:neutron_BM1} and \ref{fig:neutron_BM2}, taking $ T = 1~\text{s}$ and $ r_{\rm NS} = 10 ~\text{km}$ for large $B_{\rm max} = 10^{13}~\text{G}$ and small $B_{\rm max} = 10^{9}~\text{G}$, respectively.
Along with the results from the numerical integration of the WKB approximation~Eq.~\eqref{eq:Ph_WKB}, we show three approximations: low and high frequency approximation (Eqs.~\eqref{eq:low_high_approx}) and the stationary phase approximation (Eq.~\eqref{eq:saddle_point_approx}), if applicable. The latter indicates resonant conversion of GWs to photons and only arises for large $ B_{\rm max} $, see red line in Fig.~\ref{fig:neutron_BM1}, displaying a scaling $ \propto f^{-1/3} $.

On the other hand, in the case of small $ B_{\rm max} $ in Fig.~\ref{fig:neutron_BM2}, there is strictly speaking no resonant conversion. However, the conversion probability shows a plateau with the value $ \mathcal{P}_{h \rightarrow \gamma} \simeq B_{\rm max}^{2} r_{\rm NS}^{2} / (16 M_{P}^{2}) $~\cite{Dandoy:2024oqg}. This happens in the frequency regime with $ \vert \Delta^{\rm pla} (r_{\rm NS}) \vert r_{\rm NS} \ll 1  $ and $ \vert \Delta^{\rm vac} (r_{\rm NS}) \vert r_{\rm NS} \ll 1  $,  indicating a negligible dephasing of EM waves sourced at different locations, i.e., `quasi-resonant conversion'. In this case the conversion probability is dominated by the region of strongest magnetic field, arising at the surface of the neutron star,
\begin{align}
       f_{\rm low}^{\rm plateau} < f < f_{\rm high}^{\rm plateau} \,,
\end{align}
with
\begin{equation}
    \begin{aligned}
    f_{\rm low}^{\rm plateau} & = 2.9 \times 10^{11} ~\text{Hz} \left( \frac{B_{\rm max}}{10^{9} ~ \text{G}} \right) \left(  \frac{r_{0}}{10~\text{km}}\right) \left(  \frac{T}{1 ~\text{s}}\right)^{-1} ( 3 \cos \theta_{\gamma} (\hat{\bm{m}} \cdot \hat{\bm{r}}) - \cos \alpha )  \\
     f_{\rm high}^{\rm plateau} & = 2.1 \times 10^{17} ~\text{Hz} \left( \frac{B_{\rm max}}{10^{9} ~ \text{G}} \right)^{-2} \left(  \frac{r_{0}}{10~\text{km}}\right)^{-1} (3 ( \hat{\bm{m}} \cdot \hat{\bm{r}} )^{2} + 1  )^{-1}
\end{aligned}
\end{equation}
which we depict as dashed lines in Fig.~\ref{fig:neutron_freq}. However, we also note that, these results highly depend on the magnetic fields near the surface of the neutron star 
where GJ model may not work appropriately.
Consequently, the results are not as robust as those for resonant conversion captured by the stationary phase approximation which happens $ r_{\rm res} \gg r_{\rm NS}$, where the GJ model is believed to work relatively well.

In both cases, the highest conversion probability occurs in a rather thin hypersurface of the neutron star magnetosphere. This allows us to estimate the conversion probability for GWs incident from different directions by modeling this hypersurface as a magnetic domain in which the direction of the magnetic field is approximately uniform. 

\vspace{3mm}
In summary, we find that above the threshold~\eqref{eq:Bthreshold} for the magnetic field and for GWs in a suitable frequency range, resonant conversion occurs at some point in the neutron star magnetosphere, and this resonant conversion dominates the entire conversion probability along the line of sight.
Below this threshold, only quasi-resonant conversion can occur, and this is dominated by the region at the surface of the neutron star.
Our results are in agreement with Ref.~\cite{Dandoy:2024oqg}, and back up the strategy of~\cite{McDonald:2024nxj} of focusing only on the resonant regime, though we stress the limitations of this approach at low and high frequencies.
Our conversion probabilities are larger than those found in Refs.~\cite{Ito:2023fcr, Ito:2023nkq}, as the method of discretization employed there does not fully resolve the resonant region.
The Green's function method introduced in Sec.~\ref{sec:Green's_Function_Method} thus provides a unified and convenient framework to treat inhomogeneous magnetic domains as well as (adiabatically varying) plasma effects.

\section{Conclusions}
\label{sec:conclusions}

Using the effective-current formalism, we have reexamined how GWs convert into EM waves in magnetized media from a fully three-dimensional perspective and have identified several qualitative features that sharpen our understanding of the inverse Gertsenshtein effect.

We start with the analytically tractable case of a single, uniform magnetic domain and show that the reflected EM field can vanish at specific incident angles for certain GW polarizations. This GW analogue of the optical Brewster angle allows the magnetic domain to act as an ideal polarizer, reflecting only the EM waves associated with one of the two GW polarizations.

By extending the conventional Stokes formalism to gravitational radiation, we find that the forward-scattered photon field retains the GW Stokes vector, whereas the back-scattered component acquires a geometry-dependent polarization that can be nonzero even for an unpolarized stochastic GW background.

We then address more complex magnetic configurations with a Green-function approach that unifies earlier domain-model and $S$-matrix treatments, accommodates realistic three-dimensional geometries, and includes arbitrary photon dispersions. This framework reveals geometry-induced polarization signatures, delineates the validity of adiabatic approximations, and shows how the angular distribution in a dipolar field can produce net linear polarization from an isotropic unpolarized GW background.

Taken together, these results demonstrate that polarimetry -- expressed through a common Stokes language for GW and EM radiation -- provides a sensitive probe of GW-to-EM conversion and offers practical guidance for future laboratory experiments and multi-messenger observations.

\acknowledgments

We thank  Gongjun Choi, Sebastian Ellis and Jamie McDonald for helpful discussions. 
 C.G.C. is supported by a Ramón y Cajal contract with Ref.~RYC2020-029248-I, the Spanish National Grant PID2022-137268NA-C55 and Generalitat Valenciana through the grant CIPROM/22/69.
The authors are also grateful to the Mainz Institute for Theoretical Physics (MITP) of the Cluster of Excellence PRISMA$^{+}$ (Project ID 390831469) for its hospitality  during the completion of this work.

\appendix

\section{Notation}
\label{app:notation}

For a GW propagating with the momentum
\begin{align}
   \bm{k} = \omega \left( \sin \theta \cos \phi, \,
   \sin \theta \sin \phi, \,
   \cos \theta \right) \, ,
\end{align}
we can express the GW in the transverse traceless (TT) frame\footnote{In this paper we will be considering magnetic field configurations which are static in the TT frame. This is the case for GWs with a frequency which is much larger than the mechanical eigenfrequencies of the system sourcing the background EM field~\cite{Ratzinger:2024spd}.} as
\begin{align}
    h^{\rm TT}_{ij} = \left(  h_{+}  e_{ij}^{+}+ h_{\times}  e_{ij}^{\times} \right) e^{-i (\omega t -  \bm k \cdot {\bm r}) } \,,
    \label{eq:app_h}
\end{align}
with polarization tensors
\begin{align}
   e_{ij}^{+} =   \frac{1}{\sqrt{2}}  ( u_{i} u_{j} - v_{i} v_{j} ) \,, \quad  e_{ij}^{\times} = \frac{1}{\sqrt{2}}  ( u_{i} v_{j} + v_{i} u_{j} ) \,,
\end{align}
where we choose
\begin{equation}
    \begin{aligned}
    \label{eq:uv}
        \bm{v} & =  %\hat{\bm e}_{\phi} = (-\sin \phi , \cos \phi , 0 ) \,, \\
                    (\hat{\bm e}_z \times \bm k)/| (\hat{\bm e}_z \times \bm k) |  = (-\sin \phi , \cos \phi , 0 ) \,, \\
        \bm{u} & =  \bm{v}  \times \hat{\bm k} = (\cos \theta \cos \phi , \cos \theta \sin \phi , - \sin \theta ) \,.
    \end{aligned}
\end{equation}

The resulting polarization tensors are transverse, $e^\lambda_{ij}k_j = 0$ and normalized to $e^\lambda_{ij} (e^{\lambda'})^{ij} = \delta_{\lambda \lambda'}$ with $\lambda = +,\times$.
In Sec.~\ref{sec:Magnetic_Domain_Model} we introduce waves reflected on plane surfaces, for which the corresponding polarization tensors are obtained by the expressions above after inserting the reflected wave vector.
In particular, for a reflection at the $yz$-plane we find
\begin{equation}
    \begin{aligned}
     \bar{\bm k} & = \omega \left(- \sin \theta \cos \phi, \, \sin \theta \sin \phi, \, \cos \theta \right) \,, \\
     \bar{\bm v} & = (-\sin \phi , -\cos \phi , 0 ) \,, \\
     \bar{\bm u}&  = (- \cos \theta \cos \phi, \, \cos \theta \sin \phi, \, - \sin \theta )\,.
    \end{aligned}
\end{equation}

\paragraph{Visualizing Polarization.}

\begin{figure}
    \centering
    \includegraphics[width=0.95\linewidth]{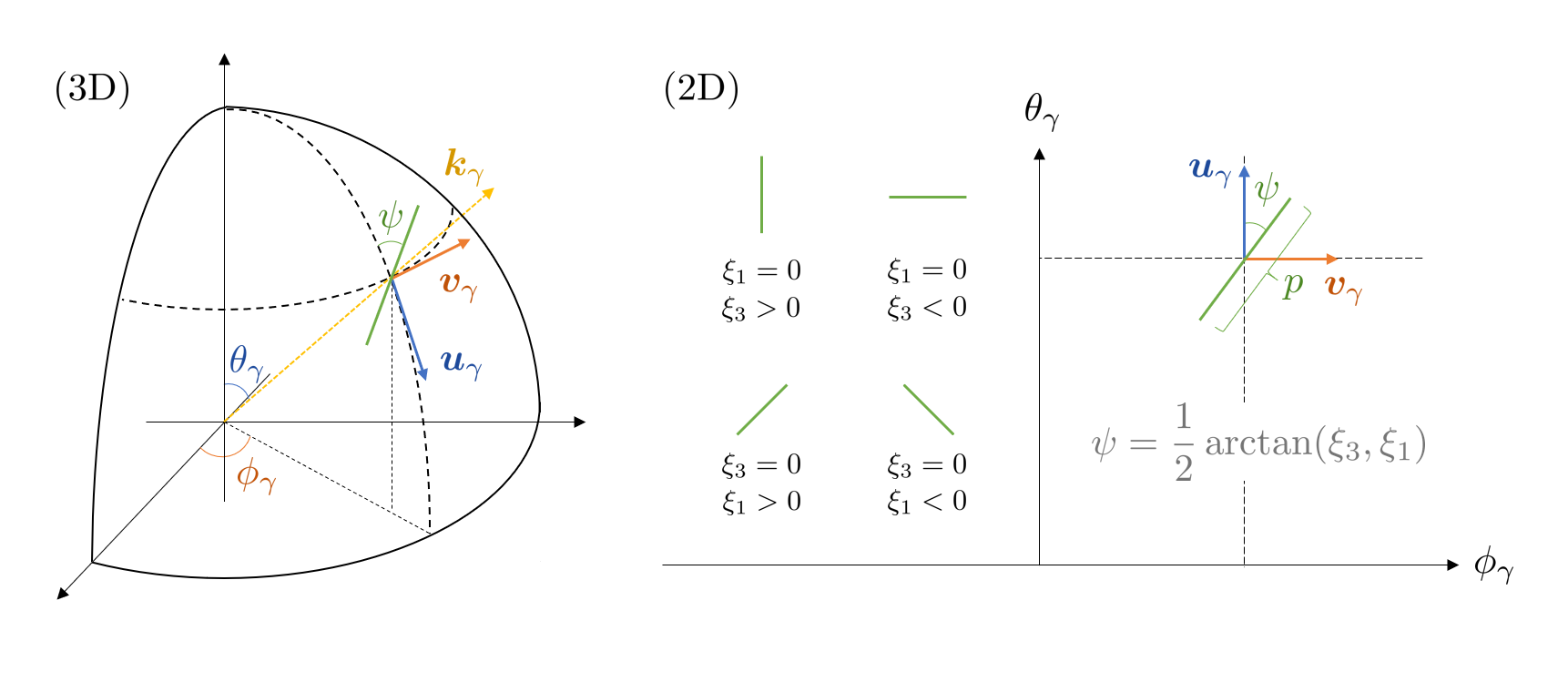}
    \caption{Schematic picture of visualizing polarization map by projecting from 3D configuration (left) to 2D ($  \phi_{\gamma}, \theta_{\gamma}$) plane (right).
    }
    \label{fig:visualization}
\end{figure}

Here, we provide the details of visualizing the polarization map on the 2D plane ($ \phi_{\gamma}, \theta_{\gamma} $) (Fig.~\ref{fig:Ref_Domain_pol}, for instance) and 3D unit sphere (Fig.~\ref{fig:SGWB_dipole})\cite{Hu:1997hv,Dodelson:2020bqr,Baumann:2022mni}. 
In our convention, for a given propagating photon direction $ \bm{k}_{\gamma} $, the transverse vector $ \bm{u}_{\gamma} $ ($\bm{v}_{\gamma} $) is obtained by varying $ \theta_{\gamma} $ ($ \phi_{\gamma} $) with $ \phi_{\gamma} $ ($ \theta_{\gamma} $) fixed, respectively, as represented in Fig.~\ref{fig:visualization}.

One of the conventional ways of representing the polarization is to use \textit{headless} vector $ \overline{\bm{p}}$, which are characterized by its magnitude and an angle.
First of all, its magnitude is determined by the degree of polarization \eqref{eq:deg_of_pol}. Therefore, for unpolarized light, the length of the vector shrinks to zero.
Second, the angle $ \psi $ with respect to $ \bm{u}_{\gamma} $ given by
\begin{align}
    \psi = \frac{1}{2} \arctan (\xi_{3} , \xi_{1})
\end{align}
where $ \arctan (x , y) \in [0,2\pi) $ is the angle between $ + \hat{\bm e}_{x} $ axis and the line connecting origin $(0,0)$ and $ (x , y) $ in a counter-clockwise way. By saying headless, we do not distinguish $ \overline{\bm{p}} $ and $ -\overline{\bm{p}} $.

As examples,
\begin{itemize}
    \item $\xi_{1} = 0 $ and $ \xi_{3} > 0$, the vector $(\xi_{3}, \xi_{1})$ is orientated along $ (1,0) $, so $\arctan(\xi_{3}, \xi_{1}) = 0$ and $\psi = 0$ (parallel to $ \bm{u} $).

    \item $\xi_{1} = 0 $ and $ \xi_{3} < 0$, the vector $(\xi_{3}, \xi_{1})$ is orientated along $ (-1,0) $, so $\arctan(\xi_{3}, \xi_{1}) = \pi$ and $\psi = \pi/2$ (parallel to $ \bm{v} $).

    \item $\xi_{3} = 0 $ and $ \xi_{1} > 0$, the vector $(\xi_{3}, \xi_{1})$ is orientated along $ (0,1) $, so $\arctan(\xi_{3}, \xi_{1}) = \pi/2$ and $\psi = \pi/4$ (parallel to $ \frac{1}{\sqrt{2}} (\bm{u} + \bm{v}) $).

     \item $\xi_{3} = 0 $ and $ \xi_{1} < 0$, the vector $(\xi_{3}, \xi_{1})$ is orientated along $ (0,-1) $, so $\arctan(\xi_{3}, \xi_{1}) = 3\pi/2$ and $\psi = 3\pi/4$ (parallel to $ \frac{1}{\sqrt{2}} (\bm{u} - \bm{v}) $).
\end{itemize}

The polarization can be represented in 3D plot (left figure of Fig.~\ref{fig:visualization}) with the vector
\begin{align}
    \overline{\bm{p}}^{\rm (3D)} = p ( \bm{u} \cos \psi  + \bm{v} \sin \psi  ) \, ,
\end{align}
or in 2D plot (right figure of Fig.~\ref{fig:visualization}) with
\begin{align}
    \overline{\bm{p}}^{\rm (2D)} = p ( \sin \psi, \cos \psi   ) \,.
\end{align}

It is also instructive to see what happens at the level of the vector potential $ \bm{A}_{h} $ in the case of the pure state. 
From the fact that $ \bm{A}_{h} \cdot \bm{k}_{\gamma} = 0  $, it is always possible to write $ \bm{A}_{h} = ( A_{u} \cdot \bm{u}_{\gamma} ) \bm{u}_{\gamma} + (A_{h} \cdot \bm{v}_{\gamma}) \bm{v}_{\gamma} $. For instance, $ \bm{A}_{h} \cdot \bm{v}_{\gamma} = 0 $ corresponds to the case when $\rho_{11} = 1 $ and $ \rho_{22} = 0$ (hence $ \xi_{3} = 1 $), so the polarization vector becomes vertical line parallel to $ \bm{u}_{\gamma} $.
Similarly, the $ \bm{A}_{h} \cdot \bm{u}_{\gamma} = 0 $ (i.e.\ $ \xi_{3} = -1 $) case is represented a the horizontal line.
In the main text, for instance in Fig.~\ref{fig:Ref_Domain_pol}, these lines are depicted in the blue and orange dashed lines respectively.

\section{Magnetic Domain Model: Technical Details}
\label{app:Derivations of Magnetic Domain Wall Model}

This appendix provides additional details on the calculations leading to the results presented in Sec.~\ref{sec:Magnetic_Domain_Model}.
See Fig.~\ref{fig:magnetic_domain} in the main text for a sketch of the setup.

\paragraph{Direct Solution to Maxwell Equations.}
The general solution of Maxwell's equations with source term $ {\bm j}_{\rm eff}$ can be obtained by combining the general homogeneous solution (solutions of the source-free Maxwell equations, i.e.\ right-moving plane wave solutions with wave vector $ \bm{k}$ and left-moving solutions with wave vector $ \bar{\bm k}$.) with a particular solution sourced by $ {\bm j}_{\rm eff}$ in the magnetized region between $x_1$ and $x_2$:
\begin{align}
    \begin{dcases}
     \bm{A}_{h}^{\rm (I)} = \bm{a}^{\rm (I)}_{L} e^{ - i ( \omega t - \bar{\bm k} \cdot \bm{r} ) }   & (x < x_{1}) \\
     \bm{A}_{h}^{\rm (II)} = \bm{b}_{L}^{\rm (II)} e^{-i (\omega t - \bar{ \bm{k} } \cdot \bm{r} ) }  +\bm{b}_{R}^{\rm (II)} e^{-i (\omega t - \bm{k} \cdot \bm{r} )} +  \bm{A}_{h}^{(p)}  & (x_{1}  < x < x_{2} ) \\
      \bm{A}_{h}^{\rm (III)} = \bm{c}^{\rm (III)}_{R} e^{-i (\omega t - \bm{k} \cdot \bm{r} )}   & (x > x_{2})
    \end{dcases} \label{eq:general_solution}
\end{align}
where we set $ \bm{a}_{R}^{\rm (I)} = \bm{c}_{L}^{\rm (III)} = 0 $ due to causality (i.e.\ we set boundary conditions such that no right-moving wave at $ x < x_{1} $ and no left-moving at $ x > x_{2} $). Given the effective bulk current~\eqref{eq:jb}, a particular solution in region (II) is given by
\begin{align}
    {\bm A}_{h}^{(p)} =
   - \frac{ B_{0} x }{2\sqrt{2} c_{\phi}} 
    e^{- i (\omega t - \bm{k} \cdot {\bm r} )}
   ( h_{+} \bm{v} - h_{\times} \bm{u} )
   =  \left[ h_{+} \begin{pmatrix}
          - s_{\phi} \\
          c_{\phi} \\
          0
      \end{pmatrix} - h_{\times}
      \begin{pmatrix}
          c_{\theta} c_{\phi} \\
          c_{\theta} s_{\phi}\\
          -s_{\theta}
      \end{pmatrix}
      \right]
   \,.
\end{align}
Note that this particular solution satisfies the Coulomb gauge condition, $ \nabla \cdot \bm{A}_{h}^{(p)} = 0$.

The remaining coefficients in Eq.~\eqref{eq:general_solution} are determined by the boundary conditions at the two interfaces, given in Eq.~\eqref{eq:boundary3}.
Explicitly, one obtains
\begin{align}
    \bm{a}_{L}^{\rm (I)} & = - \frac{i B_{0}}{4 \sqrt{2} \omega c_{\phi}^{2} s_{\theta} }  
 \left( e^{2i x_2 \omega c_{\phi} s_{\theta} } - e^{2i x_1 \omega c_{\phi} s_{\theta} } \right) \left[ h_{+} \begin{pmatrix}
          s_{\phi} \\
          -c_{2\theta} c_{\phi} \\
          c_{\theta} s_{\theta} s_{2 \phi}
      \end{pmatrix} - h_{\times}
      \begin{pmatrix}
         c_{\theta} c_{\phi} \\
         c_{\theta} s_{\phi} \\
         s_{\theta} c_{2\phi}
      \end{pmatrix}
      \right] 
      \,  \nonumber  \\
     & = \frac{i B_{0}}{4 \sqrt{2} \omega c_{\phi}^{2} s_{\theta} }  
     \left( e^{2i x_2 \bar{\bm k}\cdot {\bm n}} -  e^{2i x_1 \bar{\bm k}\cdot {\bm n}}  \right) \left[  \left(  h_{+} \bm{v} - h_{\times} \bm{u}  \right)  + 2 s_{\theta} c_{\phi} \hat{\bm{n}} \times \left(  h_{+} \bm{u} + h_{\times} \bm{u}  \right)  \right] 
\end{align}
and
\begin{equation}
    \begin{aligned}
    \bm{c}_{R}^{\rm (III)} = 
    - \frac{B_{0} (x_{2} - x_{1})}{2 \sqrt{2} c_{\phi}}
    \left[ h_{+} \begin{pmatrix}
         -s_{\phi} \\ c_{\phi} \\ 0 
    \end{pmatrix} - h_{\times} \begin{pmatrix}
        c_{\theta} c_{\phi} \\
        c_{\theta} s_{\phi} \\
        - s_{\theta}
    \end{pmatrix} \right]  =  - \frac{B_{0} (x_{2} - x_{1})}{2 \sqrt{2} c_{\phi}}
     \left(  h_{+} \bm{v} - h_{\times} \bm{u}  \right) \,, 
    \end{aligned}   
\end{equation}
which gives the reflected and transmitted waves in region (I) and (III). Plugging this into Eq.~\eqref{eq:general_solution} gives the solution in the main text, Eqs.~\eqref{eq:AT} and \eqref{eq:AR}.
Moreover, for region (II), i.e.\ within the magnetic domain, we obtain
\begin{align}
    b_{L}^{\rm (II)} = -\frac{i B_{0} e^{2 i x_{2} \omega c_{\phi} s_{\theta}} }{4 \sqrt{2} \omega c_{\phi}} \left[ h_{+} \begin{pmatrix}
        s_{\theta}^{-1}  t_{\phi} \\
        - c_{2\theta} s_{\theta}^{-1}  \\ 2 c_{\theta} s_{\phi} 
    \end{pmatrix} + h_{\times}
    \begin{pmatrix}
        t_{\theta}^{-1}  \\
        t_{\theta}^{-1}  t_{\phi}^{-1} \\
        c_{2\phi} c_{\phi}^{-1}
    \end{pmatrix}
 \right]
\end{align}
and
\begin{align}
    b_{R}^{\rm (II)} = \frac{ i B_{0}  }{4 \sqrt{2} \omega c_{\phi}} \left[ h_{+} \begin{pmatrix}
      2 i x_{1} \omega s_{\phi} + s_{\theta}^{-1} t_{\phi}  \\
       - 2 i x_{1}  \omega c_{\phi} - c_{2\theta} s_{\theta}^{-1} \\ 2 c_{\theta} s_{\phi} 
    \end{pmatrix} + h_{\times}
    \begin{pmatrix}
        2 i x_{1} \omega c_{\theta} c_{\phi} + t_{\theta}^{-1}   \\
        2 i x_{1} \omega c_{\theta} s_{\phi} + t_{\theta}^{-1} t_{\phi} \\
       - 2 i x_{1} \omega s_{\theta} + c_{2\phi} c_{\phi}^{-1}
    \end{pmatrix}
 \right] \,.
\end{align}
Terms proportional to $x_{1}$ for right-moving modes indicate resonance due to coherently transmitted waves. 
For left-moving (i.e.\ reflected) waves this enhancement is absent.

\paragraph{Green's Function Method.}
As discussed in Sec.~\ref{sec:magnetic_domains_2}, we can also study the magnetic domain model using the Green's function method.
Here we provide additional details of this calculation, especially the derivations of Eqs.~\eqref{eq:integration} and \eqref{eq:general_sol_domain}. 

In this case, the general form of the solution is (see Eq.~\eqref{eq:MD_Greens_2}),
\begin{align}
     \bm{A}_{h}
     & =  \frac{e^{- i \omega t}}{ \sqrt{2}  }  s_{\theta}    \int d^{3} \bm{r}^{\prime}   \, \left[ 
 \left(  h_{+} \bm{v} - h_{\times} \bm{u}  \right) i \omega B_{0}(x)  - \bm{n} \times (h_{+} \bm{u} + h_{\times} \bm{v} )  B_{0}^{\prime}(x) \right]  e^{i \bm{k} \cdot \bm{r}^{\prime} }  \frac{e^{i \omega \vert \bm{r} - \bm{r}^{\prime} \vert} }{ 4 \pi \vert \bm{r} - \bm{r}^{\prime} \vert }  \,.
\end{align}
To perform the integration in the $x'y'$-plane we change variables as
\begin{align}
    \bm{k} \cdot \bm{r}^{\prime} & = k_{x}x^{\prime} + k_{y} y^{\prime} + k_{z}z^{\prime} \equiv \omega \ell^{\prime}  \,,
\end{align}
and
\begin{align}
    \begin{pmatrix}
        \tilde{y}^{\prime} \\
        \tilde{z}^{\prime} 
    \end{pmatrix}
    = \frac{1}{\sqrt{1 + s_{\phi}^{2} t_{\theta}^{2}}} \begin{pmatrix}
       s_{\phi} t_{\theta} &  1  \\
        - 1  &  s_{\phi} t_{\theta}
    \end{pmatrix} 
    \begin{pmatrix}
        y^{\prime} \\ z^{\prime}
    \end{pmatrix} \equiv R  \begin{pmatrix}
        y^{\prime} \\ z^{\prime}
    \end{pmatrix}\,,
\end{align}
with $ R^{T} R = I_{2\times2} $.
The reason for performing this $SO(2)$ rotation will soon become clear below.
Also, because $B_{0}$ (and its derivative) only has $ x^{\prime} $ dependence by the setup, $ B_{0}(x^{\prime}) = B_{0} (\omega \ell^{\prime}/k_{x} ) =B_{0} ( \ell^{\prime} s_{\theta} c_{\phi} )  $ so that
\begin{align}
     \int d^{3} \bm{r}^{\prime}   B_{0}  (x^{\prime}) e^{i \bm{k} \cdot \bm{r}^{\prime} }  \frac{e^{i \omega \vert \bm{r} - \bm{r}^{\prime} \vert} }{ 4 \pi \vert \bm{r} - \bm{r}^{\prime} \vert }  
     = \int_{\ell_{1}}^{\ell_{2}}  d\ell^{\prime}  \int_{-\infty}^{\infty} d\tilde{y}^{\prime} d\tilde{z}^{\prime} ~\frac{ B_{0}(\ell^{\prime} s_{\theta} c_{\phi}) }{s_{\theta} c_{\phi}} e^{i \omega \ell^{\prime}} \frac{e^{i \omega \vert \bm{r} - \bm{r}^{\prime} \vert} }{ 4 \pi \vert \bm{r} - \bm{r}^{\prime} \vert } \,,
\end{align}
where the factor $ 1/(s_{\theta} c_{\phi}) $ comes from the Jacobian
\begin{align}
   \det \left[ \begin{pmatrix}
        \frac{k_{x}}{\omega}  & \frac{k_{y}}{\omega} & \frac{k_{z}}{\omega} \\
        0 & R_{11} & R_{12} \\
        0  & R_{21} & R_{22}
    \end{pmatrix}^{-1} \right] = \left( \frac{k_{x}}{\omega} \right)^{-1} = \left( s_{\theta} c_{\phi} \right)^{-1} \,.
\end{align}
Moreover, we can expand $ \vert \bm{r} - \bm{r}^{\prime} \vert$,
\begin{align}
    \vert \bm{r} - \bm{r}^{\prime} \vert^{2} & = (y-y^{\prime})^{2} + (z-z^{\prime})^{2} + (x-x^{\prime})^{2} \nonumber \\
    & = A y^{\prime 2} + B z^{\prime 2} + 2 C y^{\prime} z^{\prime} + 2 D y^{\prime} + 2 E z^{\prime} + F \,,
\end{align}
where
\begin{equation}
    \begin{aligned}
    A & = 1 + \tan^{2} \phi \, , \hspace{1cm}
    B = 1 + \cot^{2} \theta \sec^{2} \phi \, , 
    \hspace{1cm}
    C = \cos \theta \sec \phi \tan \phi \,, \\
    D & = - y + ( x -  \ell^{\prime} \csc \theta \sec \phi ) \tan \phi \,, \hspace{1cm}
    E  = - z + ( x -  \ell^{\prime} \csc \theta \sec \phi ) \cot \theta \sec \phi \, , \\
    F & = x^{2} + y^{2} + z^{2} - 2 x  \ell^{\prime} \csc \theta \sec \phi + \ell^{\prime} \csc^{2} \theta \sec^{2} \phi  \,.
    \end{aligned}
\end{equation}
Using the SO(2) rotated variables, the expression can be diagonalized:
\begin{equation}
    \begin{aligned}
       \vert \bm{r} - \bm{r}^{\prime} \vert^{2}    & = \lambda_{1} \tilde{y}^{\prime 2} + \lambda_{2} \tilde{z}^{\prime 2} + 2 \tilde{D} \tilde{y}^{\prime} + 2 \tilde{E} \tilde{z}^{\prime} + F \\
       & = \lambda_{1} \left( \tilde{y}^{\prime} - \frac{\tilde{D}}{\lambda_{1}} \right)^{2} + \lambda_{2}  \left( \tilde{z}^{\prime} - \frac{\tilde{E}}{\lambda_{2}} \right)^{2} + F - \frac{\tilde{D}^{2}}{ \lambda_{1}} - \frac{\tilde{E}^{2}}{ \lambda_{2}}
\end{aligned}
\end{equation}
where $ \lambda_{1} = \csc^{2} \theta \sec^{2} \phi $, $ \lambda_{2} = 1 $, $\tilde{D} = R_{11} D + R_{12} E$ and $ \tilde{E} = R_{22} E + R_{21} D $.
Also,
\begin{align}
    F - \frac{\tilde{D}^{2}}{ \lambda_{1}} - \frac{\tilde{E}^{2}}{ \lambda_{2}} = \left( x \cos \phi \sin \theta + y \sin \theta \sin \phi +  z \cos \theta  - \ell^{\prime} \right)^{2} = \left( \hat{\bm{k}} \cdot \bm{r} - \ell^{\prime} \right)^{2}  \,.
    %= \left( \hat{\bm{k}} \cdot \bm{r} - \hat{\bm{k}} \cdot \bm{r}^{\prime}  \right)^{2}
\end{align}
Now, from this expression with new variables, integration along the $ \tilde y' \tilde z' $-plane can be done as
\begin{equation}
\begin{aligned}
    & \int  d\ell^{\prime} \int_{-\infty}^{\infty} d\tilde{y}^{\prime} d\tilde{z}^{\prime} ~\frac{ B_{0} }{{s_{\theta} c_{\phi}} } e^{i \omega \ell^{\prime}} \frac{e^{i \omega \vert \bm{r} - \bm{r}^{\prime} \vert} }{ 4 \pi \vert \bm{r} - \bm{r}^{\prime} \vert }  \\
     & = \int  d\ell^{\prime}   \int_{-\infty}^{\infty} d\tilde{y}^{\prime} d\tilde{z}^{\prime} ~\frac{ B_{0} }{{s_{\theta} c_{\phi}} }  e^{i \omega \ell^{\prime}} \frac{e^{i \omega \sqrt{   \lambda_{1} \tilde{y}^{\prime 2} +  \lambda_{2} \tilde{z}^{\prime 2} +  ( \bm{k} \cdot \bm{r} - \omega \ell^{\prime} )^{2}  } } }{ 4 \pi \sqrt{   \lambda_{1} \tilde{y}^{\prime 2} +  \lambda_{2} \tilde{z}^{\prime 2} +  ( \bm{k} \cdot \bm{r} - \omega \ell^{\prime} )^{2}  } }  \\
      & = - \int d\ell^{\prime}   ~B_{0}(\ell^{\prime} s_{\theta} c_{\phi} ) e^{i \omega \ell^{\prime}}  \frac{e^{i \vert \bm{k} \cdot \bm{r} - \omega \ell^{\prime} \vert  }}{  2 i \omega } \,,
\end{aligned}
\end{equation}
yielding
\begin{equation}
\begin{aligned}
    \bm{A}_{h} 
    &  = 
-\frac{e^{- i \omega t}}{2\sqrt{2}}  s_{\theta} \int_{\ell_{1}}^{\ell_{2}} d\ell^{\prime} ~ \left[  \left(  h_{+} \bm{v} - h_{\times} \bm{u}  \right)  B_{0}(\ell^{\prime} s_{\theta} c_{\phi})  \right. \\
& \hspace{4.5cm} \left. - \frac{1}{i \omega}  \hat{\bm{n}} \times \left(  h_{+} \bm{u} + h_{\times} \bm{u}  \right)  B_{0}^{\prime}(\ell^{\prime} s_{\theta} c_{\phi})  \right]
  e^{i \omega \ell^{\prime}}  e^{i \vert \bm{k} \cdot \bm{r} - \omega \ell^{\prime} \vert  } \,. 
\end{aligned}    
\end{equation}
For the transmitted wave this gives
\begin{equation}
\begin{aligned}
    \bm{A}_{h}^{T} 
    &  = 
-\frac{e^{- i (\omega t - \bm{k} \cdot \bm{r} ) } }{2\sqrt{2}}  s_{\theta} \int_{\ell_{1}}^{\ell_{2}} d\ell^{\prime}  \left[  \left(  h_{+} \bm{v} - h_{\times} \bm{u}  \right)  B_{0}(\ell^{\prime} s_{\theta} c_{\phi})  - \frac{1}{i \omega}  \hat{\bm{n}} \times \left(  h_{+} \bm{u} + h_{\times} \bm{u}  \right)  B_{0}^{\prime}(\ell^{\prime} s_{\theta} c_{\phi})  \right] 
\\
&  = 
-\frac{e^{- i (\omega t - \bm{k} \cdot \bm{r}) }}{2\sqrt{2}}  s_{\theta} \left(  h_{+} \bm{v} - h_{\times} \bm{u}  \right)  \int_{x_{1}}^{x_{2}} \frac{dx^{\prime}}{s_{\theta} c_{\phi}} ~ B_{0}(x^{\prime}) \\ 
 &  = 
-\frac{e^{- i (\omega t - \bm{k} \cdot \bm{r})}}{2\sqrt{2}}   c_{\phi}^{-1}  \left(  h_{+} \bm{v} - h_{\times} \bm{u}  \right) \left( \int_{x_{1}}^{x_{2}} dx^{\prime} \, B_{0}(x^{\prime}) \right) \,,
\end{aligned}    
\end{equation}
while for the reflected wave we obtain
\begin{equation}
    \begin{aligned}
    \bm{A}_{h}^{R} 
    &  = 
    -\frac{e^{-i \omega t}}{2\sqrt{2}}  s_{\theta} \int_{\ell_{1}}^{\ell_{2}} d\ell^{\prime}  \left[  \left(  h_{+} \bm{v} - h_{\times} \bm{u}  \right)  B_{0}(\ell^{\prime} s_{\theta} c_{\phi})  - \frac{1}{i \omega}  \hat{\bm{n}} \times \left(  h_{+} \bm{u} + h_{\times} \bm{u}  \right)  B_{0}^{\prime}(\ell^{\prime} s_{\theta} c_{\phi})  \right]  e^{i \bar{\bm{k}} \cdot ( \bm{r} - \bm{r}^{\prime} )}   e^{ i \bm{k} \cdot {\bm r}^{\prime} }  \\ 
     &  = 
    -\frac{e^{- i (\omega t - \bar{\bm{k}} \cdot \bm{r}) }}{2\sqrt{2}}  s_{\theta}  \int_{x_{1}}^{x_{2}} \frac{dx^{\prime}}{s_{\theta} c_{\phi}} \left[  \left(  h_{+} \bm{v} - h_{\times} \bm{u}  \right)  B_{0}(x^{\prime})  - \frac{1}{i \omega}  \hat{\bm{n}} \times \left(  h_{+} \bm{u} + h_{\times} \bm{u}  \right)  B_{0}^{\prime}(x^{\prime})  \right] e^{i (\bar{\bm k} - \bm k) \cdot \bm{r}^{\prime}}  \\ 
     &  = 
    -\frac{e^{- i (\omega t - \bar{\bm{k}} \cdot \bm{r}) }}{2\sqrt{2}}  s_{\theta}
    \left[  \left(  h_{+} \bm{v} - h_{\times} \bm{u}  \right)  + 2 s_{\theta} c_{\phi}   \hat{\bm{n}} \times \left(  h_{+} \bm{u} + h_{\times} \bm{u}  \right)  \right] \int_{x_{1}}^{x_{2}} \frac{dx^{\prime}}{s_{\theta} c_{\phi}} ~  B_{0}(x^{\prime}) e^{i (\bar{\bm k} - \bm k) \cdot \bm{r}^{\prime}} 
    \\ 
     &  = 
    -\frac{e^{- i (\omega t - \bar{\bm{k}} \cdot \bm{r})}}{2\sqrt{2} }   c_{\phi}^{-1} \left[  \left(  h_{+} \bm{v} - h_{\times} \bm{u}  \right)  + 2 s_{\theta} c_{\phi} \hat{\bm{n}} \times \left(  h_{+} \bm{u} + h_{\times} \bm{u}  \right)  \right] \int_{x_{1}}^{x_{2}} dx^{\prime} ~  B_{0}(x^{\prime}) e^{2 i \omega s_{\theta} c_{\phi} x^{\prime}}  \,.
    \end{aligned}
\end{equation}

\section{$S$-matrix approach}
\label{app:S-matrix}

In this approach, the Gertsenshtein effect arises from the quantum scattering of a graviton into a photon in the presence of an external magnetic field, denoted by ${\bm B}_0$. For a comprehensive discussion, we refer the reader to Ref.~\cite{DeLogi:1977qe}. Upon canonically normalizing the graviton field, its coupling to the electromagnetic field occurs via the energy-momentum tensor
\begin{equation}
\begin{aligned}
\label{eq:LSmat}
\mathcal{L} &= \sqrt{8\pi G} \, h_{\mu\nu} T^{\mu\nu}  \\
            &= \sqrt{8\pi G} \, h_{ij} T^{ij}  \\
            &\supset \sqrt{32\pi G} \, B_0^i B_h^j h_{ij},
\end{aligned}
\end{equation}
where in the second line we have adopted the transverse-traceless gauge, and in the third line we have used the decomposition of the magnetic field as ${\bm B} = {\bm B}_0 + {\bm B}_h$, isolating the cross term proportional to $B_0 B_h$ resulting from  
\begin{align}
    T^{ij} = E^i E^j + B^i B^j - \frac{1}{2} (E^2 + B^2) \delta^{ij} \,,
\end{align}
which describes the graviton-photon conversion. 
This technique was employed in the seminal work of Raffelt and Stodolsky~\cite{Raffelt:1987im} to determine the conversion probability. While equivalent to the approach based on the effective current, the $S$-matrix formalism differs in that it does not provide a local solution for the induced electromagnetic field. Instead, it yields the probability amplitude relevant at asymptotic distances. 
Specifically, in the $S$-matrix approach, one evaluates the scattering amplitude associated with the conversion process using standard perturbation-theory techniques from quantum field theory. Applying this to Eq.~\eqref{eq:LSmat}, the $S$-matrix element takes the form
\begin{equation}
\mathcal{S} = - \frac{i}{4\pi} \sqrt{32 \pi G} \, \tilde{B}_0^i(\bm{q})\, B_h^j(\bm{k}_\gamma)\, h_{ij}(\bm{k}),
\end{equation}
where $\tilde{B}_0^i$ denotes the Fourier transform of the background magnetic field, as discussed in the main text. For a similar discussion for the case of axions, see e.g.\ Ref.~\cite{Raffelt:1987np}. Here, $\bm{q} = \bm{k}_\gamma - \bm{k}$ is the momentum transfer, $\bm{B}_h = \nabla \times \bm{A}_h$, while $\bm{A}_h (\bm{k}_\gamma)$ and $h_{ij}(\bm{k})$ correspond to the out- and in-states of standard scattering theory. Concretely, 
the scattering amplitude for producing a photon with polarization $(c_U,c_V)$ from a graviton polarized as $(c_+,c_\times)$\footnote{Note that  $|c_U|^2+|c_V|^2 =1$ and $|c_+|^2+|c_\times|^2=1$. In the main text $c_+ = h_+/(|h_+|^2+|h_\times|^2)$ and similarly for $h_\times$ . }
can be cast as
\begin{align}
\mathcal{S} &= \omega \sqrt{\frac{2 G}{\pi}} 
\left(
\begin{array}{cc}
 c_U^* & c_V^*
\end{array}
\right)
\mathcal{T}
\left(
\begin{array}{c}
 c_+ \\ c_\times
\end{array}
\right)\,,
\end{align}
with
\begin{align}
\mathcal{T} =
\frac{1}{\sqrt2}
\left(
\begin{array}{cc}
 \bm{v}_\gamma \cdot \bm{v}   & -\bm{v}_\gamma \cdot \bm{u}  \\
 - \bm{u}_\gamma \cdot \bm{v}   & \bm{u}_\gamma \cdot \bm{u}
\end{array}
\right)
\left(
\begin{array}{cc}
 \tilde{\bm{B}}_0 (\bm{q}) \cdot \bm{v} & -\tilde{\bm{B}}_0 (\bm{q}) \cdot \bm{u} \\
 \tilde{\bm{B}}_0 (\bm{q}) \cdot \bm{u}  & \tilde{\bm{B}}_0 (\bm{q}) \cdot \bm{v}
\end{array}
\right)\,.
\label{eq:Tmat}
\end{align}
This amplitude leads to the following expression for the differential cross section of the conversion process:
\begin{equation}
\frac{d\sigma}{d\Omega} = |\mathcal{S}|^2 = \frac{2 G \omega^2}{\pi} 
\left(
\begin{array}{cc}
 c_U^* & c_V^*
\end{array}
\right)
\mathcal{T}
\left(
\begin{array}{cc}
 |c_+|^2 & c_+ c_\times^* \\
 c_\times c_+^* & |c_\times|^2
\end{array}
\right)
\mathcal{T}^T
\left(
\begin{array}{c}
 c_U \\ c_V
\end{array}
\right).
\label{eq:dsigmadOmegaS}
\end{equation}
This formulation admits a clear physical interpretation. First, for an incoming plane gravitational wave, photons are emitted in all directions, with an angular distribution described by the differential cross section above.
Moreover, if the gravitational wave is in a pure polarization state, then the polarization of the resulting photon in the $(\bm{u}_{\gamma}, \bm{v}_{\gamma})$ basis is proportional to
\begin{equation}
\mathcal{T}
\left(
\begin{array}{c}
 h_+ \\ h_\times
\end{array}
\right).
\label{eq:ThAppH}
\end{equation}
This result is consistent with the findings presented in the main text via the effective-current technique. This can be proven by noting the interaction Lagrangian can be recast in the form ${\cal L} \supset A_\mu j_\text{eff}^\mu + \ldots$, which follows from Eq.~\eqref{eq:LSmat}, integrating by parts \cite{Aggarwal:2025noe}.

\section{Comparison to Axions}
\label{app:axion}

Much of the discussion in the main text can equally be applied to the case of relativistic axions \cite{Raffelt:1987im}, e.g.\ axions produced in the sun \cite{Caputo:2024oqc},  in light-shining-through-the-wall experiments \cite{Sikivie:1983ip,Graham:2015ouw,CAST:2017uph,IAXO:2019mpb}, or from the cosmic axion background \cite{Dror:2021nyr}. The resulting phenomenology is well known, this appendix serves to demonstrate that the formalism developed here for GWs correctly reproduces these results.

We consider axion with the coupling $ \mathcal{L} \ni g_{a\gamma\gamma} a F^{\mu\nu} \tilde{F}_{\mu\nu}  $ which gives an effective current
\begin{align}
    j_{\rm eff}^{\mu} = g_{a\gamma\gamma} \partial_{\nu} \left( a \tilde{\bar{F}}^{\nu\mu} \right)  && \Rightarrow &&
\begin{dcases}
     \bm{j}_{\rm eff} = g_{a\gamma\gamma} \dot{a} \bm{B}_{0} & (\mu = i) \\ \rho_{\rm eff} = - g_{a\gamma\gamma} (\nabla a ) \cdot \bm{B}_{0} & (\mu = 0) 
\end{dcases} \,,
\end{align}
where we assume a static background magnetic field.
Note that there exists an effective charge, which is absent for GW. 
For simplicity, we assume monochromatic axion field with momentum $ \bm{k} $:
\begin{align}
    a(t, {\bm r}) = a_{0} e^{-i (\omega t - \bm{k}\cdot\bm{r} ) } \,,
\end{align}
which gives
\begin{align}
    \bm{J}_{\rm eff} \equiv \bm{j}_{\rm eff}  e^{i (\omega t - \bm{k}\cdot\bm{r} ) }  = -i \omega g_{a\gamma\gamma} a_{0} \bm{B}_{0} \,.
\end{align}
In a similar fashion, we can define
\begin{align}
    \varrho_{\rm eff}  \equiv \rho_{\rm eff} e^{i (\omega t - \bm{k}\cdot\bm{r} ) } = - i \omega g_{a\gamma\gamma} a_{0} (\hat{\bm{k}} \cdot \bm{B}_{0}) \,.
\end{align}

\subsection{Magnetic Domain}

For the magnetic domain wall example discussed in Sec.~\ref{sec:Magnetic_Domain_Model} with magnetic field as in Eq.~\eqref{eq:magnetic_domain}, we have
\begin{equation}
    \begin{aligned}
    \rho_{\rm eff} & = - i g_{a\gamma\gamma} \omega a_{0} B_{0} \cos \theta e^{ -i (\omega t - \bm{k} \cdot \bm{r} ) }  [\Theta(x-x_{1}) - \Theta(x-x_{2})] , \\ 
    {\bm j}_{\rm eff} & = - i g_{a\gamma\gamma} \omega a_{0} B_{0}  e^{ -i (\omega t - \bm{k} \cdot \bm{r} ) }  [\Theta(x-x_{1}) - \Theta(x-x_{2})]  \hat{\bm{e}}_{z} \,,
    \end{aligned}
\end{equation}
with no surface current, i.e. $ \bm{K}_{1} = \bm{K}_{2} = 0  $ in Eq.~\eqref{eq:j_eff_magnetic_domain}.

The following discussion is analogous to the case of GW, except that we have to take into account the presence of the scalar potential $ \phi_{a} \equiv A_{a}^{0} $ satisfying $ \Box \phi_{a} = -\rho_{\rm eff} $, not only vector potential $ \bm{A}_{a} $, with boundary conditions imposing (i) continuity of function itself and the first derivatives at the boundaries of each domain due to the absence of the surface charge/current, and (ii) the absence of the left and right moving modes in Region I and Region III, respectively, dictated by the causality.

Following the steps outlined in Sec.~\ref{sec:Magnetic_Domain_Model} (detailed in App.~\ref{app:Derivations of Magnetic Domain Wall Model} for the  GW case), we obtain for the transmitted wave
\begin{equation}
    \begin{aligned}
    \phi_{a}^{\rm (III)} & = \frac{1}{2} g_{a\gamma\gamma} B_{0} \cot \theta \sec \phi (x_{2} - x_{1} )  e^{- i (\omega t - \bm{k} \cdot \bm{r} ) } \,, \\
    \bm{A}_{a}^{\rm (III)} & = \frac{1}{2} g_{a\gamma\gamma} a_{0} B_{0} \csc \theta \sec \phi (x_{2} - x_{1})     e^{- i (\omega t - \bm{k} \cdot \bm{r} ) } \hat{\bm{e}}_{z} \,, 
    \end{aligned}
\end{equation}
and for the reflected wave
\begin{equation}
    \begin{aligned}
    \phi_{a}^{\rm (I)} & = - \frac{i}{4 \omega} g_{a\gamma\gamma} B_{0} \cot \theta  \csc \theta \sec^{2} \phi \left( e^{2 i x_{2}  \bar{\bm{k}} \cdot \bm{n} }  - e^{2 i x_{1} \bar{\bm{k}} \cdot \bm{n} }  \right)  e^{- i (\omega t - \bar{ \bm{k} } \cdot \bm{r} ) } \,, \\
    \bm{A}_{a}^{\rm (I)} & = - \frac{i}{4 \omega} g_{a\gamma\gamma} B_{0}  \csc^{2} \theta \sec^{2} \phi \left( e^{2 i x_{2}  \bar{\bm{k}} \cdot \bm{n} }  - e^{2 i x_{1} \bar{\bm{k}} \cdot \bm{n} }  \right)    e^{- i (\omega t - \bar{ \bm{k} } \cdot \bm{r} ) } \hat{\bm{e}}_{z} \,.
    \end{aligned}
\end{equation}

In terms of electric/magnetic fields, %this yields,
\begin{equation}
    \begin{aligned}
    \bm{E}_{a}^{\rm (III)} = - \frac{i}{2}  \omega g_{a\gamma\gamma} a_{0} B_{0} \sec \phi (x_{2} - x_{1} )  e^{- i (\omega t - \bm{k} \cdot \bm{r} ) } \bm{u} \,, \\
    \bm{B}_{a}^{\rm (III)} = - \frac{i}{2}  \omega g_{a\gamma\gamma} a_{0} B_{0} \sec \phi (x_{2} - x_{1} )  e^{- i (\omega t - \bm{k} \cdot \bm{r} ) } \bm{v}  \,,
    \end{aligned}
\end{equation}
and
\begin{equation}
    \begin{aligned}
    \bm{E}_{a}^{\rm (I)} = - \frac{1}{4} g_{a\gamma\gamma} a_{0} B_{0}  \csc\theta \sec^{2} \phi  \left( e^{2 i x_{2}  \bar{\bm{k}} \cdot \bm{n} }  - e^{2 i x_{1} \bar{\bm{k}} \cdot \bm{n} }  \right) e^{- i (\omega t - \bar{\bm{k}} \cdot \bm{r} ) } \bar{\bm{u}} \,, \\
    \bm{B}_{a}^{\rm (I)} = - \frac{1}{4}  g_{a\gamma\gamma} a_{0} B_{0} \csc\theta \sec^{2} \phi  \left( e^{2 i x_{2}  \bar{\bm{k}} \cdot \bm{n} }  - e^{2 i x_{1} \bar{\bm{k}} \cdot \bm{n} }  \right)  e^{- i (\omega t - \bar{\bm{k}} \cdot \bm{r} ) } \bar{\bm{v}}  \,.
    \end{aligned}
\end{equation}
Note that in the GW case, because there is no scalar potential, the direction of the electric fields and the vector potential are the same in Coulomb gauge, in contrast to the axion case considered here.
Also, from the fact that the direction of the electric/magnetic fields are completely determined by the direction of the GW wave vector and the magnetic field, transmitted/reflected waves are always maximally polarized.

Finally, the corresponding intensities are
\begin{align}
    \mathcal{I}_{\gamma}^{T} = \frac{1}{4} \omega^{2} g_{a\gamma\gamma}^{2} a_{0}^{2} B_{0}^{2} L^{2}  \sec^{2} \phi \,,  && \mathcal{I}_{\gamma}^{R} = \frac{1}{4}  g_{a\gamma\gamma}^{2} a_{0}^{2} B_{0}^{2} \csc^{2} \theta \sec^{4} \phi \sin^{2} ( \omega L c_{\phi} s_{\theta} )  \,,
\end{align}
where we set $ x_{2} - x_{1} \equiv L$. Taking into account the intensity of the massless axion $  \mathcal{I}_{a} =  \omega^{2} a_{0}^{2} / 2  $, this reproduces the axion-photon conversion probability $ \mathcal{P}_{a \rightarrow \gamma}^{T} = \mathcal{I}_{\gamma}^{T} / \mathcal{I}_{a}  \propto B_{0}^{2} L^{2} $ for transmitted waves. Moreover, there exists reflected waves with an interference pattern similar to GW case shown in the main text.

\subsection{Localized Source}

As in Eq.~\eqref{eq:A_GF_sol}, we obtain
\begin{equation}
    \begin{aligned}
    {\bm A}_{a}({\bm r} , t)  \simeq \frac{e^{-i \omega (t - r)}}{4 \pi r} \tilde{\bm{J}}_{\text{eff} } ( \bm{k}_{\gamma} -  \bm k)\, , &&  \phi_{a}({\bm r} , t) = 
      {A}_{a}^{0}({\bm r} , t)   \simeq \frac{e^{-i \omega (t - r)}}{4 \pi r} \tilde{\varrho}_{\text{eff} } ( \bm{k}_{\gamma} -  \bm k)\, ,
    \end{aligned}
\end{equation}
implying
\begin{align}
    \bm{E}_{a} = -\frac{\partial \bm{A}_{a}}{\partial t} - \nabla \phi_{a} \simeq \frac{e^{-i \omega (t - r)} }{4\pi r} \omega^{2} g_{a\gamma\gamma} a_{0} \left( \tilde{\bm{B}} -  \hat{\bm{k}}_{\gamma} (\hat{ \bm{k}} \cdot \tilde{\bm{B}}) \right)  \,,
\end{align}
where we use the far-field approximation and neglect terms proportional to $r^{-2}$. As a consistency check, $
    \bm{k}_{\gamma} \cdot \bm{E}_{a} \propto ( \bm{k}_{\gamma} - \bm{k} ) \cdot \tilde{\bm{B}}  (\bm{k} - \bm{k}_{\gamma} ) = \bm{q} \cdot \tilde{\bm{B}} (\bm{q}) = 0
$, due to $ \nabla \cdot \bm{B} = 0 $.
In the following, let us discuss the implications of this result for the case of a dipolar magnetic field, exposed to a monochromatic axion wave and a stochastic axion background.

\begin{figure}
    \centering
    \includegraphics[width=0.5\linewidth]{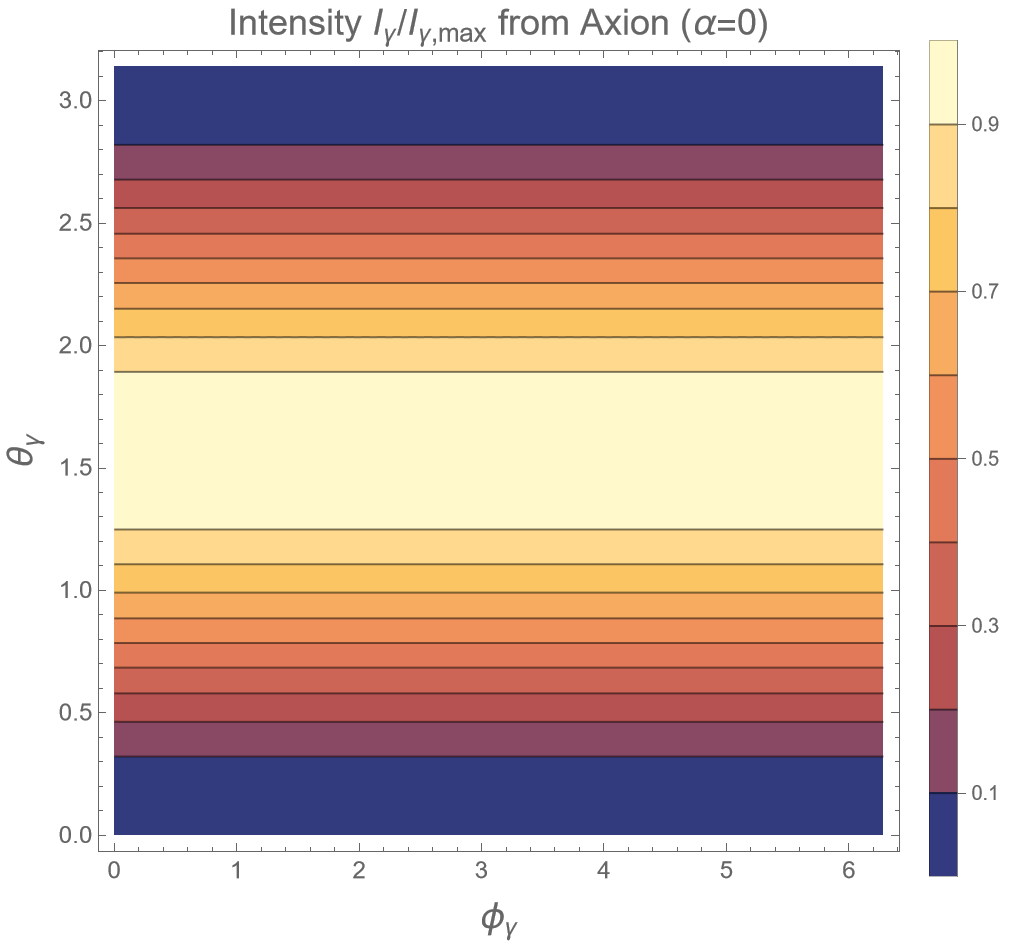}
    \includegraphics[width=0.45\linewidth]{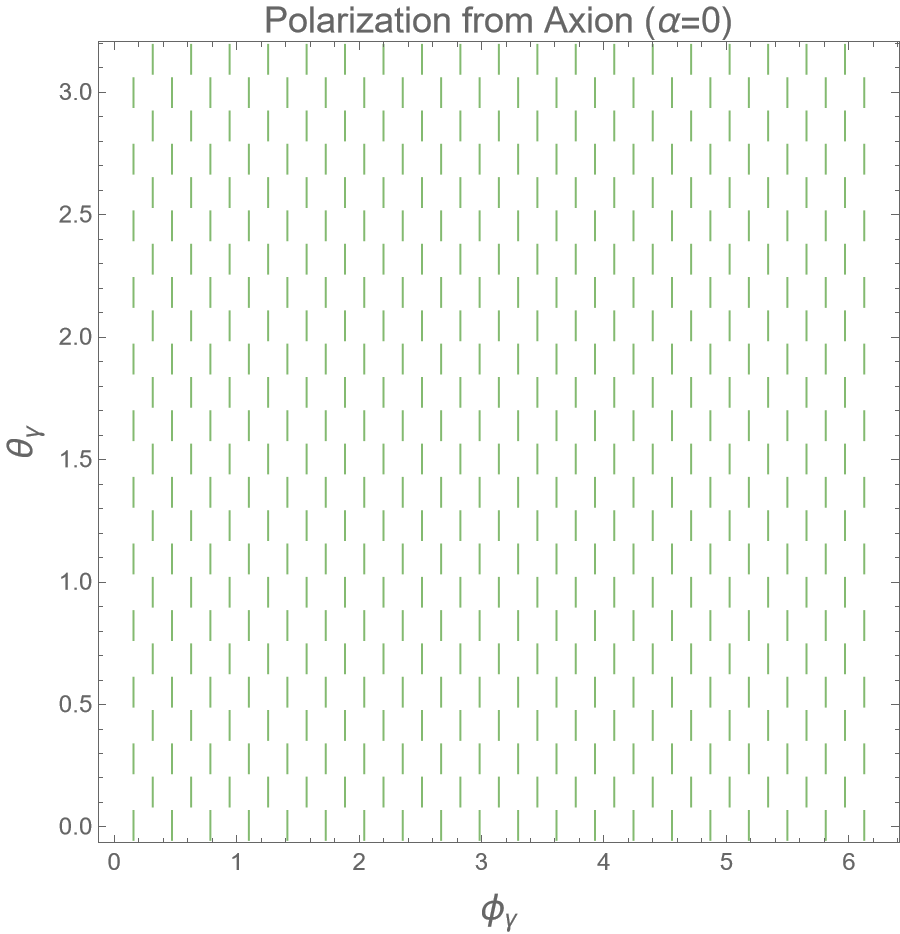}

    \includegraphics[width=0.5\linewidth]{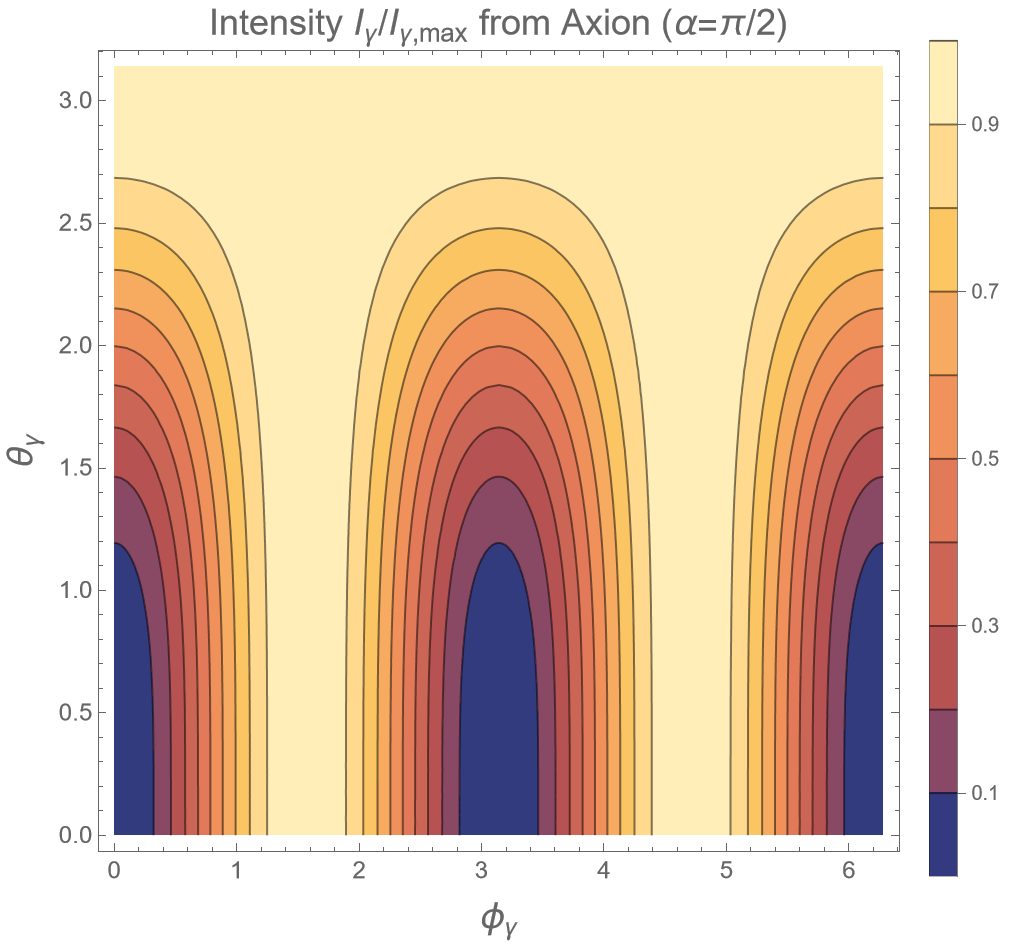}
    \includegraphics[width=0.45\linewidth]{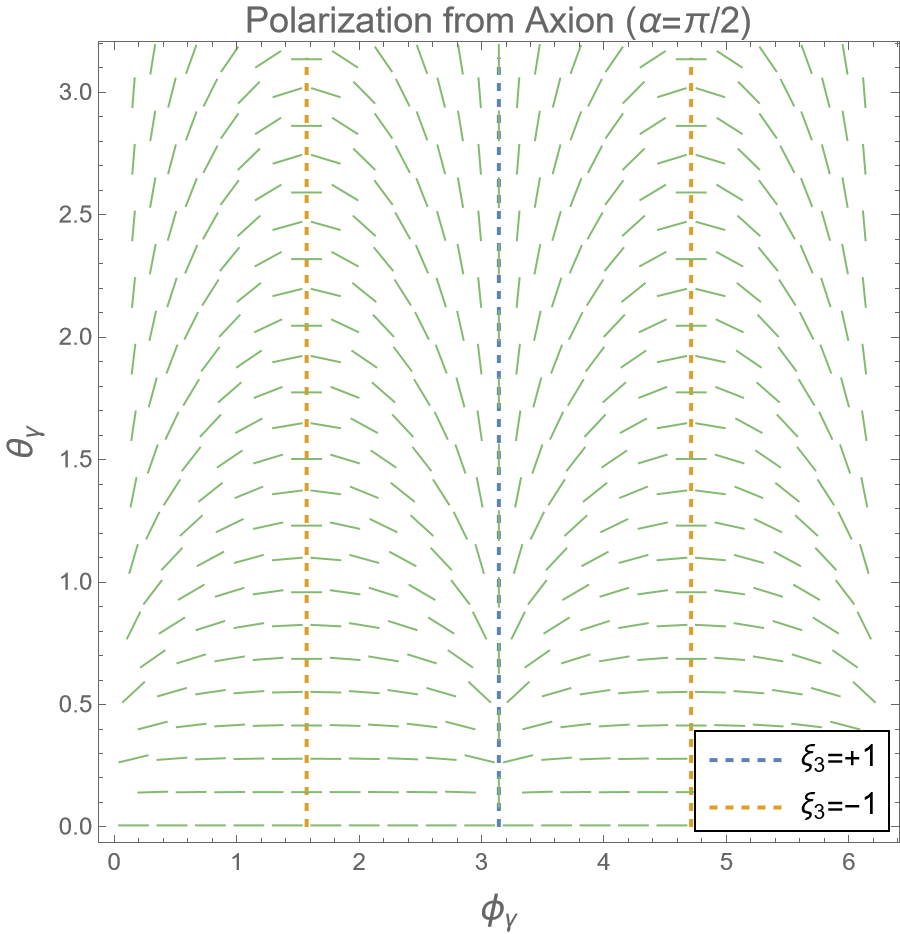}
    
    \caption{ Intensity $ \mathcal{I}_{\gamma} / \mathcal{I}_{\gamma,\max} $ (left) and polarization (right) of the induced EM fields from a monochromatic massless axion wave with $ \alpha  = 0 $ (top) and $ \alpha = \pi/ 2$ (bottom), respectively, with axion wave propagating in the $ \hat{\bm{e}}_{z}$ direction.} 
    \label{fig:dipole_axion}
\end{figure}

\paragraph{Monochromatic Axion Wave.}
Setting $ \bm{m} = m(\sin\alpha , 0 , \cos \alpha) $, $ \hat{ \bm k} = (0,0,1) $, and $ \hat{\bm k}_{\gamma } = (\sin \theta_{\gamma} \cos \phi_{\gamma}, \sin \theta_{\gamma} \sin \phi_{\gamma}, \cos \theta_{\gamma} ) $, for monochromatic axion, we have intensity
\begin{align}
    \mathcal{I}_{\gamma}=  \frac{1}{64 \pi^{2} r^{2} } a_{0}^{2} g_{a\gamma\gamma}^{2} m^{2} \omega^{4} \left[ \left( 2 s_{\theta_{\gamma}/2}^{2} c_{\phi_{\gamma}} s_{\alpha} + c_{\alpha} s_{\theta_{\gamma}} \right)^{2} + 4 s_{\alpha}^{2} s_{\phi_{\gamma}}^{2} \right]
\end{align}
and Stokes parameters
\begin{equation}
\begin{aligned}
    \xi_{1} & = \frac{ 2 s_{2\alpha} s_{\theta_{\gamma}} s_{\phi_{\gamma}} + 4 s_{\alpha}^{2} s_{\theta_{\gamma}/2}^{2} s_{2\phi_{\gamma}}  }{ ( 2 s_{\theta_{\gamma}/2}^{2} c_{\phi_{\gamma}} s_{\alpha} + c_{\alpha} s_{\theta_{\gamma}} )^{2} + 4 s_{\alpha}^{2} s_{\phi_{\gamma}}^{2}  } = \begin{dcases}
        0 & (\alpha = 0 ) \\
        \frac{s_{\theta_{\gamma}/2}^{2} s_{2\phi_{\gamma}}}{c_{\phi_{\gamma}}^{2} s_{\theta_{\gamma}/2}^{2} + s_{\phi_{\gamma}}^{2}  }& (\alpha = \pi/ 2)
    \end{dcases} \,, 
    \\ \xi_{3} & = \frac{ (2c_{\phi_{\gamma}} s_{\alpha} s_{\theta_{\gamma}/2}^{2} + c_{\alpha} s_{\theta_{\gamma}})^{2} - 4 s_{\alpha}^{2} s_{\phi_{\gamma}}^{2}  }{ ( 2 s_{\theta_{\gamma}/2}^{2} c_{\phi_{\gamma}} s_{\alpha} + c_{\alpha} s_{\theta_{\gamma}} )^{2} + 4 s_{\alpha}^{2} s_{\phi_{\gamma}}^{2}  } =
    \begin{dcases}
        1 & (\alpha = 0 )\\
 \frac{ c_{\phi_{\gamma}}^{2} s_{\theta_{\gamma}/2}^{2} - s_{\phi_{\gamma}}^{2}  }{ c_{\phi_{\gamma}}^{2} s_{\theta_{\gamma}/2}^{2} + s_{\phi_{\gamma}}^{2}  } & (\alpha = \pi/2)
    \end{dcases} \,.
\end{aligned}
\end{equation}
In Fig.~\ref{fig:dipole_axion}, we depict the intensity and the polarization map of the induced EM waves from the monochromatic massless axion wave.

\paragraph{Stochastic Axion Background.}
For this consideration, let us take $ \bm{m} = m (\cos \alpha,0, \sin \alpha) $ as before, but this time we fix the direction of the scattered photon $ \hat{\bm k}_{\gamma} = (0,0,1) $. 
Assuming an isotropic axion background,
\begin{align}
     \int d\Omega \,   \langle {E_i E_j} \rangle = \frac{m^{2} \omega^{4}}{192 \pi r^{2}} a_{0}^{2} g_{a\gamma\gamma}^{2} \begin{pmatrix}
        17 - 13 c_{2 \alpha} & 0 \\
        0 & 3 + c_{2 \alpha}  
    \end{pmatrix} \,, \nonumber
\end{align}
to obtain the following results for the differential cross section
\begin{align}
    \frac{d\sigma}{d\Omega}  \equiv r^{2} \frac{ \mathcal{I}_{\gamma}}{ \mathcal{I}_{\rm axion} }  = \frac{g_{a\gamma\gamma}^{2} m^{2} \omega^{2} }{24\pi} (5 - 3 \cos 2 \alpha)   && \text{(dipole, isotropic axion)} \,,
\end{align}
where we used $  \mathcal{I}_{\rm axion} =  \omega^{2} a_{0}^{2} / 2  $.

\begin{figure}
    \centering
    \includegraphics[width=0.5\linewidth]{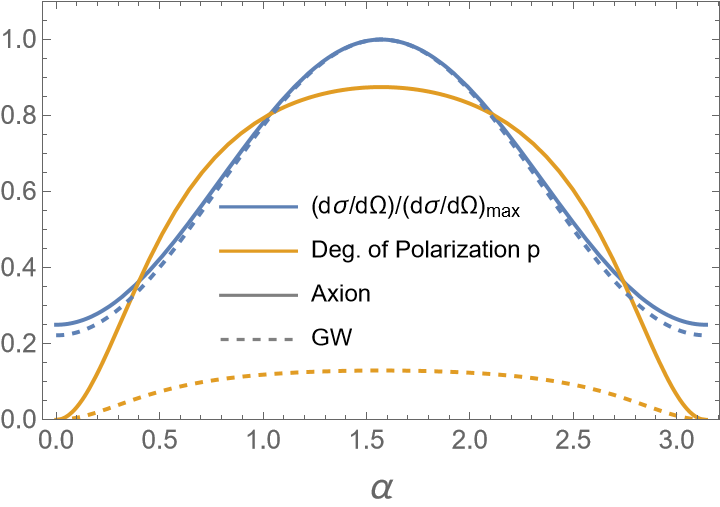}
    \caption{The differential cross section normalized by its maximum value (blue) and the degree of polarization (orange) of the scattered photon from the dipole under the presence of the isotropic massless axion background. Dotted lines are the ones for gravitational waves, see Fig.~\ref{fig:SGWB_dipole}.}
    \label{fig:SaB}
\end{figure}

The Stokes parameters and the degree of polarization become
\begin{align}
    \bm{\xi} = \left( 0 \,, 0 \,, \frac{ 7 \sin^{2} \alpha }{ 5 - 3 \cos 2 \alpha } \right) && \text{(dipole, isotropic axion)}
\end{align}
and the degree of polarization becomes $p = (\sum_{i=1}^{3} \xi_{i}^{2})^{1/2}  = \vert \xi_{3} \vert$ which has the maximum value $ 7/8 \approx 0.875$ at $ \alpha = \pi / 2$, i.e. when the axis of the dipole is perpendicular to the direction of the observer.
See Fig.~\ref{fig:SaB} for the differential cross section and the residual degree of polarization from the isotropic axion background, compared to the case of GW.

\vspace{3mm}

This discussion demonstrates the use of the Green's function formalism to the case of relativistic axions, highlighting some key similarities between GWs and axions: in both cases, the EM waves generated from a magnetized region exposed to an isotropic GW or axion background show specific angular power and polarization distributions, depending on the structure of the magnetic field. On the other hand, the spin~2 nature of the GW as opposed to the spin-0 axion also leads to key differences. For example, for a fixed B-field direction, the axion can only couple to photon polarization, whereas a GW generally couples to both.

\section{WKB Approximation}
\label{app:WKB}

In this appendix, we will discuss the WKB solution of the wave equation, with a slight generalization to the case with a source term.
In the main text, the results are used for the case when there exists a thermal mass of the photon which adiabatically changes depending on the background magnetic fields.

The equation we want to solve is the 1D wave equation with varying momentum:
\begin{align}
 \psi^{\prime\prime} +
 k^{2}(x) \psi = Q(x)
 \label{eq:wave1D}
\end{align}
where $ Q(x) $ is an arbitrary, but slowly varying source term.
The homogeneous solution can be obtained approximately using WKB approximation 
\begin{align}
    \psi_{\rm \pm} \approx A \exp \left( \pm i \int_{x_{0}}^{x} dx^{\prime} ~ k(x^{\prime}) \right)  \,.
\end{align}
This approximation holds when $ k(x) $ changes adiabatically, i.e.
\begin{align}
    \left\vert \frac{k^{\prime}(x)}{ k^{2} (x) } \right\vert  \ll 1 \,.
    \label{eq:adiabaticity}
\end{align}
Choosing a delta function source term $ \delta (x-x^{\prime}) $, we can also derive the retarded Green function:
\begin{align}
  G(x-x^{\prime}) = \frac{1}{2 i k(x^{\prime})} \exp \left( i \left\vert \int_{x^{\prime}}^{x} dx^{\prime\prime} ~ k(x^{\prime\prime}) \right\vert  \right) \,.
\end{align}
With a general source term, the particular solution is obtained by the convolution of the retarded Green function with the source term
\begin{equation}
\begin{aligned}
    \psi^{(p)}(x) & = \int_{-\infty}^{x} dx^{\prime} ~  G(x-x^{\prime}) Q(x^{\prime}) \\
    & = \int_{-\infty}^{x} dx^{\prime} ~ \frac{1}{2 i k(x^{\prime}) } \exp \left( i \left\vert \int_{x^{\prime}}^{x} dx^{\prime\prime} ~ k(x^{\prime\prime}) \right\vert  \right) Q(x^{\prime}) \,. \label{eq:WKB_sol}
\end{aligned}
\end{equation}
Plugging this back into Eq.~\eqref{eq:wave1D} we have
\begin{align}
    \psi^{\prime\prime} + k^{2}(x) \psi  = Q - \frac{ k^{\prime}}{2 i k^{2}} Q + \frac{1}{2 i k} Q^{\prime} \,.
\end{align}
The smallness of the second term is assured by the assumption of the adiabaticity of the momentum~\eqref{eq:adiabaticity}.
Now, we also have a second condition for the validity of the particular solution guaranteeing the suppression of the third term of the right-hand side:
\begin{align}
    \left\vert \frac{Q^{\prime}}{Q} \right\vert \ll k \, ,
\end{align}
i.e.\ adiabaticity of the source.

\paragraph{Example: Photon with Effective Mass}

We now turn to our example of a photon with an effective mass $ \mu  \ll \omega $ discussed in Sec.~\ref{sec:Towards_including_Medium_Effects}. In this case, we can replace $ \psi \rightarrow \bm{A}_{h} $, $ k^{2} \rightarrow \omega^{2} - \mu^{2}(x) $, and $ Q \rightarrow - \bm{j}_{\rm eff} $. Therefore, we can have $ k \simeq \omega - \frac{\mu^{2}}{2 \omega} $ in the exponent assuming small mass $ \mu \ll \omega $, and also neglect $ \mu^{2} $ in the prefactor.
Hence, we obtain
\begin{equation}
    \begin{aligned}
    \bm{A}_{h}^{T} & \approx - \frac{ e^{-i \omega t} }{2\sqrt{2}} \int_{0}^{L} dx^{\prime} \, B_{0}(x^{\prime}) e^{i \omega x^{\prime}}  \exp \left[ i \int_{x^{\prime}}^{x} dx^{\prime\prime}  \left( \omega - \frac{\mu^{2}}{2\omega} \right)     \right]  \\
    & = - \frac{ e^{-i \omega (t - x) } }{2\sqrt{2}} \int_{0}^{L} dx^{\prime} B_{0}(x^{\prime}) \exp \left( -i \int_{x^{\prime}}^{L} dx^{\prime\prime}    \frac{\mu^{2}}{2\omega}   \right)  \\
    & = - \frac{ e^{-i \omega (t - x + \varphi)  } }{2\sqrt{2}} \int_{0}^{L} dx^{\prime} B_{0}(x^{\prime}) \exp \left( i \int_{0}^{x^{\prime}} dx^{\prime\prime}    \frac{\mu^{2}}{2\omega}   \right)\,,
    \end{aligned}
\end{equation}
where we restricted the region of the integration for $ x^{\prime}$ as $[0,L]$ where the magnetic field exists, and used the form of the effective current in Eq.~\eqref{eq:effective_current_mag_domain}. Also, a phase factor $ \varphi = -\int_{0}^{L} dx^{\prime\prime} \mu^{2} / (2\omega)  $ is introduced to change the range of integration for $ x^{\prime\prime} $, reflecting the phase shift due to the finite photon mass.

The conversion probability is obtained as
\begin{align}
    \mathcal{P}_{h \rightarrow \gamma}
    = 4 \pi G  \left\vert \int_{0}^{L} dx^{\prime} \,  B_{0}(x^{\prime})  \exp \left( i \int_{0}^{x^{\prime}} dx^{\prime\prime} \, \frac{\mu^{2}(x^{\prime\prime})}{2 \omega}   \right) \right\vert^{2}\,.
    %\label{eq:Ph_WKB}
\end{align}

\paragraph{Stationary Phase Approximation} 
Sometimes, to simplify the calculation, the stationary phase approximation is used. 
To make the situation explicit, let us consider the case when $ x > x^{\prime} $.
Then, the stationary phase happens when the derivative of the exponent of Eq.~\eqref{eq:WKB_sol} happens to zero, i.e. $ \mu^{2} = 0 $.

However, in general, when we use stationary phase approximation with $ f^{\prime}(x_{\rm res}) = 0 $ and $ f^{\prime\prime}(\rm res) \neq 0 $,
\begin{equation}
    \begin{aligned}
    \int dx \, g(x) e^{i f(x)}  & = \int dx \, g(x) e^{i f(x_{\rm res}) + \frac{1}{2} f^{\prime\prime}(x_{\rm res}) (x-x_{\rm res})^{2} + \cdots  } \\
    &\approx g(x_{\rm res}) e^{i f(x_{\rm res}) }  \left( \frac{2\pi}{\vert f(x_{\rm res}) } \right)^{1/2}
    \end{aligned}    
\end{equation}
the length scale of the resonance is given by $ L_{\rm res} \sim 1 / \sqrt{ f^{\prime\prime}(x_{\rm res})  } $
while we assume adiabaticity of the prefactor, meaning that the length scale given by the change of the prefactor should be much larger than $ L_{\rm res} $, $ g(x_{\rm res})/g^{\prime}(x_{\rm res}) \gg L_{\rm res} $.
In terms of the quantities of the example, this corresponds to
\begin{align}
    \left( \frac{B_{0}^{\prime}}{B_{0}} \right)^{2} \ll \left( \frac{ \mu^{2} }{2 \omega} \right)^{\prime} \,. 
\end{align}
which is Eq.~\eqref{eq:adiabativity_magnetic_field} in the main text.

\bibliographystyle{JHEP}
\bibliography{Gert}

\end{document}